\documentclass[a4paper,11pt]{article}
 \usepackage{jheppub}
 \usepackage[T1]{fontenc}
 \usepackage{lmodern}
 
\usepackage{graphicx}
\usepackage{hyperref}
\usepackage{amssymb}
\usepackage{multirow}
\usepackage{soul}
\usepackage[low-sup]{subdepth}
\usepackage{amssymb,xcolor}
\usepackage{exscale}
\usepackage{amsmath}
\usepackage{latexsym}
\usepackage{amsfonts}
\usepackage{float}
\usepackage{braket}
\usepackage{subfigure}
\usepackage{verbatim}
\usepackage{indentfirst}
\usepackage{tikz}

\usepackage{booktabs}
\usepackage{mathrsfs}
\usepackage{epsfig}
\usepackage{dcolumn}
\usepackage{bm}
\usepackage{multirow}
\usepackage{color}

\allowdisplaybreaks

\newcommand{\lsim}{{\;\raise0.3ex\hbox{$<$\kern-0.75em\raise-1.1ex\hbox{$\sim$}}\;}}
\newcommand{\gsim}{{\;\raise0.3ex\hbox{$>$\kern-0.75em\raise-1.1ex\hbox{$\sim$}}\;}}
\newcommand{\bea}{\begin{equation}}
\newcommand{\eea}{\end{equation}}
\newcommand{\beq}{\begin{equation}}
\newcommand{\eeq}{\end{equation}}
\newcommand{\ba}{\begin{array}}
\newcommand{\ea}{\end{array}}
\newcommand{\beqa}{\begin{equation}}
\newcommand{\eeqa}{\end{equation}}
\newcommand{\beqs}{\begin{subequations}}
\newcommand{\eeqs}{\end{subequations}}

\def\baa{\begin{array}}
\def\eaa{\end{array}}

\mathchardef\minus="002D

\def\dis{\displaystyle}

\newcommand{\Fr}[2]{\mbox{$\frac{\,{#1}\,}{#2}$}}
\renewcommand{\rm}{\mathrm}
\def\leqq{\leqslant}
\def\geqq{\geqslant}
\def\({\left(}
\def\){\right)}
\def\[{\left[}
\def\]{\right]}

\def\nn{\nonumber}
\def\pd{\partial}
\def\d{\rm{d}}

\def\to{\rightarrow}
\def\ito{\!\rightarrow\!}

\def\kb{\mathbf{k}}

\newcommand{\di}{\mathrm{d}}
\newcommand{\LG}{\mathcal{L}}

\newcommand{\brnc}{\overline{N^{\rm c}_{\rm R}}}
\newcommand{\rn}{N_{\rm R}}
\newcommand{\brn}{\overline{N}_{\rm R}}
\newcommand{\td}{\text{d}}

\newcommand{\hc}{\text{ H.c. }}

\def\vs{\vspace*{1mm}}

\def\hp{\hspace*{0.3mm}}

\def\hsm{\hspace*{-0.3mm}}

\def\fNL{f_{\rm{NL}}^{}}
\def\ii{\rm{i}}
\def\HH{\mathbb{H}}
\def\NR{N_{\!R}^{}}

\title{\Large Cosmological Non-Gaussianity from Neutrino Seesaw Mechanism}

\author[a]{Jingtao You,}
\author[a]{~Linghao Song,}
\author[a,b]{~Hong-Jian He,}
\author[c]{~Chengcheng Han}

\affiliation[a]{Tsung-Dao Lee Institute \& School of Physics and Astronomy, \\
Key Laboratory for Particle Astrophysics and Cosmology,\\
Shanghai Key Laboratory for Particle Physics and Cosmology,\\ 
Shanghai Jiao Tong University, Shanghai, China}
\affiliation[b]{Department of Physics, Tsinghua University, Beijing, China;\\
Center for High Energy Physics, Peking University, Beijing, China}
\affiliation[c]{Department of Physics, Sun Yat-Sen University, Guangzhou 510275, China; \\
Key Laboratory for Particle Astrophysics and Cosmology, \\ 
Shanghai Jiao Tong University, Shanghai, China; \\
Asia Pacific Center for Theoretical Physics, Pohang 37673, Korea}

\emailAdd{119760616yjt@sjtu.edu.cn}
\emailAdd{lh.song@sjtu.edu.cn}
\emailAdd{hjhe@sjtu.edu.cn}
\emailAdd{hanchch@mail.sysu.edu.cn}

\abstract{\\  
The neutrino mass generation via conventional seesaw mechanism is realized 
at high scales around $O(10^{14})\hp$GeV with natural Yukawa couplings of $O(1)$, making the test of neutrino seesaw a great challenge.\ 
It is intriguing to note that the neutrino seesaw scale is typically around 
the upper range of the cosmological inflation scale.\ 
In this work, we propose a new framework incorporating inflation and neutrino seesaw in which 
the inflaton primarily decays into right-handed neutrinos after inflation.\ 
This decay process is governed by the inflaton interaction with the right-handed neutrinos 
that respects the shift symmetry.\ 
With the neutrino seesaw mechanism, we construct 
a new realization of the Higgs modulated reheating, 
in which the fluctuations of Higgs field 
can modulate the inflaton decays and 
contribute to the primordial curvature perturbation.\
We investigate the induced non-Gaussian signatures and demonstrate, for the first time, that such signatures provide an important means to directly probe the high scale of natural neutrino seesaw.\ We further analyze the interplay of the non-Gaussianity signatures with the low-energy neutrino experiments, and their interplay with the Higgs self-coupling measurements at the LHC and future colliders.
\\[3mm]
Phys.\ Rev.\ D\,112 (2025) 083555 [\,arXiv:2412.16033\,]. 
\\[1mm]
(Its companion paper:
Phys.\ Rev.\ D\,112 (2025) L081309 (Letter) 
[\,arXiv:2412.21045\,].) 
}

\makeatletter
\def\@fpheader{\relax}
\makeatother

\date{\today}

\begin{document} 
\maketitle
\flushbottom
\setcounter{page}{2}

\section{\hspace*{-2.5mm}Introduction}
\label{sec:1}
\label{introduction}

The discovery of neutrino oscillations has pointed to tiny 
but nonzero neutrino masses of $O(0.1)\hp$eV,
which can be naturally generated by including the right-handed neutrinos 
within the established structure of the Standard Model (SM)  
of particle physics.\ 
These right-handed neutrinos are the chiral partners of the left-handed neutrinos 
and they join together the Yukawa interactions with the Higgs doublet 
(just like any other leptons and quarks in the SM).\ 
But the right-handed neutrinos are pure singlets of the SM gauge group.\ 
As such they can naturally acquire large Majorana masses ($M_R^{}$) and realize the seesaw mechanism\,\cite{Minkowski:1977sc}\cite{GellMann:1980vs} 
to naturally generate the tiny neutrino masses $m_\nu^{}\!\thicksim\!v^2\!/M_R^{}\hp$, 
where $v\!=\!O(100)\hp$GeV 
denotes the vacuum expectation value (VEV) of the SM Higgs doublet 
and the neutrino-Higgs Yukawa couplings ($y_{\nu}^{}$) 
are set to their natural values of $\hp O(1)\hp$.\ 
This generally predicts a high scale for neutrino seesaw, 
$M_R^{}\!\thicksim\hsm\hsm v^2\!/m_\nu^{}\!=\!O(10^{14})\hp$GeV.\footnote{%
For the conventional SM setup before 1998, 
the neutrinos were assumed {\it for simplicity} to be massless and have only left-handed components 
because the SM is structured to have all the right-handed
fermions be weak singlets in each fermion family, 
where the right-handed neutrinos ($N_{\!R}^{}$) are pure gauge singlets and their absence does not affect the gauge anomaly cancellation of the SM.\ 
Weinberg realized\,\cite{weinberg-nu5} that without $\NR$, the left-handed
neutrinos can acquire small Majorana masses from a gauge-invariant dimension-5 operator ($LLHH$) 
that is suppressed by a large UV cutoff scale 
$\Lambda_{\nu}^{}\!\sim\!v^2\! /m_{\nu}^{}$, far beyond the weak scale.\ 
However this dimension-5 operator is nonrenormalizable 
and its {\it minimal UV completion} is given by the conventional
seesaw\,\cite{Minkowski:1977sc}\cite{GellMann:1980vs} with $\Lambda_{\nu}^{}\hsm\!=\!M_{\hsm R}^{}$
after adding back $\NR$ for each fermion family.\
The existence of the right-handed neutrinos is predicted by the SM structure
and provides the minimal UV completion for the dimension-5 Weinberg  
operator\,\cite{weinberg-nu5} through the seesaw mechanism 
(naturally generating the light neutrino masses), 
yet, the right-handed neutrinos point to a brand-new seesaw scale 
$\Lambda_{\nu}^{}\!\sim\!v^2\! /m_{\nu}^{}$ 
that is beyond the SM.\ 
Therefore, it is extremely important to probe the right-handed neutrinos 
as {\it the last missing piece of the SM}  
and test the neutrino mass generation via the seesaw mechanism.}\  
Hence, given the current capabilities of particle physics experiments, 
probing the natural neutrino seesaw mechanism at such high scales ($M_R^{}$) 
poses a great challenge.\

On the other hand, it is believed that the early Universe underwent an inflationary epoch, 
during which the Universe expanded exponentially over a very short period.\ Inflation not only resolves the flatness 
and horizon problems, but also seeds the primordial fluctuations that form the large-scale structures of the Universe.\ 
The energy scale of inflation could be 
as high as $O(10^{16})\hp$GeV, 
characterized by the nearly constant Hubble parameter $H_\text{inf}^{}$ 
which is typically around $10^{14}\hp$GeV, 
providing an important window 
for probing new physics at high-energy scales.\
It is intriguing to observe that both the neutrino seesaw scale 
and the Hubble parameter during inflation 
can be realized around the same scale of $O(10^{14})\hp$GeV \cite{Planck10,BICEP2:2019upn}.\
In the minimal setup, 
the inflation is triggered by a scalar inflaton field.\ 
The primordial fluctuations are generated 
by quantum fluctuations of the inflaton 
and can be directly measured through the Cosmic Microwave Background (CMB).\ 
The current CMB data indicate that these fluctuations are adiabatic and 
Gaussian\,\cite{Planck10,Planck1,Planck6,Planck9}.

However, inflation could also generate non-Gaussianity (NG) in the primordial 
perturbations\,\cite{Maldacena0210603, Meerburg:2019qqi}, 
as characterized by $n$-point ($n \!\hsm\geqq\! 3\hp$) 
correlation functions of the comoving curvature perturbation $\zeta\hp$.\ 
The primordial non-Gaussianity can arise in models of multi-field inflation or single-field 
inflation with interactions, offering an ideal opportunity to probe the relevant new physics 
at high-energy scales.\ One notable example is the ``cosmological collider'' 
method\,\cite{Nima150308043, Chen10021416, Baumann160703735, Chen160407841}, 
which aims to explore high-scale particle physics by studying the non-Gaussian properties 
of large-scale structures.\
Given the present non-observation of non-Gaussianity, the existing CMB measurements  
have already set constraints on the non-Gaussian parameter 
$\fNL\!\lesssim\hsm {O}(10)$ 
depending on the non-Gaussian shapes under consideration.\ 
The future detection of the 21{\hp}cm tomography could eventually reach the sensitivity 
down to the level of 
$\fNL \!=\hsm {O}(0.01)$ \cite{21cm,21cm2,21cm3}.\ 

\vs 

At the end of inflation, the inflaton would oscillate at the bottom of its potential and 
eventually transfer its energy to the SM particles, 
thereby reheating the Universe.\ 
This sets the stage for the transition of the Universe 
from the inflationary epoch to a radiation-dominated period, 
preceding the onset of Big Bang Nucleosynthesis.\ 
While observations of large-scale structures provide 
information about the inflation, 
the dynamics of the reheating process remain unclear.\ 
It is natural to expect that the inflaton couples directly 
to the right-handed neutrinos 
and predominantly decays into them after inflation.\ 
Then, these right-handed neutrinos can further decay into the SM particles 
through Yukawa interactions, and thus complete the reheating process.\ 
It is appealing that this approach also naturally provides an initial setup  
for the leptogenesis\,\cite{Fukugita:1986hr}, 
which generates sufficient right-handed neutrinos after reheating.

\vs 

On the other hand, during inflation, not only does the inflaton fluctuate, 
but other light scalar fields also undergo fluctuations.\ 
In particular, the Higgs boson would acquire a field value 
near the Hubble scale, which varies across different horizon patches.\ 
This variation leads to discrimination on the right-handed neutrino masses in local regions 
of the Universe via seesaw mechanism.\ In consequence, the inflaton's decay rate into 
right-handed neutrinos is modulated by the Higgs field value.\ 
This Higgs modulated reheating scenario provides a source of the primordial curvature perturbation\,\cite{Dvali0303591}.\ 
The associated non-Gaussian signatures open up 
a new window for probing the neutrino seesaw scale.\  

\vs 

In this work, we propose a new framework incorporating inflation and neutrino seesaw in which 
the inflaton primarily decays into right-handed neutrinos after inflation.\
This decay process is governed by the inflaton interaction with the right-handed neutrinos 
that respects the shift symmetry.\ 
With the neutrino seesaw mechanism, 
we construct a {\it new realization of Higgs modulated reheating,} in which the fluctuations of Higgs field 
can modulate the inflaton decays and contribute to the primordial curvature perturbations.\ 
We investigate the effects of Higgs-modulated 
reheating and the associated non-Gaussianity (bispectrum).\ 
We demonstrate the potential of our approach to probe 
the high-scale neutrino seesaw mechanism.\ We further analyze the interplay of the non-Gaussianity signatures with the low-energy neutrino experiments, and their interplay with the Higgs self-coupling measurements at the LHC and future colliders.\ 
In passing, this approach also provides a new framework of the cosmological Higgs collider (CHC),
in which the Higgs-modulated reheating is naturally realized 
by the inflaton decays into right-handed neutrinos within the neutrino seesaw.\ 
Thus, particles that couple to the Higgs field would induce cosmological collider signatures, 
which we may call the neutrino-assisted cosmological Higgs collider 
(NCHC).\footnote{%
This NCHC scenario differs from the previous cosmological Higgs collider study in the 
literature\,\cite{Lu190707390} in which the inflaton is assumed 
to couple to certain newly added singlet scalar fields 
and predominantly decay into these scalars.}\

\vs 

This paper is organized as follows.\ In Section\,\ref{sec:2}, 
we discuss the dynamics and evolution of the Higgs field 
during and after inflation.\ 
In Section\,\ref{sec:3}, 
we newly present a minimal framework 
incorporating inflation and neutrino seesaw, in which  
the inflaton decay is modulated by the Higgs boson 
through right-handed neutrinos.\
Then, we give the model realization and setup in Section\,\ref{sec:3.1} and
analyze the curvature perturbation from the Higgs-modulated reheating
through right-handed neutrinos in Section\,\ref{sec:3.2}.\ 
For Section\,\ref{sec:4}, we first study the comoving curvature perturbation 
from Higgs-modulated reheating in Section\,\ref{sec:4.1}.\  
Then, we present the systematic analysis on the three-point correlation function (bispectrum) 
of the comoving curvature perturbation in Section\,\ref{sec:4.2}.\ 
With these, we study the probe of the neutrino seesaw parameter space by using non-Gaussianity 
measurements in Section\,\ref{sec:4.3}, 
and the dependence of non-Gaussianity 
on the Higgs self-coupling in Section\,\ref{sec:4.4}.\  
Finally, we conclude in Section\,\ref{sec:5}.\ 
Appendixes\,\ref{app:A}-\ref{app:D} provide the necessary formulas and technical derivations 
to support the analyses in the main text. 

\section{\hspace*{-2.5mm}Dynamics of Higgs Field in the Early Universe}
\label{sec:2}
\label{Dynamics of the Higgs  in the early universe}

In this section, we discuss the physics of the Higgs field during and after inflation.\  
The Lagrangian density of the SM Higgs kinetic term and potential term is given by    
\begin{equation}
\label{Lagrangian_SM_Higgs}
\mathcal{L} = \sqrt{-g\,}\!\left[\hsm - g_{\mu\nu}^{}\mathrm{D}^{\mu}\mathbb{H}^{\dagger} \mathrm{D}^{\nu} \mathbb{H}
+\mu^2\mathbb{H}^{\dagger}\mathbb{H}-\lambda(\mathbb{H}^{\dagger}\mathbb{H})^2\right]\!,
\end{equation}
where $\mathbb{H}$ is the SM Higgs doublet containing four independent scalar components.\ 
Note that the Higgs self-coupling might become negative at a scale around $10^{11}$GeV depending on
the precise value of the measured top quark mass\,\cite{strumia}.\ 
Given that the Hubble parameter during inflation could be as high as 
$10^{14}\hp$GeV,  the Higgs vacuum might become unstable during inflation.\ 
But the running of Higgs self-coupling is very sensitive to the measured top quark mass.\ 
Within the $3\sigma$ range of the current top mass measurement\,\cite{PDG}, 
it is still possible to keep the Higgs coupling positive and have a value of $O(0.01)$ at the inflation scale.\footnote{%
Adding additional light scalar particle(s) to the Higgs sector at weak scale could lift 
the Higgs self-coupling to the level of ${O}(0.1)$ at the inflation scale\,\cite{HX}.\
For the current study, we will choose the minimal SM Higgs sector.}

In the following discussions, we denote the quantities at different epochs by 
using the corresponding subscripts, such as 
$A_\text{inf}^{}\hsm\equiv\! A(t\!=\!t_\text{inf}^{})$ and 
$A_\text{reh}^{}\!\equiv\! A(t\!=\!t_\text{reh}^{})$.\ 
Here $t_\text{inf}^{}$ is the physical time at the end of inflation, 
and $t_\text{reh}^{}$ is the physical time at the completion of reheating.\ 
As the effects of slow-roll parameters are fairly small, the Hubble parameter 
remains constant throughout the entire epoch of inflation.\ 
Thus, $H_\text{inf}^{}$ could be used to represent the Hubble parameter during inflation.

\vspace*{2mm}
\subsection{\hspace*{-2.5mm}Dynamics of Higgs Field during Inflation}
\label{sec:2.1}
\label{Dynamics of Higgs during Inflation}
\vspace*{1.5mm}

During inflation, the Universe exponentially expands 
and can be described by the de Sitter spacetime 
if the slow roll of the inflaton is neglected.\ 
In contrast to the inflaton, a massless spectator scalar field with self-interaction will exhibit infrared (IR) divergences in de Sitter spacetime\,\cite{Gorbenko191100022,Baumgart191209502,Cespedes:2023aal}.\ 
The framework of stochastic inflation\,\cite{STAROBINSKY1982175}\cite{Starobinsky1994} 
provides a systematic approach to deal with the IR behavior for the super-horizon mode of the massless spectator field. 

\vs 

Although the SM Higgs doublet ${\HH}$ contains four real scalar components, 
three of them correspond to the Goldstone modes 
that become the longitudinal components of the SU$(2)$ weak gauge bosons.\ 
During inflation, the fluctuation of the Higgs field is 
on the order of the Hubble parameter, 
which means the masses of the weak gauge bosons are 
also of the order of the Hubble scale and thus rather large.\ 
On the other hand, the CP-even component $h$ of the Higgs doublet ${\HH}$ 
is much lighter due to the smallness of Higgs self-coupling 
in comparison with the weak gauge coupling, $\lambda\!\ll\! g^2$.\ 
Hence for the present study we only need to deal with the light Higgs field $h\hp$.\ 
In the unitary gauge, the Higgs doublet takes the following form: 
\begin{equation}
{\HH}=\frac{1}{\sqrt{2\,}\,}\!\!
\left(\begin{aligned}
           \,0\,
           \\
           \,h\,
\end{aligned}\right)\!.
\end{equation}
Thus, the Lagrangian density of the pure Higgs sector can be expressed as follows:
\beqs
\begin{align}
& \mathcal{L} 
= \sqrt{-g\,} \!\left[\hsm-\frac{1}{2}g_{\mu\nu}^{}(\partial^{\mu}h\partial^\nu h)
-V(h)\hsm\right] \!,
\\
& V(h) = -\frac{1}{2}\mu^2 h^2\!+\!\frac{\lambda}{4}h^4 .
\end{align}
\eeqs 
In the above $V(h)$ is the Higgs potential, 
in which the quadratic mass term $\frac{1}{2}\mu^2 h^2$ 
could be omitted for a large value of $h$ during the inflation.\ 

\vs

The Higgs field $h$ can be decomposed into a long-wavelength mode ($h_{L}^{}$)  
and the short-wavelength modes 
[which include contributions above a physical 
cutoff scale $\epsilon\hp a(t) H\hp$].\ 
They are both generated by quantum fluctuations, 
\begin{equation}
\label{Higgs_h0_delta_h}
h(\mathbf{x}, t)= h_{L}^{}(\mathbf{x}, t)+\!\int\!\!\! \frac{\rm{d}^{3} k}{(2 \pi)^{3}}\hp 
\theta\hsm\big(k\!-\!\epsilon\hp a(t) H_\text{inf}^{}\big)
\!\!\left[a_{\mathbf{k}}^{}  h_{\mathbf{k}}^{}(t) e^{-\ii\hp \mathbf{k}\cdot\mathbf{x}}
\!+\!a_{\mathbf{k} }^{\dagger}  h_{\mathbf{k}}^{*}(t) e^{\ii\hp \mathbf{k}\cdot \mathbf{x}}\right]\!,
\end{equation}
where $\epsilon$ is a small parameter such that the short-wavelength modes satisfy 
the massless Klein-Gordon equation in the de Sitter space.\  
Thus, the short-wavelength modes can be solved as follows:
\begin{equation}
\label{massless_mode_function}
h_{\mathbf{k}}^{}=\frac{H_\text{inf}}{\sqrt{2k^3\,}\,}\(1\! +\ii\hp k\tau\)
e^{-\ii\hp k\tau} , 
\end{equation}
where $\hp\tau\!=\!-1/(aH)\hp$ is the conformal time.

\vs 

In this framework, the short-wavelength modes $h_{\mathbf{k}}^{}(t)$ are initially sub-horizon and correspond to the normalized modes of a massless scalar field 
in the de Sitter spacetime.\  As the Universe expands, these modes are stretched 
and eventually cross the physical cutoff $\epsilon a(t) H$, transitioning into super-horizon modes $h_{L}^{}$.\ 
The long-wavelength, super-horizon modes $h_{L}^{}$, can be effectively treated 
as a classical stochastic field, following the Langevin equation: 
\begin{equation}
\dot{h}_{L}^{}(\mathbf{x},t)=
-\frac{1}{\,3H_\text{inf}^{}\,}\frac{\partial V}{\,\partial h_{L}^{}}
+f(\mathbf{x},t) \hp.
\end{equation}
It shows that the evolution of the long-wavelength modes is driven by an effective stochastic ``force'' $f(\mathbf{x},t)$,  which is generated by the ``freezing out'' 
of short-wavelength modes:
\begin{equation}
f(\mathbf{x},t)=
\int\!\!\!\frac{\mathrm{d}^3k}{\,(2\pi)^3\,}
\delta\big(k\!-\!\epsilon a(t)H_\text{inf}^{}\big)\hp 
\epsilon a(t) H_\text{inf}^2\!
\(\!a_{\mathbf{k}}^{} h_{\mathbf{k}}^{}e^{-\ii\hp\mathbf{k}\cdot\mathbf{x}}
\!+a_{\mathbf{k} }^\dagger 
h_{\mathbf{k}}^{\ast} e^{i\mathbf{k}\cdot\mathbf{x}}\)\!,
\end{equation}
where we have used the equation 
${\td a(t)}/{\td t}\!=\!H_\text{inf}^{}\hp a(t)$, 
and the two-point correlation function of the stochastic noise 
$f(\mathbf{x},t)$ is given by 
\begin{equation}
\langle f({\bf x}_1, t_1) f({\bf x}_2, t_2)\rangle 
= \frac{\,H_\text{inf}^3\,}{\,4\pi^2\,} \delta(t_1^{}\!-\!t_2^{})\hp 
j_0^{}\big(\epsilon\hp a(t_1) H_\text{inf}|{\bf x}_1^{}\!-\!{\bf x}_2^{}|\big)
\hp , 
\end{equation}
where $j_0^{}(z) \!=\! (\sin\hsm z)/z\,$.\ 
Hence, the behavior of the Higgs field on super-horizon scales can be described 
as a classical stochastic process with a probability distribution $\rho$ 
satisfying a Fokker-Planck equation:
\begin{equation}
\label{FP_Eq_general}
\frac{\,\partial\rho[h(\mathbf{x},t)]\,}{\partial\hp t} 
= \frac{1}{\,3H_\text{inf}^{}\,}\frac{\partial}{\,\partial h\,}\! \left\{\!\rho[h(\mathbf{x},t)] \frac{\partial}{\partial h}V[h(\mathbf{x},t)]\!\right\}
+\frac{\,H_\text{inf}^3\,}{\,8\pi^2\,}\frac{\partial^2}
{\hp\partial h^2\hp}\rho[h(\mathbf{x},t)] \hp.
\end{equation}
The last term on the right-hand side represents the effect of stochastic noise, originating from the sub-horizon modes of the Higgs field as they cross the horizon.\  This term encapsulates the quantum nature of the fluctuations.

\begin{figure}[t]
\centering 
\includegraphics[width=0.75\textwidth]{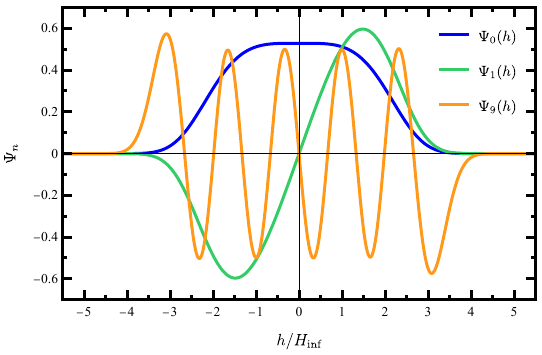}
\vspace*{-4mm}
\caption{Shape of the eigenfunction $\Psi_n^{}$ as a function of Higgs field 
$h\hp$, where we choose the eigenvalue index $n=0,1,9$ for illustration.\ 
The SM Higgs self-coupling constant is set as $\lambda\hsm =\hsm 0.01\hp$. }
\label{fig:1} 
\end{figure}

\vs 

To obtain the probability distribution $\rho(h,t)$, we expand it in terms of a set of eigenfunctions $\Psi_n(h)$ as follows:
\begin{equation}
\label{rho_expanding}
\rho(h,t) \,=\, \Psi_{0}(h)\! \sum_{n=0}^\infty\! 
a_{n}^{}\Psi_{n}^{}(h)\exp(-\Lambda_{n}t) \hp, 
\end{equation}
where $\Lambda_n$ and $\Psi_n(h)$ denote the corresponding eigenvalues and  eigenfunction of the following differential equation,
\begin{equation}
\label{diff_eq_h}
\tilde{D}_h\Psi_n(h)=-\frac{\,8\pi^2\Lambda_n\,}{H_\text{inf}^3}\Psi_n(h) \hp.
\end{equation}
In the above, the operator $\tilde{D}_h^{}$ is defined as follows:
\begin{equation}
\label{diffequ_Psi}
\tilde{D}_h=
\frac{\partial^2}{\partial h^2}-
\left[\!\(\!\frac{\partial v(h)}{\partial h}\!\)^{\!\!2}\!
-\frac{\,\partial^2 v(h)\,}{\partial h^2}\right]\!,
\end{equation} 
where $v(h)\!\equiv\![{4\pi^2}/({3H_\text{inf}^4\,})]V(h)$ and  
$V(h)\!=\!\frac{\lambda}{4}h^4$ is the SM Higgs potential.\ 
The eigenfunctions are orthonormalized as follows:
\begin{equation}
\int_{-\infty}^{+\infty}\!\!\hsm \td h\hp 
\Psi_{n}(h)\Psi_{n^{\prime}}(h) =\delta_{n,n^{\prime}} \hp.
\end{equation}
The derivation of the above equations is provided in 
Appendix\,\ref{Probability distribution for a field}.\ 
We note that all the eigenvalues are non-negative and increase with 
the index $n\hp$, and the lowest eigenvalue vanishes ($\Lambda_0\!=\!0\hp$).

In Fig.\,\ref{fig:1}, we plot the shape of the eigenfunction $\Psi_n(h)$ 
as a function of the Higgs field $h$, and we choose  
the eigenvalue index $n\!=\!0,1,9$ for illustration.\ 
The eigenvalue equation of Eqs.\eqref{diff_eq_h}-\eqref{diffequ_Psi} 
is a Sturm-Liouville problem\,\cite{SLbook} and the eigenfunction 
$\Psi_n(h)$ is proved to have $n$ zero-points and $n\!+\!1$ extreme points.\

\begin{figure}[t]
\centering  
\includegraphics[width=0.60\textwidth]{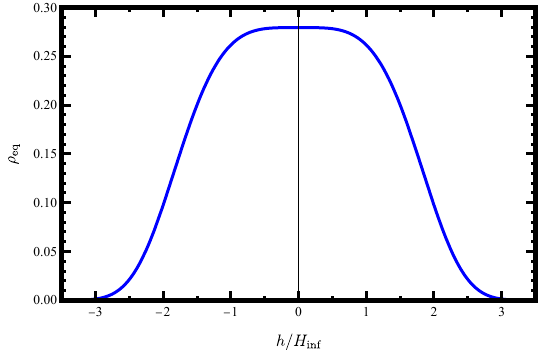}
\vspace*{-4mm}
\caption{$\!\!$Probability density distribution $\rho_{\rm{eq}}^{}$ 
		as a function of the Higgs field $h$ at the end of inflation.\ 
		The SM Higgs self-coupling constant is set as $\lambda\!=\hsm 0.01$.}
\label{fig:2}
\end{figure}

If inflation lasts for enough time, the only remaining eigenfunction 
is the ground state eigenfunction $\Psi_0^{}(h)$ corresponding to the eigenvalue $\Lambda_0^{}\!=\!0\hp$.\ Consequently, the probability distribution of the long-wavelength modes asymptotically approaches that of an equilibrium state:
\begin{equation}
\rho_{\rm{eq}}(h)=\frac{2 \lambda^{1 / 4}}{\,\Gamma(1 / 4)\,}
\!\(\!\frac{\,2\hp\pi^2\,}{3}\!\)^{\!\!\!1/ 4}\! 
\exp\!\(\!\!{\frac{\,-2 \pi^{2} \lambda h^{4}\,}{3H_\text{inf}^4}}\hsm\!\) \!, 
\end{equation}
where the normalization is imposed,
\begin{equation}
\int_{-\infty}^{+\infty}\!\!\td h\hp\rho_{\text{eq}}^{}(h) =1 \hp.
\end{equation}
In Fig.\,\ref{fig:2}, we present the equilibrium probability distribution $\rho_{\text{eq}}^{}(h)$ 
as a function of the Higgs field $h$ in the unitary gauge.

The root-mean-square value of the Higgs field $h$ can be derived as follows:
\begin{equation}
\label{hsquaremean} 
\bar h= \sqrt{\langle h^2 \rangle}= 
\left[\int_{-\infty}^{+\infty}\!\!\td h\hp h^2 \rho_{\text{eq}}^{}(h)  \right]^{\!\!1/2}\!\!\simeq\hp 0.363\!
\(\!\!\frac{\,H_\text{inf}^{}~}{\,\lambda^{1/4}\,}\!\!\)\!.
\end{equation}
For the present analysis, the SM Higgs self-coupling constant 
is set as $\lambda\!=\!0.01\hp$, corresponding to
$\bar{h}\!\simeq\!1.15H_\text{inf}^{}\hp$.\ 
Thus, the Higgs field in our Universe can be approximated by a uniform background 
$\bar h\!\simeq\! H_\text{inf}$ 
combined with Gaussian quantum fluctuations 
$\delta h(\mathbf{x},t)$ around the background $\bar h$, namely, 
\begin{equation}
    h(\mathbf{x},t)=\bar h(t)+\delta h(\mathbf{x},t) \hp,
\end{equation}
which enables the application of the mean-field (MF) approximation in the subsequent analysis. In order to obtain the mode functions of the quantum fluctuation $\delta h(\mathbf{x},t)$, we expand Eq.\eqref{Lagrangian_SM_Higgs} by substituting $ h(\mathbf{x},t)=\bar{h}(t)+\delta h(\mathbf{x},t)$ to obtain the Lagrangian for $\delta h$ 
\begin{equation}
\label{Lagrangian_SM_Higgs_delta}
\mathcal{L}(\delta h)  \rightarrow \sqrt{-g\,}
\left[-\frac{1}{2} \partial_{\mu}\delta h \partial^{\mu}\delta h\!+\!\frac{1}{2}\mu^2(\delta h)^2\!-\!\frac{3}{2}\lambda \bar{h}^2
(\delta h)^2 \!-\! (\lambda \bar{h})\delta h^3\!-\!\frac{\lambda}{4}\delta h^4\right] \!.
\end{equation}
In the MF approximation, the field $\delta h$ is taken to be the linear solution to the equation of motion.\ 
The mass of the Higgs fluctuation is $m_{h}^2\!=\!3\lambda \bar{h}^2\!-\!\mu^2\simeq 0.4 \lambda^{1/2} H_\text{inf}^2$, which means the mass of the fluctuation $\delta h$ is suppressed by the SM Higgs self-coupling constant $\lambda$.\
In this case, the mass of the Higgs fluctuation 
could easily satisfy 
$m_{h}^2\!\ll\! H_{\rm{inf}}^2$ with a small $\lambda\hp$.\

For convenience, we use the approximation that the fluctuation $\delta h$ is massless and thus the mode functions of $\delta h(\mathbf{k})$ are also the solution for the massless Klein-Gordon equation in the de Sitter spacetime, as shown in Eq.\eqref{massless_mode_function}.

\vspace*{1.5mm}
\subsection{\hspace*{-2.5mm}Evolution of Higgs Field after Inflation}
\label{sec:2.2}
\vspace*{1mm}

\begin{figure}[t]
\centering  
\includegraphics[width=0.75\textwidth]{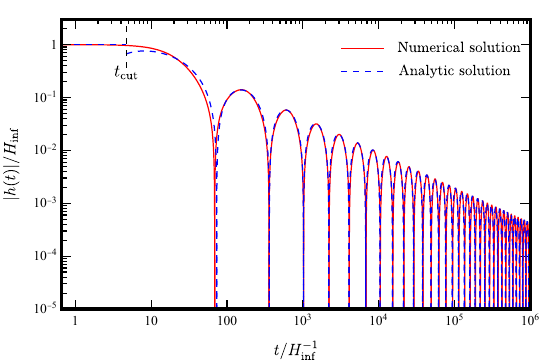}
\vspace*{-3mm}
\caption{Evolution of background value of the Higgs field 
$|h(t)|$ after inflation, where we set an initial value 
$h_{\text{inf}}^{}\hsm=\! H_{\text{inf}}^{}\hp$.\ 
In this plot, the red solid curve represents the numerical solution 
and the blue dashed curve denotes the analytic solution.}
\label{fig:3}
\end{figure}

After inflation, if the inflaton potential is quadratic around the bottom of 
the potential, the inflaton would oscillate and behave like the cold matter 
($w\!=\!0$) having a mass 
$m_{\phi}^{}\!\thicksim\!{O}(1\!-\!\!10)H_\text{inf}^{}\hp$.\ 
Consequently, the Universe will expand as 
$\hp a(t)\!\hsm\thicksim\! t^{2/3}$, from which the Hubble parameter is given by 
$H\!=\!{2}/{(3\hp t)}$.\ Using the Lagrangian (\ref{Lagrangian_SM_Higgs}) and considering only the quartic term of the Higgs potential, 
we find that the evolution of the super-horizon mode of the Higgs field 
after inflation is described by the following Klein-Gordon equation:
\begin{equation}
\label{evolutionofHiggs}
\ddot{h}(t)+\frac{2}{\,t\,} \dot{h}(t)+\lambda h^{3}(t)=0 \hp.
\end{equation}
We solve Eq.\eqref{evolutionofHiggs} numerically to determine the evolution of the Higgs field $h\hp$.\ 
The results are presented in Fig.\,\ref{fig:3},
where for illustration we set the Higgs self-coupling constant $\lambda\!=\!0.01\hp$ 
and choose an initial value 
$h_{\text{inf}}^{}\!=\! H_{\text{inf}}^{}\hp$.\ 
In this plot, the red solid curve represents the numerical solution, 
whereas the blue dashed curve denotes the analytic solution.

Besides, we present a semi-analytical solution to Eq.\eqref{evolutionofHiggs} 
which is given in Appendix\,\ref{app:A}.\  
In the following, we derive the analytical formulas for the evolution of 
the Higgs field $h(t)$ in the case of $\hp h_\text{inf}^{}\!>\!0\,$,
\begin{equation}
\label{h0 theoretical solution formal}
h(t)=\left\{
\begin{array}{ll}
\! h_{\text{inf}}^{}\,, &  t\leqq t_{\text{cut}}^{}\hp, 
\\[1mm]
\!
A\hp H_{\text{inf}}^{}\!\displaystyle  \(\!\hsm\frac{h_{\text{inf}}^{}}{\,H_{\text{inf}}^{}\hp\lambda\,}
\!\)^{\!\hsm\frac{1}{3}}
\!\!\(H_\text{inf}\,t\)^{-\frac{2}{3}}\!
\cos\!\Big(\!\lambda^{\frac 1 6}h_\text{inf}^{\frac 1 3}\hp
\omega\hp t^{\frac 1 3}\!+\theta\hsm\Big)\hsm,~~~~ & t > t_{\text{cut}}^{} \hp,
\end{array}\right.
\end{equation}
where the relevant parameters are given as follows:
\beqs 
\begin{align} 
& t_{\text{cut}}^{}= \dis
\frac{\,\sqrt{2\,}\,}{\,3\sqrt{\lambda\,}\hp h_{\text{inf}}^{}\,} \hp,
~~~~ A= \!\(\!\frac{2}{\,9\,}\!\)^{\!\!\frac{1}{3}}\! 5^{\frac 1 4} 
\simeq 0.9\hp,
\\
\hspace*{-10mm}
& \dis\omega =\frac{\,\Gamma^2(3/4)\,}{\sqrt{\pi\,}}
6^{\frac{1}{3}} 5^{\frac 1 4}
\simeq 2.3\hp,
~~~~\theta=-3^{-\frac 1 3}2^{\frac 1 6}\dis\omega\!-\hsm\arctan 2\simeq\! -2.9 \hp.
\end{align}
\eeqs
One can readily derive the solution for the case 
$h_\text{inf}^{}\!<\!0\,$.
In Fig.\,\ref{fig:3}, we compare the numerical solutions with our analytic solution.\  
We find that for $\hp t \!\gg\! t_{\rm{cut}}^{}$ the analytic solution agrees with 
the numerical results very well.\
This result indicates that after inflation the Higgs field would oscillate in 
its quartic potential $\frac{1}{4}\lambda h^4$  and the amplitude of these oscillations will decrease 
due to the Hubble friction term in Eq.\eqref{evolutionofHiggs}.\

\section{\hspace*{-2.5mm}Higgs-Modulated Reheating Using Right-handed Neutrinos}
\label{sec:3} 
\label{Higgs Modulated Reheating through Right-handed Neutrino Channel}

In this section, we present a minimal realization incorporating inflation and neutrino seesaw, in which  
the inflaton decay is modulated by the Higgs boson through right-handed neutrinos.\
We will give the model realization and setup in Section\,\ref{sec:3.1} and
study the Higgs-modulated reheating through 
right-handed neutrinos in Section\,\ref{sec:3.2}.\

\subsection{\hspace*{-2.5mm}Model Realization and Setup}
\label{sec:3.1} 
\label{the model}

In addition to the particle content of the standard model (SM), 
we introduce the scalar field inflaton $\phi$ and right-handed neutrinos $N_R$.\  
The relevant Lagrangian is given as follows: 
\begin{equation}
\begin{aligned}
\label{Total_Lagrangian}
\Delta\LG & = \sqrt{-g\,}\hp 
\bigg[\!-\frac{1}{2}\hp\partial_{\mu}\phi\hp\partial^{\mu}\phi\hp
-\!V(\phi)+\brn\ii\partial\!\!\!/\rn + 
\frac{1}{\Lambda} \pd_\mu \phi\, \brn \gamma^\mu \gamma^5 \rn 
\\
& ~\quad+\!\(\!\hsm -\frac{1}{2} M \brnc \rn
\!- y_\nu  \, \overline{L}_{\rm L} \widetilde{\mathbb{H}} \rn \hsm +\!\!\hc\!\!\!\hsm\)\!\hsm\bigg],
    \end{aligned}
\end{equation}
where $\HH$ is the SM Higgs doublet and 
$\widetilde{\mathbb{H}}= i \sigma_2 {\HH}^*$ with $\sigma_2^{}$ 
as the second Pauli matrix.\ 
In Eq.\eqref{Total_Lagrangian}, $V(\phi)$ is the inflaton potential and its concrete form is irrelevant 
to the following discussion.\  
After inflation, the potential $V(\phi)$ is assumed to be 
dominated by the inflaton mass term under which the inflaton $\phi$ will oscillate.\ 
In the above, we have suppressed the flavor indices for the SM leptons and right-handed neutrinos.\ Each left-handed lepton doublet 
$L_{\rm L}\!\!=\! (\nu_{\rm L}^{},\hp e_{\rm L}^{})^T$ 
interacts with 
a right-handed neutrino $\rn$ through the Yukawa coupling $y_\nu^{}$, 
which is generally a complex matrix.\ 
Because of the shift symmetry, 
the inflaton $\phi$ couples to the right-handed neutrinos through a unique dimension-5 effective operator (with cutoff $\Lambda\hp$).\footnote{%
As a demonstration, this dimension-5 operator 
in Eq.\eqref{Total_Lagrangian} 
can be induced from a UV model with an approximate 
global $U(1)_{B-L}^{}$ symmetry 
that is spontaneously broken by a new scalar field $\Phi$ with a $U(1)_{B-L}^{}$ charge $-2$ and 
having a VEV, 
$\left<\Phi\right>\hsm =\!f\hp$.\ 
After spontaneous symmetry breaking (SSB), 
the Yukawa interaction
$(y_x^{}\overline{N^c_R}N_R\Phi\!+\hsm\rm{H.c.})$ 
will generate a Majorana mass $M\!=\!y_x^{}f$
for the right-handed neutrino $N_R^{}\hp$, where   
$f\hsm =\hsm M/y_x^{}\!=\!O(10)M\hp$ for a 
natural Yukawa coupling $y_x^{}\!=\hsm O(0.1)$.\  
The inflaton $\phi$ emerges as a pseudo-Nambu-Goldstone boson from $\Phi$ and has its mass generated by soft breaking of the $U(1)_{B-L}^{}$ symmetry, and its interactions with fermions are dictated by their U(1) charges.\ 
The residual shift symmetry enforces that the  $\phi\,$-$N_R^{}$-$N_R^{}$ coupling  
takes the form of a dimension-5 operator.\ 
Its cutoff scale $\Lambda\!=\!f\hp$ is given by the 
$U(1)_{B-L}^{}$ breaking scale $f$, which is about a factor 
of $O(10)$ of the $N_R^{}$ mass scale as shown above 
because the $N_R^{}$ mass originates from the SSB of 
$U(1)_{B-L}^{}$.}

To maintain the perturbative unitarity of the theory during inflation requires 
$\Lambda$ to be no less than $60 H_\text{inf}$, namely,  
$\Lambda \hsm\!\geqq\!\hsm (\dot{\phi})^{1/2} \hsm\!\simeq\hsm 60 H_\text{inf}\hp$.\ 
The shift symmetry plays a key role for maintaining the flatness of the inflaton potential 
throughout inflation and is widely realized in models, such as 
the natural inflation\,\cite{Freese:1990rb} and axion monodromy inflation\,\cite{Silverstein:2008sg}\cite{McAllister:2008hb}, 
or other models with inflation driven by pseudo-Nambu-Goldstone bosons.\ 
The inflaton may couple to SM fermions or gauge bosons through higher-dimensional operators, 
allowing it to decay into SM particles which are usually suppressed.\ 
Consequently, the inflaton is expected to primarily decay into right-handed neutrinos.
	
For simplicity, we will focus on the case of one generation of fermions for the present study.\ 
In this case, the neutrino mass matrix is shown as follows:
\begin{equation}
\label{eq:L-Mnu}
{\mathbf{M}_\nu} =\!
\begin{pmatrix}\!\!
0 & \frac{\,y_\nu^{} h\,}{\sqrt{2\,}\,}\!
\\[1.6mm] 
\!
\frac{\,y_\nu^{} h\,}{\sqrt{2\,}\,} & M \!
\end{pmatrix}\!.
\end{equation}
By diagonalizing the neutrino seesaw mass matrix, 
we derive the neutrino mass-eigenstates of $\nu$ and $N$ as follows:\footnote{%
For the parameter space, we consider that the value of $y_{\nu}^{}h/M$ always 
satisfies the condition $|y_{\nu}h/M| \!\ll\! 1\hp$
at the time of reheating.}
\begin{equation}
\label{mass_nu}
m_{\nu}^{}\simeq -\frac{~y_\nu^2 h^2\,}{2 M},~
\quad
M_{N}^{} \simeq M\!+\!\frac{~y_\nu^2 h^2\,}{2M},
\end{equation}
for $M\!\gg\!|y_\nu h|\,$.\ 
The mixing angle $\theta$ for diagonalizing ${\mathbf{M}_\nu}$ is given by 
\begin{equation}
\label{eq:theta-mix} 
\tan\hsm\theta =\frac{\sqrt{2}y_{\nu}^{}h}{\,\sqrt{M^2\!+\!2y_{\nu}^2h^2\,}\!+\!M\,}
\simeq \frac{y_\nu^{}h}{\,\sqrt{2\,}M\,} \,. 
\end{equation}
In the above, we see that the heavy neutrino mass-eigenvalue is shifted 
by an amount of 
$\hp\frac{\,y_\nu^2 h^2}{2M}\hp$ relative to $M\hp$.\ 
This lifting effect is crucial for our mechanism to work, 
as we are actually probing the seesaw scale of  
the heavy neutrino mass eigenvalue.

\vs

In the Lagrangian \eqref{Total_Lagrangian}, the inflaton is coupled to the right-handed neutrino $N_R^{}$ through a dimension-5 operator 
$\frac{1}{\Lambda} \pd_\mu^{} \phi \brn \gamma^\mu \gamma^5 \rn\hp$.\ 
After inflation, the inflaton decays through this operator until the reheating completes.\ 
The inflaton should decay into the mass eigenstates ($\nu$ and $N$) instead of chiral eigenstates ($\nu_L^{}$ and $N_R^{}$).\ 
So, in terms of mass eigenstates, the relevant interaction vertices from this dimension-5 operator take the following form:
\begin{equation}
\frac{\,\cos^2\!\theta\,}{\Lambda} \pd_\mu^{}\hsm\hp\phi \hp\overline{N} \gamma^\mu \gamma^5 N
\!+\!\frac{\,\sin^2\!\theta\,}{\Lambda} 
\pd_\mu^{} \phi\hp \overline{\nu} \gamma^\mu \gamma^5 \nu
\hsm -\!\(\!\hsm\frac{\,\sin\!2\theta\,}{2\Lambda} \pd_\mu^{}\phi\hp\overline{N} \gamma^\mu \gamma^5 \nu\!+\hsm\rm{h.c.}\!\!\)\!.
\end{equation}
Hence, there are three decay channels $\phi\ito NN, N\nu,\nu\nu\hp$.\ 
Thus, we compute the decay rates of the inflaton as follows:\footnote{In our practical calculation, we include all the kinematic factors.}
\beqs 
\label{eq:phi-decay}
\begin{align}
\Gamma (\phi\ito NN) & =
\frac{\,m_{\phi}^{}M_N^2\cos^4\!\theta~}{\,4\hp\pi\Lambda^2\,}   
\!\(\!1\!-\!\frac{\,4M_N^2\,}{m_{\phi}^2}\!\)^{\!\!\!1/2}\!,
\\
\Gamma (\phi\ito N\nu) & =
\frac{\,m_{\phi}^{}M_N^2(\sin\!2\theta)^2~}{\,32\hp\pi\Lambda^2\,}\!\!   
\(\!1\!-\!\frac{M_N^2}{m_\phi^2}\!\)^{\!\!2}\hsm ,
\\
\Gamma (\phi\ito \nu\nu) & =
\frac{\,m_{\phi}^{}m_\nu^2\sin^4\!\theta~}{\,4\pi\Lambda^2\,}   
\!\!\(\!1\!-\!\frac{\,4m_\nu^2\,}{m_{\phi}^2}\!\)^{\!\!\!1/2}\!,
\end{align}
\eeqs 
where the last decay rate $\Gamma (\phi\ito \nu\nu)$ is suppressed by
light neutrino mass factor $m_\nu^2$ and is thus fully negligible.\
Since the mixing angle $\theta \!\simeq\!\frac{y_\nu^{} h}{\,\sqrt{2\,}M\,}\!\ll\! 1$,
we see that the inflaton decay is dominated by the channel
$\phi\ito NN$.\  Consequently, if we neglect the kinematic factors 
in the above formula, 
the total decay rate of the inflaton can be approximated as follows
:
\begin{equation}
\label{decay_rate}
\begin{aligned}
\Gamma
&\simeq\frac{\,m_{\phi}^{}M^2\,}{4\pi\Lambda^2}\!\!
\left[\hsm 1\!+\frac{1}{4}\!\(\!\!\frac{\,y_\nu^{}h\,}{M}\!\!\)^{\!\!2}\right]\!.
\end{aligned}
\end{equation}
In our setup, the reheating occurs instantaneously at the time 
$\Gamma \!\!=\!\! H(t_\text{reh}) \!\!=\!\! {2}/{(3\, t_\text{reh})}$.\ 
Equation \eqref{decay_rate} shows that the decay of the inflaton through the right-handed neutrino is modulated 
by the Higgs field, and then the Higgs fluctuation 
would induce the curvature perturbation.

Note that our scenario differs from the inflaton decays through the SM fermion channel, 
where $\Gamma \!\propto\! m_f^2 \!\propto\! (y_f^{}h)^2$.\ 
Since the Higgs field value decreases after inflation, the decay width of inflaton 
$\Gamma \!\propto\! h^2$ would decrease even faster than the Hubble parameter $H(t)$, 
preventing the completion of reheating\,\cite{Lu190707390}.\ 
However, unlike the SM fermions, the mass of the right-handed neutrino 
is mainly contributed by the Majorana mass $M$ instead of the Higgs field value $h$.\
This feature prevents the inflaton decay rate $\Gamma$ from fast decreasing with $h$.\ 
Thus, a viable Higgs-modulated reheating can be realized.\ 
We note that the conventional seesaw mechanism has the seesaw scale $M$ typically around $10^{14}$\,GeV, 
which is comparable to the Hubble scale $H_\text{inf}^{}$ during inflation.\ 
We consider the parameter space of $M_{\!N}^{} \!\!<\!\! m_\phi^{}/2\hp$, 
and thus the inflaton decaying into two heavy neutrinos is generally kinetically allowed.\ 
In our setup, the dimension-5 operator discussed above causes the inflaton to decay predominantly into right-handed neutrinos after inflation.\  If the inflaton couples to 
the SM fermions via dimension-5 operators and under the shift symmetry, the corresponding decay rates are suppressed by the fermion masses 
which depend on the Higgs field value (that decreases quickly after inflation).\ 
Couplings between the inflaton and SM gauge bosons (via operators such as $\phi F^{\mu\nu} \tilde{F}_{\mu\nu}$) can be forbidden if the shift symmetry is anomaly-free with respect to the SM gauge group, i.e., the sum of the anomaly parts of fermion triangle loops (containing $\phi$ and two SM gauge bosons as external lines) vanishes.\
For the inflaton coupling with the SM Higgs, it may induce additional decay of the inflaton into pairs of the SM Higgs boson, which could become a dominant channel if this coupling would be large enough (and its consequence was discussed in \cite{Lu190707390}).\ 
In this work, we consider a different scenario, in which the inflaton-Higgs derivative coupling is negligibly small.

\vs

In passing, we note that our model differs from the literature\,\cite{Karam200614404},  
where a right-handed neutrino is introduced and its mass is modulated solely 
due to the Dirac mass term and the curvature perturbation is 
from the kinematic blocking of inflaton decays.\
It assumes that the Dirac mass  $y_h^{}h\!\!\gg\!\! M$ with negligible Majorana mass $M$
of the right-handed neutrino.\ After the Higgs vacuum expectation value decreases below a certain threshold, 
the inflaton can decay.\ 
This is a different scenario of Higgs-modulated reheating and 
the resultant NG originates from this blocking effect.\ 
In contrast, our model has the modulation arise from the neutrino seesaw mechanism.\ 
Additionally, the model of \cite{Karam200614404} 
has the inflaton couple to the right-handed neutrino 
via a dimension-4 operator without shift symmetry, 
which would induce a large Planck-scale Majorana mass term 
for the right-handed neutrino and could make the inflaton decay difficult.\  
In our model, the inflaton has derivative coupling with the right-handed neutrinos
under the shift symmetry, so it does not directly contribute to the Majorana mass 
of the right-handed neutrinos.\ Hence, our present work has proposed {\it a new scenario of 
Higgs-modulated reheating} and can test {\it high scale} seesaw mechanism.\

\vs 

We note that the derivative coupling 
between the inflaton and heavy neutrino can also induce
cosmological collider signals during the inflation\,\cite{Chen:2018xck}\cite{Hook:2019zxa}.\ 
For the present study, our primary focus 
is on the predictions for the local type $f_{\text{NL}}^{}$ 
which is generated from the Higgs-modulated reheating through neutrino seesaw.\ 
This differs from the conventional cosmological collider signals
generated by the inflaton correlation functions during inflation
(which does not invoke Higgs-modulated reheating).

\subsection{\hspace*{-2.5mm}Higgs-Modulated Reheating}
\label{sec:3.2}
\label{Higgs Modulated Reheating}

In our model, the decay rate of the inflaton is influenced by the SM Higgs field.\ 
Fluctuations of the Higgs field value (as generated during inflation) cause variations 
in the inflaton's decay rate across different Hubble patches.\ 
These variations perturb the local expansion history, seeding large-scale inhomogeneity and anisotropy 
in the Universe through these Higgs fluctuations.\ 
The $\delta N$ formalism\,\cite{Starobinsky:1982ee, Salopek:1990jq, Comer:1994np, Sasaki:1995aw, Sasaki:1998ug, Wands:2000dp, Lyth:2003im, Rigopoulos:2003ak, Lyth:2004gb} 
can be used to compute these fluctuations.\ 
The number of e-folds of the cosmic expansion after inflation can be derived as follows:
\begin{align}
N({\bf x}) & =\int\!\! \di\ln a(t) 
=\int\limits_{t_\text{inf}}^{t_\text{reh}({\bf x})}\!\!\!\!\! \di t\hp H(t)
\,+\int\limits_{t_\text{reh}({\bf x})}^{t_{\rm f}}\!\!\!\!\! \di t\hp  H(t)
\nn\\
&= 
\int\limits_{\rho_\text{inf}}^{\rho_\text{reh}(h({\bf x}))}
\!\!\!\!\! \td\rho\,
\frac{\,H\,}{\dot{\rho}} ~+
\int\limits_{\rho_\text{reh}(h({\bf x}))}^{\rho_{\rm f}^{}}
\!\!\!\!\! \td\rho\, 
\frac{\,H\,}{\dot{\rho}} \,,
\label{eq:N(x)}
\end{align}
where $a(t)$ is the scale factor, $\rho(t)$ is the total energy density of the Universe 
at the time $t$,
$t_\text{inf}^{}$ is the physical time at the end of inflation, 
$t_\text{reh}^{}$ is the physical time at which reheating occurs,
$\rho_{\rm{f}}^{}$ is a reference energy density and 
$t_{\rm f}^{}$ is the reference time where the energy density $\rho=\rho_{\rm{f}}$
after the completion of reheating.\

\vs 

For the present study, we consider the Universe as a perfect fluid, 
both before and after the completion of reheating.\ 
At the end of inflation ($t_\text{inf}^{}$), we assume the inflaton's decay rate $\Gamma$ 
is significantly smaller than the Hubble scale.\ 
Reheating is completed at a subsequent time $t_\text{reh}^{}$ when the Hubble parameter satisfies 
$H(t_{\rm{reh}}^{})\!=\! \Gamma_{\rm {reh}}\hp$.\ 
As discussed in the previous section, during the period before reheating completion 
($t_\text{inf}^{} \!<\! t \!<\! t_\text{reh}^{}$), the inflaton oscillates 
near the minimum of a quadratic potential.\ 
This corresponds to the matter-dominated Universe, 
where the pressure $p\!=\!0$ and thus the equation of state parameter 
$w\!=\!p/\rho\!=\!0\hp$.\ 
Throughout this stage, the Universe expands as $a \!\thicksim\! t^{2/3}$, 
with the Hubble parameter $H \!=\! {2}/{(3\hp t)}\hp$.\ 
After reheating completes ($t \!>\! t_\text{reh}^{}$), 
the Universe becomes radiation-dominated, which means that the equation of state parameter is $w\!=\!1/3$ and the scale factor behaves as 
$a \!\thicksim\hsm t^{1/2}$, with the Hubble parameter given by $H \!=\! {1}/{(2{\hp}t)}\hp$.\
Here, the right-handed neutrinos decay fast enough after being produced.\footnote{%
We note that the decay width of the right-handed neutrino is $\Gamma_N^{} \!=\! \frac{\,y_\nu^2 M_N}{8\pi}$.\ 
Comparing $\Gamma_N^{}$ with the Hubble parameter at reheating 
($H_\text{reh}$), we find that in most of the parameter space that can be probed in the near future from non-Gaussianity, this relation ($\Gamma_N\!>\! H_\text{reh}$) can be satisfied.}

\vs 

As a result, the state of the Universe changes before and after the completion of reheating.\ 
For a fluctuation $\delta\Gamma_\text{reh}(\mathbf{x})$ in the decay rate, 
there will be variations in the local reheating time $t_\text{reh}^{}(\mathbf{x})$ 
across different Hubble patches.\ 
These variations translate into differences in the expansion history among these patches, 
which can be quantitatively expressed as fluctuations in the number of e-folds 
$\hp\delta N(\mathbf{x},t)\hp$ of local expansion after inflation.

\vs 

On the other hand, the comoving curvature perturbation after reheating, $\zeta_h(\mathbf{x},t)$, 
is equal to the $\delta N(\mathbf{x},t)$ of cosmic expansion among different Hubble patches 
under the uniform energy density gauge,

\begin{equation}
\zeta_h^{}(\mathbf{x},t)=\delta N(\mathbf{x},t)
=N(\mathbf{x},t)\!-\!\langle N(\mathbf{x},t) \rangle \hp.
\end{equation}

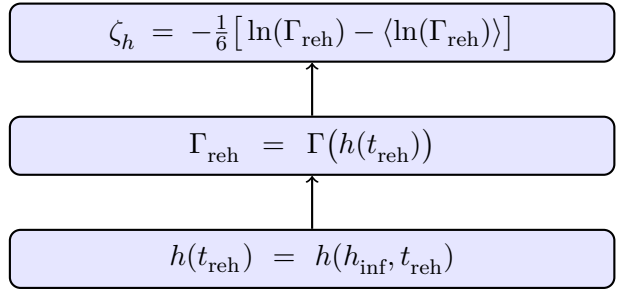
\begin{figure}[H] %
\centering 
\begin{tikzpicture}[node distance=1.5cm, auto, thick,
every node/.style={rectangle, rounded corners, draw=black, fill=blue!10, text width=20em, 
text centered, minimum height=2em},every arrow/.style={-{Latex[length=2mm, width=1.5mm]}}]
\node (h_treh) {$h(t_{\text{reh}^{}}) \!\!=\!\! h(h_{\text{inf}}^{}, t_{\text{reh}}^{})$};
\node (Gamma_reh) [above of=h_treh] {$\Gamma_{\text{reh}}^{} \!\!=\!\! \Gamma\big(h(t_{\text{reh}}^{})\big)$};
\node (zeta_h) [above of=Gamma_reh] {$\zeta_h^{} \!=\! -\frac{1}{6} \big[\ln(\Gamma_{\text{reh}}) \!-\! \langle \ln(\Gamma_{\text{reh}}) \rangle\big]$};
\draw[->] (h_treh) -- (Gamma_reh);
\draw[->] (Gamma_reh) -- (zeta_h);
\end{tikzpicture}
\caption{Schematic plot showing how the comoving curvature perturbation $\zeta_h^{}$ sourced from the Higgs-modulated reheating is a function of the Higgs field $h_\text{inf}^{}$ during the inflation, $\zeta_h^{}\!=\hsm\zeta_h(h_\text{inf}^{})\hp$.} 
\label{fig:4} 
\end{figure}

Since the Higgs fluctuation causes the fluctuation of the inflaton decay rate 
$\delta\Gamma_\text{reh}^{}(\mathbf{x})$ in the Higgs-modulated reheating, 
we can express the comoving curvature perturbation as a function of the local Higgs fluctuation 
$h(\mathbf{x},t_\text{reh}^{})$.\ 
By utilizing the continuity equation 
\beq 
\dot{\rho} + 3H(1\!+\!w)\rho = 0\hp,
\eeq 
we integrate Eq.\eqref{eq:N(x)} and derive the local e-folding number, 
\begin{equation}
N({\bf x}) 
=-\frac{1}{\,3(1\!+\!w_1^{})\,}
\ln\!\frac{\,\rho_\text{reh}^{}\hsm\big(h({\bf x})\hsm\big)\,}{\rho_\text{inf}^{}}
-\frac{1}{\,3(1\!+\!w_2^{})\,}\ln\! 
\frac{\rho_f^{}}{\,\rho_\text{reh}^{}\hsm\big(h({\bf x})\hsm\big)\,} \,,
\end{equation}
where $w_1^{}\!=\!0$ is the equation of state parameter 
before reheating is completed,\footnote{%
Our approach also applies to the general case of 
$w_1^{}\hsm \neq\hsm {1}/{3}\,$.} and 
the equation of state parameter after reheating is $w_2^{}\hsm\!=\hsm\!1/3\hp$.\  
Utilizing the first Friedmann equation $3H^2M_{p}^2 \!=\!\rho\,$  
and noting that reheating completes when 
$H(t_{\rm{reh}}^{})\!=\! \Gamma_{\rm {reh}}\hp$ is reached, 
we can derive the following comoving curvature perturbation after reheating 
($t \!>\! t_\text{reh}$){\hp}:
\begin{equation}
\label{zeta_Higgs}
\begin{aligned}
\zeta_h^{}({\bf x},t> t_\text{reh}^{})
& =\delta N({\bf x})=N({\bf x})\!-\!\langle N({\bf x})\rangle 
\\
& =-\frac{1}{\,12\,}
\big[\!\ln \rho_\text{reh}^{}({\bf x})\!-\!\langle \ln\rho_\text{reh}^{}({\bf x}) \rangle \big]
\\
&=-\frac{1}{\,6\,}
\big[\!\ln(H_\text{reh})\!-\!\langle \ln(H_\text{reh}^{})\rangle \big]
\\
&=-\frac{1}{\,6\,}
\big[\!\ln(\Gamma_\text{reh}^{})\!-\!\langle \ln(\Gamma_\text{reh}^{})\rangle \big] 
\hp.
\end{aligned}
\end{equation}
In this scenario, we derive a relationship between curvature perturbation 
$\zeta_h^{}(\mathbf{x})$ from Higgs-modulated reheating and 
the Higgs field $h_\text{inf}^{}$ during inflation, 
as illustrated in Fig.\,\ref{fig:4}.\ 

In the case where the decay of the right-handed neutrino $N$ is significantly delayed (corresponding to a much smaller Yukawa coupling between the Higgs and $N$), the right-handed neutrino begins to dominate the Universe shortly after the inflaton decays. Since the mass of $N$ is close to that of the inflaton, the Universe quickly becomes matter dominated and remains so until $N$ decays into Standard Model particles. In this regime, the decay rate of $N$ is largely independent of the Higgs vacuum expectation value, and thus the modulation effect is suppressed. We note here that, in most of the parameter space accessible to near-future experiments, the decay rate of $N$ remains larger than the Hubble rate at reheating. Therefore, we neglect this effect in our present analysis.

\vspace*{2mm}

In Section\,\ref{sec:2}, we have demonstrated that the value of the Higgs field at the completion of reheating,  $h(t_\text{reh}^{})$, is determined by the initial value of $h_\text{inf}^{}\,$.\ 
Given that the inflaton's decay rate $\Gamma$ is a function of the Higgs value $h(t_\text{reh})$ 
from Eqs.\eqref{mass_nu} and \eqref{eq:phi-decay}, 
and that the comoving curvature perturbation $\,\zeta\hp$ depends on the decay rate $\Gamma$, 
we can thus establish a relationship between the comoving curvature perturbation 
$\zeta$ and the Higgs field $h_\text{inf}^{}$ during inflation.\ 

Note that, at the time of the reheating occurring, the value of the Higgs field 
after inflation becomes an oscillatory function of its initial value,
\begin{equation}
h(t_\text{reh}^{},h_\text{inf}^{}) \,\thicksim\, \(h_\text{inf}^{}\)^{\hsm\frac{1}{3}}
\cos\!\hsm\(\!\omega_\text{reh}^{}h_\text{inf}^{\hsm\frac{1}{3}} \hsm+\hsm\theta \!\)\!,
\end{equation}
with the oscillating frequency estimated as 
\begin{equation}
\label{omega_reh_def}
\omega_\text{reh}^{} = \lambda^{\frac{1}{6}}t_\text{reh}^{\frac{1}{3}}\hp\omega \,.
\end{equation}
When $t_\rm {reh}$ is large, the oscillation frequency can become very high.\ 
Note that $\zeta_h^{}$ is a function of $h^2$ and can be expanded into the form 
$A\!+\! B h^2/M^2\!+\hsm {O}(h^4/M^4)$,  
which includes a factor $\cos^2(\omega_\rm {reh} h_\text{inf}^{{1}/{3}}+\theta)$.\ 
Since $h_\text{inf}^{}$ varies across different Hubble volumes 
and $\zeta_h^{}$ is highly sensitive to $h_\text{inf}^{}$, 
averaging over a sufficiently large volume allows the factor 
$\cos^2(\omega_{\rm{reh}}^{} h_\text{inf}^{{1}/{3}}\!+\theta)$ to be effectively treated as ${1}/{2}$\,\cite{Suyama:2013dqa}. Consequently, in the subsequent calculations, we directly set  $\cos^2(\omega_{\rm{reh}}^{} h_\text{inf}^{{1}/{3}}\!+\hsm\theta) \!\to\! {1}/{2}\,$.  

Combined with the inflaton fluctuation $\delta \phi$ during inflation, 
the total comoving curvature perturbation can be written as follows:
\begin{equation}
\label{zeta-h}
\zeta \,=\, \zeta_\phi^{} + \zeta_h^{} \,,
\end{equation}
where  $\zeta_\phi^{}$ is generated by the inflaton fluctuation  $\delta \phi\,$,
\beq 
\zeta_\phi^{} \,\simeq\, -\frac{\,H_\text{inf}^{}\,}{\dot{\phi_0}} \delta \phi(\mathbf{x}) \,, 
\eeq 
and $\zeta_h^{}$ originates from the Higgs-modulated reheating.\ 
Because these two components are generated at different times and are independent of each other, 
the power spectrum of $\,\zeta\hp$ contains both contributions:  
\beq 
\mathcal{P}_{\zeta} = \mathcal{P}_{\zeta}^{(\phi)} + \mathcal{P}_{\zeta}^{(h)},
\eeq
where $\mathcal{P}_{\zeta}^{(\phi)}$ denotes the contribution induced by inflaton fluctuations,
\beq 
\mathcal{P}_{\zeta}^{(\phi)} = 
\(\!\!\frac{\,H\,}{\hp\dot{\phi}\,}\!\)^{\!\!2} \!\mathcal{P}_\phi^{} =\(\!\!\frac{\,H_\text{inf}\,}{\hp\dot{\phi}\,}\!\)^{\!\!2}\!
\frac{\,H_\text{inf}^2\,}{\,4\pi^2\,}\,.
\eeq 
For convenience, we define $R$ as the square root of the ratio between the power spectrum
of the Higgs-modulated reheating and that of  
the comoving curvature perturbation $\zeta\,$,
\begin{equation}
\label{rh}
R \,\equiv\(\!\!\frac{~\mathcal{P}_{\zeta}^{(h)}}{\mathcal{P}_{\zeta}^{(o)}}
\!\)^{\hsm\!\!\!1/2} ,
\end{equation}
\\[-3mm]
where 
$\mathcal{P}_{\zeta}^{(o)}\!\simeq\! 2.1\!\times\! 10^{-9}$ 
is the observed curvature perturbation\,\cite{Planck1}\cite{Planck6}.\ 
To be consistent with observation, we should require 
$R\! < \!1\hp$.

In the literature various modulated reheating models were studied, 
which often assume that all primordial perturbations originate from Higgs-modulated reheating 
($R \!=\! 1$)\,\cite{Karam200614404}\cite{Wands10040818,Ichikawa:2008ne,DeSimone12106618,Karam210302569}.\  
The existing cosmological observations require  $R \!<\! 1\hp$.\  
In this work, we compute $\mathcal{P}_{\zeta}^{(h)}$ through the mean-field method and impose $R \!<\! 1$ for our parameter space.

Moreover, the modulated reheating can provide a source of primordial non-Gaussianity (NG).\ 
Primordial non-Gaussianity is characterized by the three-point correlation function of $\zeta\hp$, 
which is also called the bispectrum 
$\left\langle \zeta_{\mathbf{k}_{1}} \zeta_{\mathbf{k}_{2}} \zeta_{\mathbf{k}_{3}} \right\rangle $.\
This primordial non-Gaussianity leaves an imprint on the CMB anisotropy 
and large-scale structure observations.\  Different physics effects involved 
during inflation lead to different shape functions of the primordial NG, 
and several templates of shape functions are measured 
by using the Planck-2018 data to identify signals 
of potential new physics\,\cite{Planck1,Planck9,Planck10}.\ 
For instance, the local non-Gaussianity of the Bardeen potential $\Phi$ is given by 
\begin{equation}
\label{fnl_def_PLANCK_formal}
\langle \Phi_{\mathbf{k}_1}\Phi_{\mathbf{k}_2}\Phi_{\mathbf{k}_3}\rangle'_{\text {local }}
=2 A^{2} f_{\mathrm{NL}}^{\mathrm{local}}
\biggl\{\!\frac{1}{\,k_{1}^{3} k_{2}^{3}\,}
+\frac{1}{\,k_{2}^{3} k_{3}^{3}\,}
+\frac{1}{\,k_{3}^{3} k_{1}^{3}\,}
\!\biggr\} .
\end{equation}
In the above, $\langle \Phi_{\mathbf{k}_1}\Phi_{\mathbf{k}_2}\Phi_{\mathbf{k}_3} \rangle'$ 
is defined as the 3-point correlation function excluding 
the $\delta$ function of momentum conservation, 
$\langle \Phi_{\mathbf{k}_1}\Phi_{\mathbf{k}_2}\Phi_{\mathbf{k}_3} \rangle 
\!=\! \left(2\pi\right)^3\!\delta^3\hsm(\mathbf{k}_{1}\!+\!\mathbf{k}_{2}\!+\!\mathbf{k}_3)\langle \Phi_{\mathbf{k}_1}\Phi_{\mathbf{k}_2}\Phi_{\mathbf{k}_3} \rangle'$. 

\vs

For studying the modulated reheating, the comoving curvature perturbation $\zeta\hp$ 
is expanded as a function of the fluctuation of the Higgs field during the inflation $\hp\delta h_\text{inf}\hp$ 
around its mean value $\bar{h}\,$,
\begin{equation}
\zeta_h\big(\delta h_\text{inf}(\mathbf{x})\big)=\delta N\big(\delta h_\text{inf}(\mathbf{x})\big)
=N^{\prime} \delta h_\text{inf}+\frac{1}{2}N^{\prime\prime}(\delta h_\text{inf})^2+\cdots\cdots,
\end{equation} 
where $N^{\prime}$ and $N^{\prime\prime}$ denote the first and second derivatives of the e-folding number $N$ with respect to the Higgs field $\delta h_\text{inf}$, evaluated at its mean value,
\begin{equation}
N^{\prime}=\left.\frac{\td N}{\td h_\text{inf}}\right|_{\bar{h}}, \hspace*{7mm}
N^{\prime\prime}=\left.\frac{\td^2 N}{\td h_\text{inf}^2}\right|_{\bar{h}}.
\end{equation}
This perturbative approach is also referred to as the mean-field method and 
has been commonly used to calculate the $n$-point correlation functions of curvature perturbation.\ 
Using this expansion, we can determine the amplitude of curvature perturbations
$\mathcal {P}_\zeta \!=\! {N^\prime}^2 \mathcal {P}_{\delta  h}$ 
and the primordial local non-Gaussianity \cite{Wands10040818,Ichikawa:2008ne,DeSimone12106618,Karam210302569,Litsa201111649}.

\section{\hspace*{-2.5mm}Probing Neutrino Seesaw Using Primordial Non-Gaussianity}
\label{sec:4} 
\label{Probing Seesaw Mechanism using Primordial non-Gaussianity}

As discussed in Section\,\ref{sec:3.2}, 
for the scenario of Higgs-modulated reheating, the fluctuation of the Higgs field can contribute to 
the comoving curvature perturbation $\zeta$ with a fraction $R\hp$.\ 
Moreover, the Higgs-modulated reheating will also generate primordial non-Gaussianities (NG), 
which could be detected by the CMB or large-scale structure observations.\ 
In this section, we investigate the primordial local non-Gaussianity arising from 
the Higgs-modulated reheating.\ We will demonstrate that the primordial non-Gaussianity provides 
a viable approach to probe the neutrino seesaw scale in this framework.\ 
For this purpose, we compute the 2-point and 3-point correlation functions of 
the comoving curvature perturbation from the Higgs-modulated reheating $\zeta_h^{}\hp$.\ 
In this section, the Hubble scale during inflation $H_\text{inf}^{}$ is abbreviated as $H$,  
and the notation of the Higgs field value during inflation $h_\text{inf}$ is simplified as $h\,$,
which differs from the Higgs field value $h(t)$ after inflation.

\vspace*{1mm}
\subsection{\hspace*{-2.5mm}Comoving Curvature Perturbation from Higgs-Modulated Reheating}
\label{sec:4.1}

As discussed in the Section\,\ref{sec:3.2}, the comoving curvature perturbation $\zeta$ from 
the Higgs-modulated reheating depends on the logarithm of decay rate at the completion of 
reheating $\ln (\Gamma_\rm{reh}^{})$, as described by Eq.\eqref{zeta_Higgs}. 

The relation between $\Gamma_\text{reh}^{}$ and the Higgs field $h_\text{inf}^{}$ during inflation 
is shown in Fig.\,\ref{fig:4}, namely, 
$\Gamma_\text{reh}^{}\hsm\!=\!\Gamma(h_\text{reh}^{}(t_\text{reh}^{},h_\text{inf}^{}))$.\ 
In the mean-field approximation, we expand the comoving curvature perturbation to the order of 
$\hp\delta h_\text{inf}^2\,$,
\begin{equation}
\label{eq:zeta-expand}
\zeta_h^{}(\mathbf{x}) = -\frac{1\!-\!3w}{\,6(1\!+\!w)\,}\!\!
\left[\hsm\frac{\,\Gamma'_0\,}{\,\Gamma_0^{}}\hp\delta h_\text{inf}^{}(\mathbf{x})
+\frac{\,\Gamma_0^{}\Gamma''_0\!-\!\Gamma'_0 \Gamma'_0\,}{2\Gamma_0^2}
\hp\delta h_\text{inf}^2(\mathbf{x})\hsm\right]\!,
\end{equation}
where we have defined,
\beqs
\label{eq:Gamma0-p-pp}
\begin{align}
\Gamma_0 &= \left.\Gamma_\text{reh}\right|_{h_\text{inf}({\bf x})=\bar{h}}^{}\,,
\\[1mm]
\Gamma'_0
&= \left.\frac{\,\d\Gamma_\text{reh}\,}{\d h_\text{inf}}\right|_{h_\text{inf}({\bf x})=\bar{h}}
=\left.\frac{\,\d\Gamma_\text{reh}^{}\,}{\d h_\text{reh}^{}}
\frac{\,\partial h_\text{reh}^{}\,}{\partial h_\text{inf}^{}}\right|_{h_\text{inf}({\bf x})=\bar{h}}^{} ,
\\[1mm]
\Gamma''_0
&= \left.\frac{\,\d^2\Gamma_\text{reh}\,}{\d h_\text{inf}^2}\right|_{h_\text{inf}({\bf x})=\bar{h}}
=\left.\frac{\,\d^2\Gamma_\text{reh}\,}{\d h_\text{reh}^2}
\hsm\!\(\!\!\frac{\,\partial h_\text{reh}^{}\,}{\partial h_\text{inf}} \!\!\)^{\!\!2}
\right|_{h_\text{inf}({\bf x})=\bar{h}}^{} \!\!+\! \left.
\frac{\,\d\Gamma_\text{reh}^{}\,}{\d h_\text{reh}^{}}\!
\(\!\!\frac{\,\partial^2 h_\text{reh}^{}\,}{\partial h_\text{inf}^2} \!\!\)\!
\right|_{h_\text{inf}({\bf x})=\bar{h}}^{} .
\end{align}
\eeqs

After inflation the inflaton potential is quadratic, so the Universe is matter-dominated, 
implying $w\hsm =\hsm 0\hp$.\ 
We can establish the relation between the curvature perturbation and the Higgs fluctuation as follows: 
\begin{equation}
\label{zetaHiggsAppendix}
\zeta_h^{}(\mathbf{x}) = 
-\frac{1}{\,6\,}\!\!\left[\!\frac{\,\Gamma'_0\,}{~\Gamma_0^{}\,} 
\delta h_\text{inf}^{}(\mathbf{x}) \!+\! \frac{\,\Gamma_0^{}\Gamma''_0\hsm -\hsm \Gamma'_0 \Gamma'_0\,}
{2\hp\Gamma_0^2} \delta h_\text{inf}^2(\mathbf{x})\hsm\right]\!
\equiv z_1^{}\delta h_\text{inf}^{}(\mathbf{x})\!+\!\frac{1}{\,2\,} z_2^{} \delta h_\text{inf}^2(\mathbf{x}) \hp,
\end{equation}
where $z_1^{}$ and $z_2^{}$ are the linear and second-order coefficients:  
\begin{equation}
\label{z1z2def}
z_1^{}=-\frac{1}{\,6\,}\frac{\,\Gamma'_0\,}{\Gamma_0^{}} \hp,  
\hspace*{8mm}
z_2^{} = -\frac{1}{\,6\,} \!\hsm\left[\!\frac{\Gamma''_0}{\,\Gamma_0^{}\,} \hsm -\! 
\(\!\!\frac{\,\Gamma'_0\,}{\Gamma_0^{}}\!\!\)^{\!\!2}\right] \!. 
\end{equation}
In the following, since we only deal with Higgs fluctuations $\delta h_\text{inf}^{}$ during the inflation, 
we will omit the subscript ``inf'' and directly use the notation $\delta h$ for convenience, 
i.e., $\delta h\equiv \delta h_\text{inf}^{}\hp$.\ 

\vs 

From Eq.\eqref{zetaHiggsAppendix}, we further derive the corresponding form in Fourier space:
\begin{equation}
\label{z_h_momentum}
\begin{aligned}
\zeta_h^{}(\mathbf{k}) &= 
\int\!\! {\d}^{3}\hp\mathbf{x}\hp\zeta(\mathbf{x}) e^{-\ii\hp\mathbf{k}\cdot\mathbf{x}}
=z_1^{} \delta h(\mathbf{k}) + \frac{\,z_2^{}\,}{\,2\,}\hsm\!\int\!\!{\d}^{3}\hp\mathbf{x}\hp  
\delta h^{2}(\mathbf{x})\hp e^{-\ii\hp\mathbf{k}\cdot\mathbf{x}}
\\
& = z_1^{}\delta h(\mathbf{k}) + \frac{\,z_2^{}\,}{\,2\,}\hsm\!\int\!\!{\d}^{3} \mathbf{x}\hp
\frac{\,{\d}^{3} \mathbf{k}_1^{}\,}{\,(2\pi)^{3}\,} \frac{\,{\d}^{3}\mathbf{k}_2^{}\,}{\,(2\pi)^{3}\,}
\delta h(\mathbf{k}^{}_1)\delta h(\mathbf{k}^{}_2) e^{\ii (\mathbf{k}^{}_1+\mathbf{k}^{}_2-\mathbf{k})\cdot\mathbf{x}}
\\
& = z_1^{} \delta h(\mathbf{k})+\frac{\,z_2^{}\,}{\,2\,}\hsm\!\int\!\! 
\frac{\,{\d}^{3}\mathbf{k}_1^{}\,}{\,(2 \pi)^{3}\,}\hp\delta h(\mathbf{k}_1^{})\delta h(\mathbf{k}\!-\!\mathbf{k}^{}_1)\,.
\end{aligned}
\end{equation}
Thus, we derive the 2-point correlation function of $\hp\zeta$ from the Higgs-modulated reheating to the leading order,
\begin{equation}
\langle \zeta_{\mathbf{k}^{}_1}^{}\zeta^{}_{\mathbf{k}_2^{}}
\rangle_h^{} = z_1^2 \langle \delta h_{\mathbf{k}^{}_1}\delta h_{\mathbf{k}^{}_2}^{}\rangle \hp,
\end{equation}
from which we obtain the power spectrum of $\hp\zeta\,$, 
\begin{equation}
\mathcal{P}_\zeta^{(h)}=z_1^2 \mathcal{P}_{\delta h}^{}
= \frac{~z_1^2H^2\,}{\,4\pi^2\,} ~ .
\end{equation}

\vs 

From Eq.\eqref{rh}, we use $R^2$ to reflect the ratio between 
the contribution from Higgs-modulated reheating 
and the comoving curvature perturbation $\zeta\hp$.\ 
So, in this case, we have 
\begin{equation}
R = \(\!\!\!\frac{~\mathcal{P}_{\zeta}^{(h)}\,}{\mathcal{P}_{\zeta}^{(o)}}\!\!\)^{\hsm\!\!\!1/2}
\!=\, |z_{1}^{}|\!\(\!\! \frac{~\mathcal{P}_{\delta h}^{}\,}
{\mathcal{P}_{\zeta}^{(o)}}\!\!\)^{\hsm\!\!\!1/2} ,
\end{equation}
where $\mathcal{P}_{\zeta}^{(o)}\!\simeq\! 2.1\!\times\! 10^{-9}$ 
is the observed curvature perturbation\,\cite{Planck1}\cite{Planck6}.\
Thus, we require that the ratio $R^2$ be less than unity, i.e., 
$R\!=\!\big(\mathcal{P}_{\zeta}^{(h)}\!/\mathcal{P}_{\zeta}^{(o)}\big)^{\!1/2} \!\!<\! 1\hp$.\

\subsection{\hspace*{-2.5mm}Three-Point Correlation Function of Curvature Perturbation} 
\label{sec:4.2}
\label{From Higgs correlation functions to the Shape Function of zeta}
\vspace*{1mm}

In this subsection, we derive the three-point correlation function of the comoving curvature perturbation 
$\hp\zeta\hp$ contributed by the Higgs-modulated reheating,  
$\zeta_h^{}\!=\!z_1^{}\delta h+\frac{1}{2} z_2^{} \delta h^2$.\ 
The three-point correlation function of $\hp\zeta\hp$ from modulated reheating 
$\langle \zeta_{\mathbf{k}_1}^{}\zeta_{\mathbf{k}_2}^{}\zeta^{}_{\mathbf{k}_3}\rangle_h^{}$ 
consists of two parts: 
\\[-5mm]
\begin{equation}
\label{3ptz_h}
\langle \zeta_{\mathbf{k}_1}^{}\zeta_{\mathbf{k}_2}^{}\zeta_{\mathbf{k}_3}^{}\rangle_h^{}
\hp =\hp z_{1}^3\langle \delta h_{\mathbf{k}_1}^{}\delta h_{\mathbf{k}_2^{}} \delta h_{\mathbf{k}_3}^{}
\rangle \!+\! z_1^2z_2\langle\delta h^4\rangle
(\mathbf{k}_1^{}, \mathbf{k}_2^{}, \mathbf{k}_3^{}) \,.
\end{equation}
On the right-hand side of the above formula, 
the first term $z_{1}^3\langle \delta h_{\mathbf{k_1}}\delta h_{\mathbf{k_2}}\delta h_{\mathbf{k_3}}\rangle$ is the three-point correlation function of the Higgs fluctuation 
$\delta h(\mathbf{k})$ generated by the self-interactions 
of the Higgs field.\ 
The second term arises from replacing one $\delta h(\mathbf{k})$ by the nonlinear term 
$\frac{1}{2}\hp z_2^{}\hp\delta h^2$, 
which exists even if the Higgs fluctuation 
$\delta h(\mathbf{k})$ is purely Gaussian.

\vs 

\begin{figure}[t]
\centering
\includegraphics[width=0.3\textwidth]{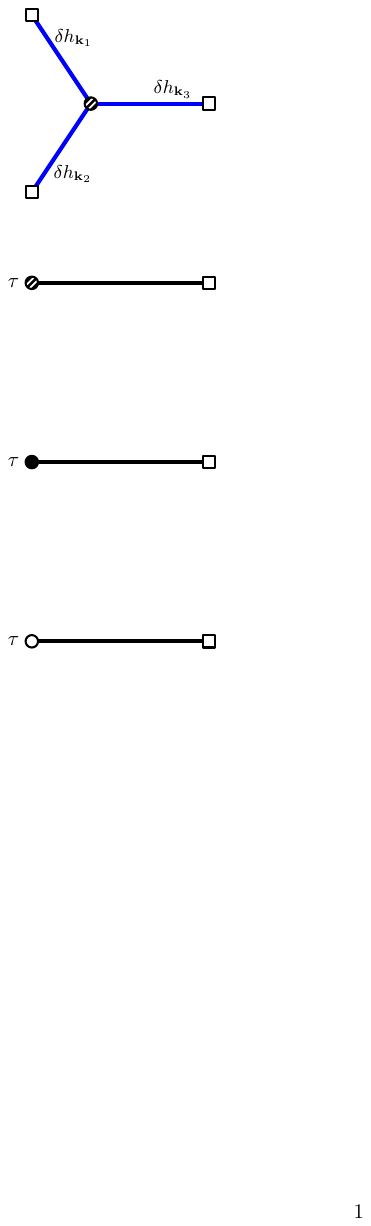}
\vspace*{-3mm}
\caption{\hspace*{-1mm}
Three-point correlation function of Higgs fluctuation 
$\delta h$ from the Higgs self-interaction 
$\Delta\mathcal{L}\!=\!-\sqrt{-g\,}[(\lambda {\hp}\bar{h})\delta h^3]\,$.\ 
The Higgs propagator is depicted by a blue solid line 
with a dot and a square at its endpoints, representing a bulk-to-boundary propagator, 
which includes one ``plus type'' and one ``minus type''.\ The square at one end of the propagator indicates 
its boundary point ($\tau_f^{}\!\to\! 0^-$).\ 
The shaded dot at the vertex means that contributions 
from both plus- and minus-type propagators must be summed.}   
\label{fig:5}
\end{figure}

As discussed in Section\,\ref{sec:2.1}, the Higgs field could be treated as a massless scalar boson 
in de Sitter spacetime during inflation.\ 
Because of the SM Higgs self-interaction term, 
$\Delta\mathcal{L}\!=\!-\sqrt{-g\,}[(\lambda {\hp}\bar{h})\delta h^3]$, 
the three-point correlation function of $\delta h$ is presented as the diagram in Fig.\,\ref{fig:5}.\ 

According to the Schwinger-Keldysh (SK) 
path integral formalism\,\cite{Weinberg2005}\cite{Chen170310166}, 
we can compute the three-point correlation function of $\delta h$ via the following integral: 
\begin{equation}
\begin{aligned}
\label{3pth_SM0}
\langle \delta h_{{\kb_1}^{}}\delta h_{{\kb_2}}^{}\delta h_{{\kb_3}}^{}\rangle'(\tau_f^{})
&=-\ii\hp 3!\lambda\hp \bar{h} \!\int_{-\infty}^{\tau_f^{}}\!\!\!\d \tau\hp  
a^4 \!\hsm\left[\prod _{i=1}^3 \!G_{+}^{}({\kb}_i^{},\tau) \!-\! \prod _{i=1}^3 \!G_{-}^{}({\kb}_i^{},\tau)\right] 
\\
& = 12\lambda\hp\bar{h}\hp 
{\mathrm{Im}\!\(\!\int_{-\infty}^{\tau_f} \!\!\!\d\tau\hp a^4 
\prod _{i=1}^3\! G_{+}^{}({\kb}_i,\tau) \!\hsm\)},
\end{aligned}
\end{equation}
where $\lambda$ is the SM Higgs self-coupling constant, $\bar{h}$ is the uniform Higgs background during inflation, and the $G_{\pm}(\mathbf{k}_i,\tau)$ 
is the bulk-to-boundary propagator of massless scalar in the SK path integral defined in 
Appendix\,\ref{SK path integral for massless scalar}.\ 
In Eq.\eqref{3pth_SM0}, 
$\langle \delta h_{{\kb_1}^{}}\delta h_{{\kb_2}}^{}\delta h_{{\kb_3}}^{}\rangle'$ 
is defined as the three-point correlation function without including the $\delta$ function of momentum conservation, 
$\langle \delta h_{{\kb_1}^{}}\delta h_{{\kb_2}}^{}\delta h_{{\kb_3}}^{}\rangle 
\!=\! \left(2\pi\right)^3\!\delta^3\hsm(\mathbf{k}_{1}^{}\!+\!\mathbf{k}_{2}^{}\!+\!\mathbf{k}_3^{})
\langle \delta h_{{\kb_1}^{}}\!\delta h_{{\kb_2}}^{}\!\delta h_{{\kb_3}}^{}\rangle'$.\ 
We derive the integral in the last line of Eq.\eqref{3pth_SM0} 
to the leading order of $\tau_f^{}$ as follows: 
%
\begin{equation}
\begin{aligned}
\label{int  A1 A}
& \mathrm{Im}\!\(\!\int_{-\infty}^{\tau_f}\!\!\!\d\tau\hp a^4
\prod_{i=1}^3 \hsm\!G_{+}^{}({\kb}_i,\tau)\!\!\)
\\
&=\mathrm{Im}\! \int_{-\infty}^{\tau_f}\!\!
\frac{\d\tau}{\,(H\tau)^4\,} \frac{H^6}{\,8\hp k_1^3k_2^3k_3^3\,}
\!\(\prod _{i=1}^3 (1 \!-\hsm \ii\hp k_{i}^{}\tau) \!\!\)\! e^{\ii\hp (k_1^{}+k_2^{}+k_3^{})\hp\tau}
\\
&=\frac{H^2}{\,24\hp k_1^3k_2^3k_3^3\,} 
\hsm\biggl\{\! 
(k_1^3\!+\!k_2^3\!+\!k_3^3)\!\Big[ \!\ln (k_t|\tau_f|)\!+\!\gamma\!-\!\frac{4}{3}\Big]\!
\!+\!k_1^{}k_2^{}k_3^{} \!-\! \sum_{a\neq b}\!k_a^2k_b^{} 
\!\biggr\} ,
\end{aligned}
\end{equation}
where $\tau_f^{}\!\!\to\! 0^-$ is the conformal time when the inflation ends, 
$\hp\gamma\hsm\!\simeq\!\hsm 0.577\hp$ is the Euler-Mascheroni constant, and wavenumber 
$\,k_t^{}\!=\! k_1^{}\!+\hsm k_2^{}\!+\hsm k_3^{}$
is around the scale of the present observable Universe.\ 
In the above correlation function, the leading contribution is given by the logarithmic term of 
$\ln (k_t^{}|\tau_f^{}|)$ and there is no inverse power term of $k_t^{}|\tau_f^{}|$ as proved in 
Ref.\,\cite{Weinberg2005}.\

Thus, we further derive the three-point correlation function of $\delta h$ as follows: 
\begin{equation}
\label{3pth_SM1}
\begin{aligned}
\langle \delta h_{{\kb_1}}^{}\delta h_{{\kb_2}}^{}\delta h_{{\kb_3}}^{}\rangle' 
&= \frac{1}{\,2\,}\lambda\hp \bar{h} H^2\biggl\{\!\hsm 
\(\!\frac{1}{\,k_1^3k_2^3\,}\!+\!\mathrm{2\,perm.}\!\)\!\!\(\!\!-N_e\!+\!\gamma\!-\!\frac{4}{\,3\,}\!\)
\\
& \hspace*{19mm}
+\!\frac{1}{\,k_1^2k_2^2k_3^2\,} \!-\!\(\!\!\frac{1}{\,k_1 k_2^2 k_3^3\,}
\!+\hsm\mathrm{5\,perm.}\!\)\!\hsm\biggr\} \hp,
\end{aligned}
\end{equation}
where $N_e$ is the number of e-folds of the expansion from the time 
at which the fluctuation mode with momentum $\,k_t^{}$ 
first passed outside the horizon [\,${k_t^{-1}}\!\hsm\simeq\! {1}\hsm /(a_k^{} H)\hp$] 
until the end of inflation.\
The e-folding number $N_e$ is derived as follows:
\begin{equation}
N_e = \ln\!\frac{\,a_\text{end}^{}\,}{a_k^{}}
=\ln\!\frac{\,-(\hsm H\tau_f^{}\hsm)^{-1}\,}{\,{k_t^{}}/{H}\,}
=-\ln (k_t^{}|\tau_f|) \thicksim 60 \hp. 
\end{equation}

For the second term on the right-hand side of Eq.\eqref{3ptz_h}, it can be expressed as a four-point correlation function of $\delta h$, 
and to the leading order it is given by the product of two two-point correlation functions:
\begin{align}
\label{3pth_NL}
& z_1^2z_2^{} \langle\delta h^4\rangle\hsm 
({\kb_1^{}},{\kb_2^{}},{\kb_3^{}})
= \frac{\,z_{1}^2z_2^{}\,}{2}\!\!
\int\!\!\!\frac{\,{\d}^{3}{\kb_0}\,}{(\,2\pi)^{3}\,}
\langle\delta h({\kb_1^{}}) \delta h({\kb_2^{}})\delta h({\kb_0^{}})
\delta h({\kb_3^{}\hsm\!-\!\kb_0^{}})\rangle \!+\!(\text{2\,perm.})
\nn \\
&=\frac{\,z_{1}^2z_2^{}\,}{2}\!\! 
\left[\hsm\int\!\!\!\frac{\,{\d}^{3}{\kb_0^{}}\,}{\,(2 \pi)^{3}\,}
\langle\delta h({\kb_1^{}}) \delta h({\kb_0^{}})\rangle
\langle\delta h({\kb_2^{}}) \delta h({\kb_3^{}\hsm\!-\!\kb_0^{}})\rangle 
\!+\! ({\kb_1^{}}\!\!\leftrightarrow\!{\kb_2^{}})\hsm\right] \hsm\!+\!  (2\,\rm{perm.})
\nn\\
&=\frac{\,z_{1}^2z_2^{}\,}{2}\!
\bigg[\hsm\!\int\!\! {{\d}^{3}{\kb_0}}
\left(2\pi\right)^3\!\delta^3({\kb_1^{}}\!+\!{\kb_0^{}}) \delta^3({\kb_2^{}}
\!+\!{\kb_3^{}}\!-\!{\kb_0^{}}) \frac{H^4}{\,4k_1^3k_2^3\,}
\!+\!(\kb_1^{}\!\!\leftrightarrow\!{\kb_2^{}}) \!\bigg] \hsm\!+\!(2\,\rm{perm.})
\nn\\
&= (2\pi)^3\delta^3 ({\kb_1^{}}\!+\hsm {\kb_2^{}}\!+\!{\kb_3^{}})
z_{1}^2z_{2}^{}\!\hsm\left[\!\frac{H^4}{\,4k_{1}^3k_2^3\,}\!+\! (2\,\rm{perm.)}\hsm\right] \!.
\end{align}

Combined with the part arising from the nonlinear rate, we derive the three-point correlation function of 
the comoving curvature perturbation $\hp\zeta\hp$ from Higgs-modulated reheating as follows:
\begin{equation}
\begin{aligned}
\label{3pttotal_meanfield}
\langle\zeta_{{\kb_1}}^{}\zeta_{{\kb_2}}^{}\zeta_{{\kb_3}}^{}\rangle_h^\prime
=&\frac{\,z_1^3\,}{2}{\lambda\hp\bar{h} H^2}\hsm 
\biggl\{\!\biggl[\!\frac{1}{\,k_1^3 k_2^3\,}\!+\!(\rm{2\,perm.})\!\biggr]\!\hsm 
\Bigl(\!-\!N_e\!+\!\gamma\!-\!\frac{4}{\hp 3\hp}\hsm\Bigr) 
\!+\!\frac{1}{\,k_1^2 k_2^2 k_3^2\,} 
\\
& \hspace*{5.4mm}
-\!\biggl[\!\frac{1}{\,k_1 k_2^2 k_3^3\,}\!+\!(\rm{5\,perm.})\!\biggr]\!\hsm\biggr\}
\!+\!\frac{\,z_1^2z_2^{}\,}{4}H^4\biggr[\!\frac{1}{\,k_1^3 k_2^3\,}\!+\!(2\,\rm{perm.})\!\biggr].
\end{aligned}
\end{equation}
In the above, we have included the non-Gaussianity contributions from both the nonlinear term and the Higgs self-interactions, whereas the previous studies only include the former\,\cite{Wands10040818,Ichikawa:2008ne,DeSimone12106618,Karam210302569}.\

\vspace*{1mm}
\subsection{\hspace*{-2.5mm}Probing the Seesaw Scale through Local Non-Gaussianity}
\label{sec:4.3}
\vspace*{1mm}

In the previous subsection, we derived the three-point correlation function  
$\langle\zeta_{{\kb_1}}^{}\zeta_{{\kb_2}}^{}\zeta_{{\kb_3}}^{}\rangle_h^\prime$ 
for the curvature perturbation.\ 
For this subsection, we further compute the local non-Gaussianity
$f_\text{NL}^{\text{local}}$ originating from the Higgs-modulated reheating 
in our model, with which we study the probe of the seesaw mechanism in this framework.\

The three-point correlation function 
$\langle\zeta_{{\kb_1}}^{}\zeta_{{\kb_2}}^{}\zeta_{{\kb_3}}^{}\rangle_h^\prime$ 
contributes to three distinct classes of non-Gaussian shape templates
(local, equilateral, and orthogonal types)
according to Refs.\,\cite{Planck9}\cite{Planck10}.\ 
We present a systematic analysis of the specific contributions of $\langle\zeta_{{\kb_1}}^{}\zeta_{{\kb_2}}^{}\zeta_{{\kb_3}}^{}\rangle_h^\prime$ 
to these templates as in Appendix\,\ref{app:B}.\ 
We evaluate the amplitude of a given non-Gaussian shape template, parameterized by $f_\text{NL}^{i}$, 
which is expressed as a function of the model parameters,
\begin{equation}
f_\text{NL}^{i}=f_\text{NL}^{i}\hsm 
(M,y_\nu^{},\lambda,H_{\rm{inf}}^{},m_\phi^{},\Lambda),
\end{equation}
where ``$i$'' represents the type of non-Gaussian shape template. 
For the present analysis, we choose a set of relevant parameters having benchmark values, 
as shown in Table\,\ref{tab:1}.\ 
The amplitude of the comoving curvature perturbation power spectrum 
$\hp\mathcal{P}_{\zeta}^{}\hp$ is taken as, 
$\ln(10^{10} \mathcal{P}_{\zeta}^{}) \hsm\simeq\hsm 3.047\hp$, 
according to the Planck-2018 data\,\cite{Planck1}\cite{Planck6}.\  
The SM Higgs self-coupling is set to be $\lambda\!=\!0.01\hp$.\ 

\begin{table}[H]
\centering
\begin{tabular}{c||c|c|c|c|c|c}
\hline\hline
Parameters & $\mathcal{P}_{\zeta}^{}$ & $N_e$ & $H_{\rm{inf}}^{}$ & $m_\phi^{}$ & $\Lambda$ & $\lambda$ 
\\
\hline
Values & $2.1\!\hsm\times\hsm\! 10^{-9}$ & 60 &
$(1,\hp 3)\!\hsm\times\hsm\!10^{13}\hp\text{GeV}$ 
& $40H_{\rm{inf}}^{}$ & $60H_{\rm{inf}}^{}$ & $0.01$ 
\\
\hline\hline
\end{tabular}
\label{tab_para}
\caption{The relevant parameters with benchmark values chosen for the present analysis.}
\label{tab:1}
\end{table}

\vs

We find that the non-Gaussianity predicted by our model mainly belongs to the local type.\ 
This is due to the following reasons.\ First, in the three-point correlation function arising from the Higgs self-coupling, 
as described in Eq.\eqref{3pth_SM1}, the local type is amplified by the number of e-folds $N_e$, 
whereas the other two shapes do not receive such enhancement.\ Second, as shown in Eq.\eqref{3pth_NL}, 
the contribution from the nonlinear term exclusively generates local-type non-Gaussianity.\
Hence, for the present study, we primarily focus on the magnitude of local non-Gaussianity (NG) 
as predicted by our model, which can be approximately expressed by the following:\footnote{%
The accurate formula is given in Eq.\eqref{fNL_complete_result} of Appendix\,\ref{app:B}.}
\begin{equation}
\label{fnllocal_appro}
f_{\mathrm{NL}}^{\mathrm{local}} \simeq 
-\frac{10}{3}\frac{z_1^3H^3}{\,(2\pi)^4 \mathcal{P}_{\zeta}^{2}\,}\!
\(\!\!\frac{\,\lambda\hp\bar{h}\,}{\,2H\,}N_e \!-\!\frac{\,z_2^{}H\hp\,}{4z_1^{}}\!\)\!.
\end{equation}
It is found that the local-type non-Gaussianity $f_\rm{NL}^{}$ arising from the nonlinear 
term\,\cite{Wands10040818,Ichikawa:2008ne,DeSimone12106618}\cite{Litsa201111649} 
is given by
\begin{equation} 
\label{fnlotherpaper}
f_{\mathrm{NL}}^{\mathrm{local}}(\rm{NT}) = \frac{5z_2^{}}{\,6z_1^2\,}
= 5\biggl(\! 1\!-\!\frac{\,\Gamma_0''\hp\Gamma_0^{}\,}{\Gamma_0^{\prime\,2}}\!\biggr).
\end{equation}
We note that this result is compatible with the scenario where the Higgs fluctuation 
is the only source of primordial fluctuation, i.e., $R\!=\!1\hp$.\ 
Additionally, these studies assumed the absence of intrinsic non-Gaussianity in the Higgs field 
due to its self-coupling 
$\Delta\mathcal{L}\!=\!-\sqrt{-g\,}(\lambda \bar{h})\delta h^3$, 
whose contribution, however, could be significant
as will be shown below.\  
For the present analysis, we incorporate both the intrinsic non-Gaussianity of the Higgs self-interactions and 
the nonlinear term.\ 
By assuming $R\!=\!1$ and purely Gaussian $\delta h$, 
our result will reduce to Eq.\eqref{fnlotherpaper} as shown 
in Appendix\,\ref{app:B}. %
However, in the general case, Eq.\eqref{fnllocal_appro} includes two distinct contributions: 
one originating from the Higgs three-point correlation function induced by Higgs self-interactions, 
and another from the nonlinear term.

\vs

In the following, we will demonstrate that these contributions 
are significant for sample inputs of
the Higgs self-coupling $\lambda\!\!=\hsm\!0.01$ and the e-folding $N_e\!\!=\!60\hp$.\  
For illustration, we provide three sets of representative benchmark points 
where the seesaw scale is chosen as 
$M\!=\!(3,\hp 7,\hp 14)H_{\rm{inf}}^{}$, respectively.\  
For each given seesaw scale $M$, we input the sample values of the Higgs-neutrino Yukawa coupling as 
$y_\nu^{}\!=\!(0.3,\hp 0.6)\hp$, respectively.\
With these, we evaluate the contributions to the local non-Gaussianity $f_{\mathrm{NL}}^{\mathrm{local}}$
by the Higgs self-interaction and nonlinear term,
which are presented and compared in Table\,\ref{tab:2} for the three sets of benchmarks of
the neutrino seesaw scale $M$ and Yukawa coupling $y_{\nu}^{}\hp$.\
For comparison, we also vary the Higgs self-coupling as $\lambda\!=\!0.02$ 
and present the corresponding contributions in the parentheses of each entry for the ratio $R$
and the non-Gaussianities 
$f_{\mathrm{NL}}^{\rm{local}}$(HSC),
$f_{\mathrm{NL}}^{\rm{local}}$(NT), and 
$f_{\mathrm{NL}}^{\rm{local}}$(total), as shown in Table\,\ref{tab:2},
where HSC, NT, and total stand for the contributions by the Higgs self-coupling,
the nonlinear term, and their sum, respectively.\ 
This demonstrates that, in the neutrino seesaw parameter space, 
the Higgs self-interaction gives the dominant contribution to non-Gaussianity, 
whereas the nonlinear term provides a sizable but subdominant contribution.\ 
It also shows that the non-Gaussianity measurements 
are sensitive to the size of the Higgs self-coupling 
$\lambda$ at the seesaw scale.

\begin{table}[t]
\centering
\resizebox{\textwidth}{!}{
\begin{tabular}{c||cc|cc|cc}
\hline\hline
 Benchmarks & $A_1$ & $A_2$ & $B_1$ & $B_2$ & $C_1$ & $C_2$ \\ 
\hline\hline
$M/H_\text{inf}$& 3 & 3 & 7 & 7 & 14 & 14 \\ 
\hline
$y_\nu$& 0.3 & 0.6 & 0.3 & 0.6 & 0.3 & 0.6 \\ 
\hline
$R$\!& 0.02\,(0.01) & 0.08\,(0.06) & 0.03\,(0.02) & 0.10\,(0.07) & 0.03\,(0.02) & $0.13\,(0.09)$ \\ 
\hline
$f_{\mathrm{NL}}^{\rm{local}}$(HSC)& 0.037\,(0.019) & 2.38\,(1.19) & 0.07\,(0.036) & 4.55\,(2.28) & $-0.15\,(-0.08)$ & $-9.8\,(-4.9)$ \\ 
\hline
$f_{\mathrm{NL}}^{\rm{local}}$(NT)& 0.008\,(0.003) & 0.49\,(0.17) & 0.01\,(0.005) & 0.93\,(0.33) & $-0.03\,(-0.01)$ & $-2.0\,(-0.7)$\\ 
\hline
$f_{\mathrm{NL}}^{\rm{local}}$(total) & 0.045\,(0.021) & 2.86\,(1.36)& 0.09\,(0.041) & 5.49\,(2.61)& $-0.18\,(-0.09)$ & $-11.8\,(-5.6)$ \\ 
\hline\hline
\end{tabular}
}
\vspace*{-1mm}
\caption{Comparison of major contributions to the non-Gaussianity 
arising from the Higgs self-interaction and the nonlinear term 
for three sets of benchmark points with specific neutrino seesaw scale $M$  
and the Higgs-neutrino Yukawa coupling $y_\nu^{}$.\
The ratio $R$ of Eq.\eqref{rh} is presented in the 4th row.\  
The predicted values of $f_{\mathrm{NL}}^{\rm{local}}$(HSC),
$f_{\mathrm{NL}}^{\rm{local}}$(NT) and 
$f_{\mathrm{NL}}^{\rm{local}}$(total) are shown in the 5th, 6th and 7th rows respectively,
which corresponds to the contributions of the Higgs self-coupling (HSC), the nonlinear term,
and their sum.\ 
In each entry, the number outside (inside) the parentheses corresponds to 
the input of Higgs self-coupling
$\lambda\!=\!0.01\,(0.02)$.\  
}
\label{two_contributions}
\label{tab:2}
\end{table}

\vs 

In the seesaw mechanism, the mass of the light neutrino $\nu$ 
is determined by the Majorana mass $M$ 
and the neutrino-Higgs Yukawa coupling $y_\nu^{}$ through the formula:
\begin{equation}
\label{mass_left_nu}
m_\nu^{} = \frac{~y_\nu^2\hp v^2\,}{\,2M\,}\thicksim {O}(0.05)\hp\text{eV}\,,
\end{equation}
where $v\!\simeq\!246\hp${GeV} is the vacuum expectation value of the Higgs field 
after electroweak symmetry 
breaking.\footnote{%
In Eq.\eqref{mass_left_nu}, $m_\nu^{}$
is the mass of the light neutrino $\nu$ at the electroweak scale.\ 
During and after the inflation, 
we can include the renormalization-group (RG) running effect 
for this neutrino mass at the corresponding Hubble scale of ${O}(10^{13})$GeV, 
which is about 30\% larger than its low-energy value 
at the electroweak scale\,\cite{Antusch0305273}\cite{He11042654}.}\ 
The neutrino oscillation data give, 
$\Delta m_{31}^{2}\hsm\simeq 2.5 \!\times\! 10^{-3}\,${eV}$^{2}$ and 
$\Delta m_{21}^{2}\hsm\simeq\!7.5\hsm\times\hsm 10^{-5}\,${eV}$^{2}$ \cite{nuMassFit}, 
which require at least one of the light neutrino masses to be  
$m_\nu^{} \!\gtrsim\hsm 0.05\hp${eV}.\ 
With this, we can estimate the neutrino seesaw scale to be around $10^{14}\hp${GeV} for 
the Yukawa coupling $y_\nu^{}\!=\!{O}(1)\hp$.\ 
On the other hand, cosmological measurements based on the CMB alone already set upper bounds 
on the sum of the light neutrino masses, approximately 
$\sum\!m_\nu^{} \!\lesssim\! 0.26\hp${eV} \cite{Planck6}.\
When combined with observations of large-scale structures, 
this bound can be tightened to $O(0.1)\hp${eV}{\hp}.\
For instance, the eBOSS Collaboration\,\cite{nuSum1} 
placed a 95\% upper bound $\sum\!m_\nu^{} \!\hsm\lesssim\hsm 0.10\hp${eV} 
and the DES Collaboration\,\cite{nuSum2} set a constraint 
$\sum\!m_\nu^{} \!\lesssim\hsm 0.13\hp${eV} at 95\%\,C.L.\ 
Given the existing cosmological upper bounds on 
the neutrino mass sum $\sum\!m_\nu^{}$
and two mass-squared differences $(\Delta m_{3\ell}^{2},\hp \Delta m_{21}^{2})$
measured by oscillation data, 
we can determine the largest light neutrino mass for the normal mass ordering (NO) and
inverted mass ordering (IO) from the following conditions:
\beqs 
\label{eq:NOm3-IOm2}
\begin{align}
\label{eq:NO-m3}
\rm{NO}\!:~~~ & 
\sum\!m_\nu^{} =m_3^{}\!+\!\sqrt{m_3^2\!-\!\Delta m_{31}^2\,}+\!\sqrt{m_3^2\!-\!\Delta m_{31}^2\!+\!\Delta m_{21}^2\,}\,,
\\
\label{eq:IO-m2}
\rm{IO}\!:~~~ & 
\sum\!m_\nu^{} =m_2^{}\!+\!\sqrt{m_2^2\!-\!\Delta m_{21}^2\,} + \sqrt{m_2^2\!-\!|\Delta m_{32}|^2\,} \,,
\end{align} 
\eeqs 
where $m_3^{}$ is the largest mass for the normal ordering and $m_2^{}$ is the largest mass for  
the inverted ordering.\

If we choose the weaker bound on the light neutrino mass sum 
$\sum\!m_\nu^{} \hsm\!\!\lesssim\!\! 0.26\hp${eV} 
by CMB alone\,\cite{Planck6}, we can solve the conditions \eqref{eq:NOm3-IOm2} to obtain the largest light neutrino mass
to be around $0.1\hp${eV} for either NO or IO.\ 
In comparison, choosing the stronger bound $\sum\!m_\nu^{} \!\lesssim\! 0.13\hp${eV} \cite{nuSum2}, 
we find from the conditions \eqref{eq:NOm3-IOm2}
that the largest light neutrino mass is 
$m_3^{}\hsm\simeq\hsm 0.06\,${eV} for the NO and 
$m_2^{}\simeq\hsm 0.05\,${eV} for the IO.\

\begin{figure}[t]
\vspace*{-5mm}
\centering
\begin{subfigure}
\centering
\hspace{0.79cm}\includegraphics[width=0.92\textwidth]{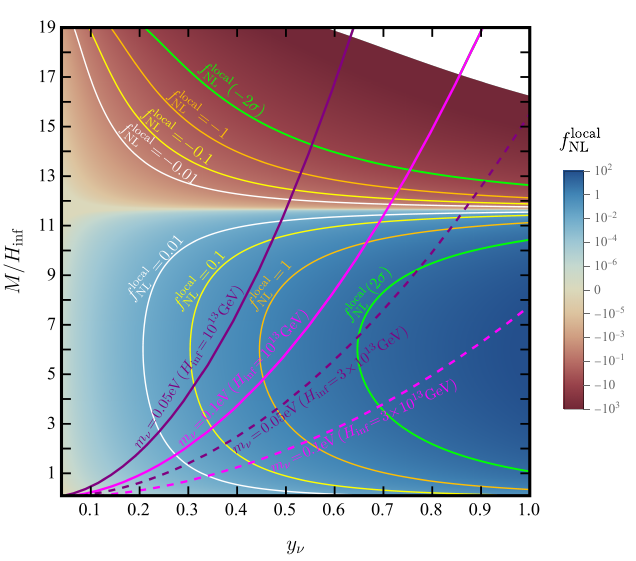}
\end{subfigure}\\[0em]
\begin{subfigure}
\centering
\includegraphics[width=0.7\textwidth]{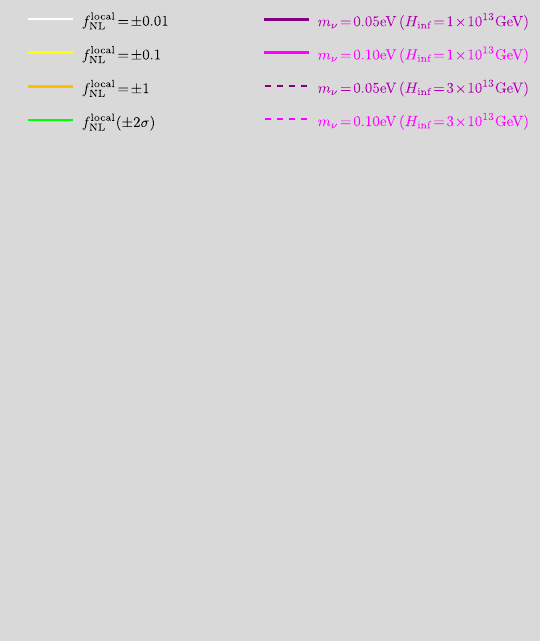}		
\end{subfigure}
\caption{Prediction of the non-Gaussianity $f_{\rm{NL}}^{\rm{local}}$ 
from the seesaw parameter space of heavy neutrino mass scale $M$ versus Yukawa coupling $y_\nu^{}\hp$, where the SM Higgs self-coupling is input as $\lambda\!=\!0.01\hp$, and the Hubble parameter 
during inflation is set as $H_\text{inf}^{}\!=\!10^{13}\text{GeV}$ and 
$3\!\times\! 10^{13}\text{GeV}$, respectively.\ The blue region represents the parameter space 
with positive $f_{\mathrm{NL}}^{\mathrm{local}}$, whereas the red region represents the 
parameter space with negative $f_{\mathrm{NL}}^{\mathrm{local}}$.\ 
The green contour depicts the $2\sigma$ bound based on Planck-2018 data, 
corresponding to $-11.1 \hsm\leqq\hsm f_{\mathrm{NL}}^{\mathrm{local}}\!\leqq\hsm 9.3\hp$.\ 
The colored region satisfies the requirement of $R\!<\! 1\hp$ and the uncolored region
in the upper-right corner corresponds to $R\!\geqq\! 1\hp$. The seesaw predictions for the light neutrino mass of 
$m_\nu^{}\!\!=\!0.1{\hp}$eV (pink curves) and 
$m_\nu^{}\!\!=\!0.05{\hp}$eV (purple curves), and  
for the Hubble parameter $H\!\!=\!\!10^{13}\hp$GeV (solid curves)
and $H\!=\!3\times\!10^{13}\hp$GeV (dashed curves) are given. }
\label{fig:6}
\vspace*{-1.5mm}
\end{figure}

\begin{figure}[t]
\vspace*{-5mm}
\centering
\begin{subfigure}
\centering
\hspace{0.79cm}\includegraphics[width=0.92\textwidth]{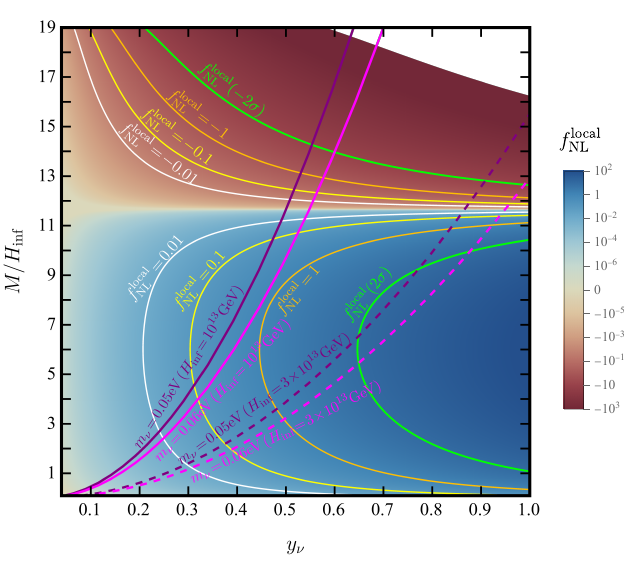}
\end{subfigure}\\[0em]
\begin{subfigure}
\centering
\includegraphics[width=0.7\textwidth]{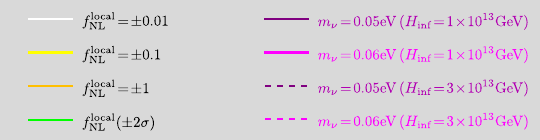}		
\end{subfigure}
\caption{Prediction of the non-Gaussianity $f_{\rm{NL}}^{\rm{local}}$ 
from the seesaw parameter space of heavy neutrino mass scale $M$ versus Yukawa coupling $y_\nu^{}\hp$, where the SM Higgs self-coupling is input as $\lambda\!=\!0.01\hp$, and the Hubble parameter 
during inflation is set as $H_\text{inf}^{}\!=\!10^{13}\text{GeV}$ and 
$3\!\times\! 10^{13}\text{GeV}$, respectively.\ The seesaw predictions for the light neutrino mass of 
$m_\nu^{}\!\!=\!0.06{\hp}$eV (pink curves) and 
$m_\nu^{}\!\!=\!0.05{\hp}$eV (purple curves), and  
for the Hubble parameter $H\!\!=\!\!10^{13}\hp$GeV (solid curves)
and $H\!=\!3\times\!10^{13}\hp$GeV (dashed curves) are given.}
\label{fig:7}
\end{figure}

To probe the seesaw parameter space in our model, we compute the non-Gaussianity 
$f_{\rm{NL}}^{\rm{local}}$ 
for different values of the seesaw scale $M$ and neutrino-Higgs Yukawa coupling $y_\nu^{}$,
and set the SM Higgs self-coupling $\lambda\!=\!0.01\hp$.\ 
We present our numerical findings in Fig.\,\ref{fig:6} and Fig.\,\ref{fig:7}.\ 
The colored region satisfies the requirement $R \!\!<\!\! 1\hp$, 
whereas the white region in the upper-right corner of each plot corresponds to 
$R\!>\! 1\hp$ and is excluded.\
The region with blue color corresponds to $f_{\mathrm{NL}}^{\mathrm{local}}\!>\!0$, 
and the red regions represent  $f_{\rm{NL}}^{\rm{local}}\!<\!0\hp$.\ 

In Figs.\,\ref{fig:6} and \ref{fig:7}, the green contours describe the existing $2\hp\sigma$ bounds,  
$-11.1 \!\leqq\! f_{\mathrm{NL}}^{\mathrm{local}}
 \!\leqq\! 9.3\hp$, 
as given by the Planck-2018 data\,\cite{Planck9}.\ 
Moreover, we display contours for 
$f_\text{NL}^{}\!\!=\!\pm1,\,\pm0.1,$ $\pm0.01$, 
plotted as orange, yellow, and white curves, respectively.\  
These contours represent the potential sensitivity reaches of 
the ongoing and future observations, 
such as those from DESI\,\cite{DESI:2016fyo}, CMB-S4\,\cite{Abazajian:2019eic}, 
Euclid\,\cite{Euclid},  SPHEREx\,\cite{SPHEREx:2014bgr}, LSST\,\cite{LSST}, 
and SKA\,\cite{SKA} experiments.\ 
We present the seesaw predictions for the light neutrino mass of 
$m_\nu^{}\!\!=\!0.1 \,(0.06){\hp}$eV (pink curves) and 
$m_\nu^{}\!\!=\!0.05{\hp}$eV (purple curves), and  
for the Hubble parameter $H\!\!=\!\!10^{13}\hp$GeV (solid curves)
and $H\!=\!3\times\!10^{13}\hp$GeV (dashed curves), 
where we have included the renormalization-group
running effects for the light neutrino mass at the Hubble scale.\ 
It shows that a larger value of the Hubble parameter will shift the pink and purple curves 
toward the right-hand-side region 
with larger Yukawa coupling $y_\nu^{}\hp$.\

For the local-type non-Gaussianity (NG) 
$f_{\mathrm{NL}}^{\mathrm{local}}\!\!>\hsm\!0\hp$, 
we see from Fig.\,\ref{fig:6} 
that the existing measurements of Planck-2018 (shown as the $2\sigma$ green contours)
already have sensitivity to probe 
the case of a light neutrino mass around $m_\nu^{}\!\!=\!0.05\hp\text{eV}$ for Hubble parameter 
$H\!\!=\! 3\hsm\!\times\hsm\!10^{13}\hp$GeV, as shown by the purple dashed curve;
whereas for a larger light neutrino mass around $m_\nu^{}\!\!=\!0.1\hp\text{eV}$ 
with $H\!\!=\!\! 3\!\times\!10^{13}\hp$GeV, a large portion of the parameter region in our model
is already excluded by the Planck-2018 data as shown by the pink dashed curve.\   
Then, for a smaller Hubble parameter $H\!\!=\!\!10^{13}\hp$GeV, Fig.\,\ref{fig:6} shows that
for inputting the light neutrino mass of range $m_\nu^{}\!=\!(0.05\!-\!0.1)\hp$eV,  
our seesaw predictions (purple and pink solid curves) have significant parameter space 
consistent with the current bound of Planck-2018 
(the $2\sigma$ green contour),  
but they can be further probed by the
improved non-Gaussianity measurements of future experiments as shown by the 
(orange,\,yellow,\,white) contours.\ 
For instance, probing the case of a light neutrino mass $m_\nu^{} \!=\! 0.05\hp\text{eV}$ 
and Hubble parameter $H\!=\!10^{13}\hp$GeV (purple solid curve)
requires a sensitivity to $f_{\rm{NL}}^{\rm{local}}\!\thicksim\! 0.1$ or even 
$f_{\rm{NL}}^{\rm{local}}\!\thicksim\! 0.01$, 
depending on the seesaw mass scale $M$.\ 
In contrast, the case of $m_\nu^{} \hsm\!=\! 0.1\hp${eV} 
(pink solid curve) can be more effectively probed in the near future.\ 

\vs 

Then, for the case of $f_{\rm{NL}}^{\rm{local}}\hsm\!<\!0\hp$, Fig.\,\ref{fig:6} shows 
that the existing bounds of Planck-2018 measurements 
(shown as the $\hsm -2\sigma$ green contour) 
have already excluded a large portion of the seesaw parameter space in our model.\
For instance, for a light neutrino mass $m_\nu^{}\!=\!0.1\hp$eV\,($0.05\hp$eV)
and Hubble parameter $H\!=\!10^{13}\hp$GeV, the parameter space with seesaw scale 
$M\!\gtrsim\!13 H_{\rm{inf}}^{}$ 
($M\!\!\gtrsim\!\!14 H_{\rm{inf}}^{}$) 
is excluded by the Planck-2018 data in our model.\ 
The future measurements of non-Gaussianity under planning will be able to extensively probe 
the seesaw parameter space in our model with $M\!\gtrsim\!12H_{\rm{inf}}^{}\hp$
in the case of $f_{\rm{NL}}^{\rm{local}}\hsm\!<\!0\hp$.\ 

\vspace*{0.5mm}

In parallel, we further present the seesaw predictions in Fig.\,\ref{fig:7}
for choosing the largest light neutrino mass $m_{\nu}^{}\!\!=\!0.06\,$eV [representing normal ordering (NO)
of light neutrino masses] or  $m_{\nu}^{}\!=\!0.05\,$eV [representing the inverted ordering (IO)], 
as depicted by the pink curves and purple curves, respectively.\  
We see that, for $H\!=\!3\!\times\!10^{13}\hp$GeV  (dashed curves), 
the existing constraints of Planck-2018 (green contours)
already have sensitivity to probe a part of the seesaw parameter space in our model 
with the largest light neutrino mass
corresponding to the NO versus IO of light neutrinos.\ 
Figure\,\ref{fig:7} further shows that, 
for a smaller Hubble parameter 
$H\!=\!10^{13}\hp$GeV (solid curves), the future non-Gaussianity measurements
with sensitivities of $f_{\rm{NL}}^{\rm{local}}\!\lesssim\! 0.1$
can sensitively probe the seesaw predictions in our framework with   
$m_{\nu}^{}\!=\!0.06\,$eV (pink solid curve) versus  
$m_{\nu}^{}\!=\!0.05\,$eV (purple solid curve).\

\vspace*{1mm}

From Figs.\,\ref{fig:6}  and \ref{fig:7}, we see that the cosmological non-Gaussianity measurements
are rather sensitive to the value of light neutrino mass scale $m_\nu^{}\hp$,  
which is also constrained by the low-energy neutrino experiments, 
especially the determination of the light neutrino mass ordering 
(as will be measured by the neutrino oscillation experiments 
JUNO\,\cite{JUNO:2015zny} and DUNE\,\cite{DUNE:2016hlj}) and the determination of light neutrino mass scale
(by the on-going and 
future neutrinoless double-$\beta$ decay experiments\,\cite{Dolinski:2019nrj}).\ 
Hence, our analyses of 
Figs.\,\ref{fig:6} and \ref{fig:7} 
demonstrate the important interplay
between the light neutrino mass determinations 
by the low-energy experiments 
and the high-scale cosmological measurements 
on the non-Gaussianity in our model. 
 
\vs  
 
As a final point,  
we note in Figs.\,\ref{fig:6} and \ref{fig:7}
that the non-Gaussianity
$f_{\rm{NL}}^{\rm{local}}$ has sign flip around the value of 
$M\!\!\simeq\!\!12 H_{\rm{inf}}^{}$.\ This can be understood as follows.\ 
We note that in the parameter space considered in this study, the sign of 
the non-Gaussianity is mainly determined by the contribution from the 
Higgs self-interaction.\ As shown in Eq.\eqref{fnllocal_appro}, 
it is linked to the coefficient $z_1^{}$, which depends on the derivative 
of the inflaton decay width.\ 
Since the Higgs field during inflation enters the decay width through neutrino seesaw 
as shown in Eq.\eqref{mass_nu}, the derivative 
$\Gamma_0'\!=\!{\td \Gamma_\text{reh}^{}}/{\td h_\text{inf}^{}}$ is given by\footnote{%
	In fact, a quantity (with dependence on the Higgs field $h$) could be converted into a function of 
	the right-handed neutrino mass $M_N^{}$, including the left-handed neutrino mass $m_\nu^{}$ and the mixing angle $\theta$.}
\begin{equation}
	\Gamma_0'=\frac{\td \Gamma_\text{reh}}{\,\td M_N^{}\,}
	\frac{\,\td M_N^{}\,}{\td h_\text{inf}} \,,
\end{equation}
where $\Gamma_0'\!\propto\hsm\!z_1^{}$ according to Eq.\eqref{z1z2def}
and the derivative 
$\frac{\,\td M_N^{}\,}{\td h_\text{inf}^{}}
\!\hsm\simeq\hsm\!\frac{\,y_\nu^2 h_\text{reh}^{}\,}{M^2}
\frac{\,\td h_\text{reh}\,}{\td h_\text{inf}}$
is nonzero.\ Hence, the condition $z_1^{}\!\!=\!0\hp$ requires 
$\frac{\,\td \Gamma_\text{reh}\,}{\td M_N}\!=\!0\hp$.\  
At the time of reheating, we have $M_N^{}\!\simeq\! M$, 
at which we compute the derivative,
\begin{equation}
	\frac{\td \Gamma_\text{reh}}{\td M_N}
	=\frac{\,m_\phi^{}M\,}{\,8\pi\Lambda^2~}\!\!\left[\!
	\(\!\!1\!-\!\frac{\,M^2\,}{m_\phi^2}\!\)^{\hsm\!\!2}\!\!
	\(\!\!1\!-\!\frac{\,4M^2\,}{m_\phi^2}\!\)^{\hsm\!\!\frac{1}{2}}
	\!\!-\!\frac{\,8M^2\,}{m_\phi^2}\hsm\right]\!\!\!
	\(\hsm\!1\!-\!\frac{\,4M^2\,}{m_\phi^2}\!\)^{\hsm\!\!-\frac{1}{2}} . 
\end{equation}
From the condition $\frac{\,\td \Gamma_\text{reh}\,}{\td M_N}\!=\!0\hp$, we 
solve the ratio of the seesaw scale $M$ over the inflaton mass $m_\phi^{}$ 
as follows:
\begin{equation}
	\label{sign_transition_line}
	\frac{M^{}}{\,m_\phi^{}\,}\simeq 0.29 \,.
\end{equation}
Thus, for the input of $m_\phi^{}\!=\!40H_{\rm{inf}}^{}$, the sign-transition point
$z_1^{}\!=\!0$ corresponds to $M\!\simeq\!11.6H_{\rm{inf}}^{}\hp$.\
This nicely explains that the transition between $f_{\rm{NL}}^{\rm{local}}\!>\!0\hp$
and $f_{\rm{NL}}^{\rm{local}}\!<\!0\hp$ 
in Figs.\,\ref{fig:6} and \ref{fig:7}
happens around the horizontal line of $M\!\simeq\!12H_{\rm{inf}}^{}\hp$.

\vspace*{1mm}
\subsection{\hspace*{-2.5mm}Dependence of Non-Gaussianity on the Higgs Self-Coupling}
\label{sec:4.4}
\vspace*{1mm}

\begin{figure}[t]
\vspace*{-5mm}
\centering
\begin{subfigure}
\centering
\hspace{0.79cm}\includegraphics[width=0.92\textwidth]{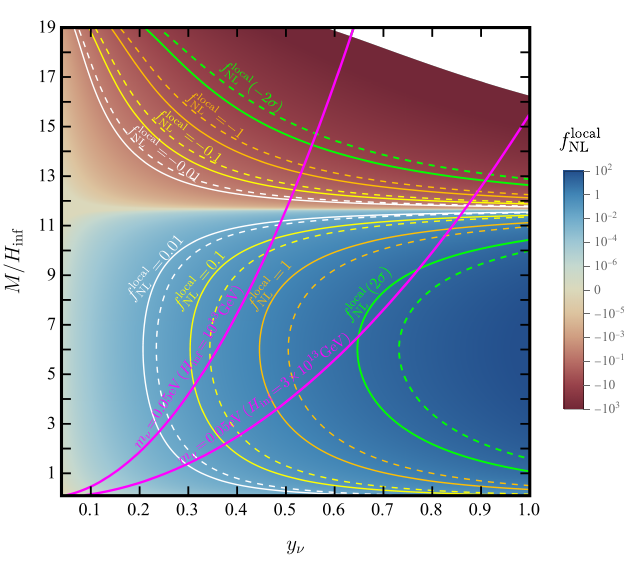}
\end{subfigure}\\[0em]
\begin{subfigure}
\centering
\includegraphics[width=0.7\textwidth]{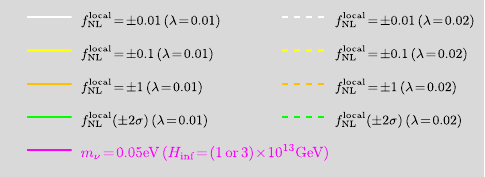}		
\end{subfigure}
\caption{Prediction of the non-Gaussianity $f_{\rm{NL}}^{\rm{local}}$ 
from the seesaw parameter space of heavy neutrino mass scale $M$ 
versus the Yukawa coupling constant $y_\nu^{}\hp$, 
where we input the SM Higgs self-coupling  
$\lambda\!=\!0.01\hp$ (solid curves) and $\lambda\!=\!0.02\hp$ (dashed curves), 
and the Hubble parameter during inflation is set as 
$H_\text{inf}^{}\!=\!10^{13}\text{GeV}$ and 
$3\!\times\! 10^{13}\text{GeV}$, respectively.\ 
The contours in solid curves represent the given bounds of non-Gaussianity $f_{\rm{NL}}^{\rm{local}}$ 
with Higgs self-coupling $\lambda\!=\!0.01\hp$, 
whereas the contours in dashed curves correspond to the bounds 
with Higgs self-coupling $\lambda\!=\!0.02\hp$.\
The seesaw predictions are presented by pink curves for the light neutrino mass of 
$m_\nu^{}\!\!=\!0.05{\hp}$eV, with the Hubble parameter $H\!\!=\!\!10^{13}\hp$GeV and $H\!=\!3\times\!10^{13}\hp$GeV, respectively.}
\label{fig:8}
\vspace*{-2mm}
\end{figure}

In this subsection, we analyze the dependence of non-Gaussianity (NG) on
the Higgs self-coupling ($\lambda$).\

As shown in Eq.\eqref{fnllocal_appro}, the local-type non-Gaussianity 
$f_{\mathrm{NL}}^{\mathrm{local}}$ receives contributions from 
both the Higgs self-coupling (HSC) term and the nonlinear term (NT).\
In Table\,\ref{tab:2}, we have presented numerically their individual contributions 
and their sum for comparison.\  
It shows that the Higgs self-coupling produces the dominant contribution to non-Gaussianity
(as shown in the 5th row), and the nonlinear term gives the subdominant contribution
(as shown in the 6th row).\ To examine the sensitivity of $f_{\mathrm{NL}}^{\mathrm{local}}$ 
to the Higgs self-coupling contribution in Table\,\ref{tab:2},
we vary the Higgs self-coupling from $\lambda\!=\!0.01$ to  $\lambda\!=\!0.02\hp$.\ 
It shows that, in each entry of Table\,\ref{tab:2}, 
the number outside (inside) 
the parentheses corresponds to the input of Higgs self-coupling
$\lambda\!=\!0.01\,(0.02)$.\ 
We see that increasing the $\lambda$ value by a factor 2 generally causes the 
reduction of both $f_{\mathrm{NL}}^{\mathrm{local}}$(HSC) and 
$f_{\mathrm{NL}}^{\mathrm{local}}$(NT) by about a factor of $\frac{1}{2}$ and $\frac{1}{3}\hp$,
respectively.\ These features can be explained by the following simple scaling behaviors:
\beq
\label{eq:fNL-lambda-0}
f_{\mathrm{NL}}^{\mathrm{local}}(\mathrm{HSC}) \propto \lambda^{-1}, 
\hspace*{8mm}
f_{\mathrm{NL}}^{\mathrm{local}}(\mathrm{NT}) \propto \lambda^{-\frac{3}{2}},
\eeq
which we have derived in Appendix\,\ref{app:C}.\ 
(We note that the above scaling behaviors are derived by using the mean-field approximation
which are valid for the parameter space under consideration, but not for arbitrarily 
small or large coupling $\lambda\,$.)\ 
These features are also reflected in Fig.\,\ref{fig:8}, 
where the Higgs self-coupling $\lambda\!=\!0.01$ corresponds to the 
$f_{\mathrm{NL}}^{\mathrm{local}}$ contours given by the solid curves,
and the value $\lambda\!=\!0.02$ corresponds to the $f_{\mathrm{NL}}^{\mathrm{local}}$ 
contours given by the dashed curves.\ 
(The $f_{\mathrm{NL}}^{\mathrm{local}}$ contours in solid curves are
the same as those in Figs.\,\ref{fig:6} and \ref{fig:7}.)\
In Fig.\,\ref{fig:8}, we show that, 
by inputting a larger Higgs self-coupling value 
$\lambda\!=\!0.02\,$,
the non-Gaussianity contours (given by dashed curves) 
impose weaker bounds 
on the seesaw parameter space of $(M,\,y_\nu^{})$ 
as compared to the contours (in solid curves) 
with a smaller coupling $\lambda\hsm\!=\!0.01\hp$.\footnote{%
This is in contrast to the conventional collider probe of the Higgs self-coupling $\lambda\hp$,
where a larger $\lambda$ value always produces stronger signals of the di-Higgs production\,\cite{He:2015spf}.}\   
The above analyses show that the measurements of non-Gaussianity
$f_{\mathrm{NL}}^{\mathrm{local}}$ are sensitive to the probe of the Higgs self-coupling $\lambda$
at the seesaw scale, which is quantitatively connected to the low-energy values of $\lambda$ 
(measured by the LHC and future high-energy colliders\,\cite{He:2015spf}) via the renormalization-group evolution.\ 
Hence, this also demonstrates the important interplay on probing the Higgs self-coupling $\lambda$ 
between the high-scale cosmological non-Gaussianity measurements and the TeV-scale collider measurements.\

\vspace*{1mm}

\begin{table}[t]
\centering
\begin{tabular}{c||cc|cc|cc}
\hline\hline
 Benchmarks & $A_1$ & $A_2$ & $B_1$ & $B_2$ & $C_1$ & $C_2$ \\ 
\hline\hline
$M/H_\text{inf}$& 3 & 3 & 7 & 7 & 14 & 14 \\ 
\hline
$y_\nu$& 0.3 & 0.6 & 0.3 & 0.6 & 0.3 & 0.6 \\ 
\hline
$R(\Lambda\hsm\!=\hsm 60H_\text{inf})$ & 0.021 & 0.083 & 0.026 & 0.10 & 0.033 & 0.13 \\ 
\hline
$R(\Lambda\hsm\!=\hsm 80H_\text{inf})$ & 0.010 & 0.039 & 0.012 & 0.048 & 0.015 & 0.062 \\ 
\hline
$R(\Lambda\hsm\!=\hsm 100H_\text{inf})$ & 0.005 & 0.021 & 0.007 & 0.027 & 0.009 & 0.034 \\ 
\hline
$f_{\mathrm{NL}}^{\rm{local}}(\Lambda\hsm\!=\hsm 60H_\text{inf})$ & 0.045 & 2.9 & 0.086 & 5.5 & $-0.18$ & $-12$ \\ 
\hline
$f_{\mathrm{NL}}^{\rm{local}}(\Lambda\hsm\!=\hsm 80H_\text{inf})$ & 0.004 & 0.29 & 0.009 & 0.55 & $-0.018$ & $-1.2$ \\ 
\hline
$f_{\mathrm{NL}}^{\rm{local}}(\Lambda\hsm\!=\hsm 100H_\text{inf})$ & 0.001 & 0.048 & 0.001 & 0.092 & $-0.003$ & $-0.20$ \\ 
\hline
\hline
\end{tabular}
\caption{Comparison of the ratio $R$ and the non-Gaussianity in our framework with different cutoff $\Lambda$
for three sets of benchmark points with specific neutrino seesaw scale $M$  
and the Higgs-neutrino Yukawa coupling $y_\nu^{}$.\
The ratio $R$ of Eq.\eqref{rh} with 3 benchmark cutoffs is presented in the 4th, 5th and 6th rows.\  
The predicted non-Gaussianity $f_{\mathrm{NL}}^{\rm{local}}$ with different cutoffs $\Lambda$ 
are shown in the last three rows.\ 
}
\label{tab:3}
\end{table}

\vs 

In passing, we comment on the effect of the cutoff scale $\Lambda$ 
for the $\phi\hp$-$N_R^{}$ interaction of Eq.\eqref{Total_Lagrangian}.\ 
It can affect the time of reheating completion $t_{\text{reh}}^{}$.\ 
A smaller $\Lambda$ leads to a larger inflaton decay rate and thus smaller $t_{\text{reh}}^{}$,
allowing less time for the Higgs field value to decrease 
[cf.\ Eq.\eqref{h0 theoretical solution formal}]  
and causing larger fluctuations, thereby increasing the non-Gaussianity.\
These are shown in Table\,\ref{tab:3}.\ 
Hence, further understanding of the UV dynamics of inflation for the determinations
of $H_{\rm {inf}}^{}$, $m_\phi^{}$, and $\Lambda$ 
would be beneficial for 
a definitive probe of the seesaw parameter space.\

\section{\hspace*{-2.5mm}Conclusions}
\label{sec:5} 
\label{Conclusion}

The conventional seesaw mechanism provides the most appealing resolution to the origin of tiny masses of active neutrinos.\  
However, the natural seesaw scale is as high as $10^{14}\,$GeV, 
posing a great challenge for experimental tests at particle colliders.\ 
Based on the fact that the natural neutrino seesaw scale 
can be around the upper range of the inflation scale, 
we proposed a new framework incorporating inflation and neutrino seesaw in which the inflaton primarily decays into heavy right-handed neutrinos.\
The inflaton couples to the right-handed neutrinos through an effective interaction that respects the shift symmetry.\ 
With the neutrino seesaw, 
we construct a {\it new realization of Higgs modulated reheating,} in which the fluctuations of Higgs field 
can modulate the inflaton decays and
contribute to the primordial curvature perturbation,
leading to non-Gaussian signatures at large scales.\ 
This provides, for the first time, an important means to directly probe the neutrino seesaw mechanism in the early Universe by measuring the non-Gaussian signatures.\
Moreover, it is appealing that this scenario also naturally provides 
an initial setup for the leptogenesis of matter-antimatter asymmetry
where sufficient right-handed neutrinos are generated after reheating.\ 
This approach further provides a new framework of 
the cosmological Higgs collider, in which the Higgs-modulated reheating 
is naturally realized by the inflaton decays into right-handed neutrinos through neutrino seesaw.\

In Section\,\ref{sec:2}, we studied the dynamics and evolution of the Higgs field during and after inflation.\ 
During the inflation, large quantum fluctuations 
of the Higgs field 
are generated due to the high scale of inflation $H_\text{inf}^{}\!=\!O(10^{13}\!-\!10^{14})\hp${GeV}.\ 
If the inflation lasts long enough, 
the distribution of the Higgs field 
would finally reach an equilibrium state and 
the average of the Higgs field takes 
a value around the Hubble scale, 
as shown in Eq.\eqref{hsquaremean}.\ 
After inflation, the value of the Higgs field $h(t)$ 
would oscillate and decrease.\ 
We also gave a semi-analytic formula 
\eqref{h0 theoretical solution formal} 
for the evolution of the Higgs field after inflation, 
which fits well with the exact numerical calculations as demonstrated in Fig.\,\ref{fig:3}.

\vs 

In Section\,\ref{sec:3}, we presented a new approach 
that incorporates both the inflation 
and neutrino seesaw mechanism.\  
The inflaton $\phi$ and the right-handed neutrino $N_{\!R}^{}$ 
can couple together through a unique dimension-5 operator 
as in Eq.\eqref{Total_Lagrangian} that respects the shift symmetry.\ 
In this approach, 
the inflaton primarily decays into right-handed neutrinos after inflation.\ 
With the neutrino seesaw mechanism, the Higgs field can influence the reheating process 
through the modulation of inflaton decays into right-handed neutrinos.\ 
In consequence, the primordial curvature perturbations are generated by fluctuations of the Higgs field.\ 
We established a relation \eqref{zeta_Higgs} 
between the curvature perturbations and the Higgs fluctuations, 
as illustrated in Fig.\,\ref{fig:4}.\
A further expanded formula for this relation up to $O(\delta h_\text{inf}^2)$
is given in Eq.\eqref{eq:zeta-expand}.\

\vs 

In Section\,\ref{sec:4}, we investigated the primordial local non-Gaussianity 
arising from the Higgs-modulated reheating through the inflaton decays into 
right-handed neutrinos within the seesaw mechanism.\  
We demonstrated that our method provides observable signatures of primordial non-Gaussianity, 
which can be used to probe the neutrino seesaw mechanism.\ We performed a full analysis for 
the two-point and three-point correlation functions of the comoving curvature perturbation 
$\zeta^{}_h$ as contributed by the Higgs fluctuations in our framework.\  
We found that the primordial non-Gaussianity is contributed by both 
the Higgs self-interaction term and the nonlinear term,
where the former can be dominant and was not considered in the literature.\ 
We presented the predictions for the local non-Gaussianity $f_\rm{NL}^\text{local}$ 
from the parameter space of neutrino seesaw, as shown in Figs.\,\ref{fig:6} and \ref{fig:7}.\ 
We found that, for fairly modest values of Higgs self-coupling, 
the sensitivities to the non-Gaussianity of 
$f_\rm{NL}^\text{local}\hsm\!=\!O(1)$ 
could probe the seesaw scale 
$M\!\hsm\thicksim\hsm\!10^{13}\hp$GeV  
and can also have important interplay
with probing the light neutrino mass scale 
and mass ordering in the low-energy neutrino experiments.\ 
We further studied the dependence of the non-Gaussianity on the Higgs self-coupling $\lambda$
(as shown in Fig.\,\ref{fig:8}).\ 
We found that the non-Gaussianity measurements are also
sensitive to the SM Higgs self-coupling $\lambda$ 
at the neutrino seesaw scale and thus provide complementary probes
of $\lambda$ to the on-going LHC collider experiment.\ 

\vs

In the near future, combining the neutrino data 
(such as those from the oscillation experiments JUNO\,\cite{JUNO:2015zny} and DUNE\,\cite{DUNE:2016hlj}, 
and the neutrinoless double-$\beta$ decay experiments\,\cite{Dolinski:2019nrj}) 
with the improved cosmological non-Gaussianity measurements 
(such as those from  DESI\,\cite{DESI:2016fyo}, CMB-S4\,\cite{Abazajian:2019eic}, 
Euclid\,\cite{Euclid},  SPHEREx\,\cite{SPHEREx:2014bgr}, LSST\,\cite{LSST}, and SKA\,\cite{SKA} experiments)
is expected to provide a more sensitive probe of the high-scale neutrino seesaw mechanism through our approach.\
The main idea and results of this work are summarized 
in the companion Letter paper\,\cite{lett}.

\vspace*{5mm}
\noindent 
{\large\bf Acknowledgments}
\\[1mm] 
We thank Xingang Chen, Misao Sasaki, Zhong-Zhi Xianyu, and Yi Wang for useful discussions.\  The research of H.\,J.\,H., L.\,S.\ and J.\,Y.\ was supported in part by the National Natural Science Foundation of China (Grant Nos.\,12175136, 12435005 and 11835005) and by the Shenzhen Science and Technology Program (Grant No.\ JCYJ20240813150911015).\ C.\,H.\ acknowledges support from the Sun Yat-Sen University Science Foundation, the Fundamental Research Funds for the Central Universities at Sun Yat-sen University under Grant No.\,24qnpy117, the National Key R{\&}D Program of China under Grant 2023YFA1606100, the National Natural Science Foundation of China under Grants No.\,12435005, and  the Key Laboratory of Particle Astrophysics and Cosmology (MOE) 
of Shanghai Jiao Tong University.\
This work was supported in part by the State Key Laboratory of Dark Matter Physics at Shanghai Jiao Tong University.

\newpage
\appendix

\noindent
{\Large\bf Appendix}	

\vspace*{-2mm}

\section{\hspace*{-1.5mm}Evolution of Higgs Field after Inflation}
\label{app:A}
\label{A Detailed Discussion about Higgs Evolution after inflation}

In this Appendix, we provide further technical derivations for the evolution of the Higgs field after inflation
to support the analysis of 	Section\,\ref{sec:2}.\ 
For the sake of convenience, we rescale all physical quantities  
according to their dimensions by the Hubble parameter $H_\text{inf}^{}$ 
during the inflation.\ Thus, they become dimensionless 
under the following rescaling:\footnote{%
In this rescaling, we have also included a numerical factor $\frac{\,3\,}{2}$
for convenience, such that the initial time
$t_0^{}$ defined below Eq.\eqref{eq:a-H} will simply become $t_0^{}\!=\! 1\hp$.} 
\begin{equation}
\label{eq:dimensionless}
\dis 
h\rightarrow \big(\!\Fr{3}{2}H_\text{inf}^{}\big)\hp h \hp,~~~~
t\rightarrow  \big(\!\Fr{3}{2}H_\text{inf}^{}\big)^{\!-1} t \hp,~~~~ 
\dot{h} \rightarrow \big(\!\Fr{3}{2}H_\text{inf}^{}\big)^{\!2} \dot{h}\,.
\end{equation}
After inflation and before the completion of reheating, we assume the Universe is matter-dominated.\ 
Hence, the scale factor $a(t)$ and the Hubble parameter $H(t)$ take the following forms:
\begin{equation}
\label{eq:a-H}
a(t)=\dis \!\bigg(\!\frac{t}{\,t_0^{}\,}\!\bigg)^{\!\!2/3},
~~~~~
H(t)=\frac{2}{\,3\hp t\,} \,,
\end{equation}
where $t_0^{}$ represents the initial time of matter domination, and the Hubble parameter at $t_0^{}$ 
satisfies $H(t_0^{})\!=\!\frac{2}{\,3\hp t_0^{}} \!=\! H_{\inf}^{}\hp$.\ 
The equation of motion for the Higgs field $h(t)$ is given by
\begin{equation}
\label{eom_Higgs}
\ddot{h}(t)+\frac{\,2\,}{t} \dot{h}(t)+\lambda h^{3}(t)=0 \hp.
\end{equation}
We can define the conformal time $\tau$ during the matter-dominated stage as follows:  
\begin{equation}
\tau=\int\!\!\! \frac{\d t}{\,a(t)\,} = 3\hp t^{1/3} \hp.
\end{equation}
The definition of conformal time used here differs from that used in the calculation of correlation functions.\ 
During inflation, conformal time is always negative, with values approaching zero, which corresponds to 
an infinitely distant future.\ However, in the present context of a different cosmic expansion, conformal time 
is positive.\ These two definitions can be related by a simple constant shift.\ Thus, we can express the 
scale factor $a$ in terms of conformal time $\tau$ as $\hp a(\tau) \!=\! \frac{1}{\,9\,}\tau^2$.\ 
Based on this conformal time, we define the following: 
\begin{align}
\varphi &\equiv a\hp h=t^{2/3}h=\frac{1}{\,9\,}\tau^2 h \,,
\nn\\
\varphi' &\equiv \frac{\,\d\varphi\,}{\,\d\tau\,}\hp,  \hspace*{6mm}
\varphi''\equiv \frac{\d^2\varphi}{\,\d\tau^2\,}  \hp, 
\\
a' &\equiv\frac{\d\hp a}{\,\d\tau\,}\hp,  \hspace*{6mm}  
a''\equiv\frac{\,\d^2a\,}{\,\d\tau^2\,} \,. 
\nn 
\end{align}
For the following derivations, we will use the two relations involving derivatives of the scale factor $a(\tau)\,$,
\begin{equation}
\frac{\,a'\,}{a}\!=\!\frac{2}{\,\tau\,}\hp, \hspace*{6mm}
\frac{~a''\hp}{a}\!=\!\frac{2}{\,\tau^2\,} \hp.
\end{equation}
We then obtain the equation of motion for the redefined field $\varphi(\tau)$,
\begin{equation}
\label{eof_varphi}
\varphi''(\tau)-\frac{\,a''}{a}\varphi(\tau)+\lambda\hp\varphi(\tau)^3=0 \hp .
\end{equation}
For any relevant quantity $A\hp$, we denote $A_0$ as $A(t\!=\!t_0^{})\!=\!A(\tau\!=\!\tau_0^{})$, 
where $t_0^{}$ and $\tau_0^{}$ denote the initial physical time and its corresponding conformal time, respectively.\ 
We set the initial conditions as follows:
\begin{align}
& t_0=1\hp,~~~~ \tau_0=3\hp,~~~~ 
a_0^{}=1 \hp,
\nn\\
& h_0^{} = h_\text{inf}^{}\hp,~~~~ 
\dot{h}_0^{} = 0 \hp,~~~~ 
\varphi_0^{} = h_0^{}\hp,~~~~ \varphi_0'=\frac{2}{\,3\,}h_0^{}\hp.
\end{align}
In the above, 
$t_0^{}\!=\!1$ implies $H(t_0^{})\!=\!H_\text{inf}^{}$ and $h_0^{}$ 
is the initial value of $h(t)$ which is determined at the end of inflation.\ 
We take $h_0^{}\!>\!0$ as an example for the following calculation, but the results can be readily generalized 
to the case of $h_0^{} \!<\! 0\hp$.\ 
When the conformal time is small, the term $-\frac{2}{\,\tau^2\,}\varphi$ dominates, and the other term $\lambda \varphi^3$ 
can be neglected for small $\lambda\hp$.\  Hence, we could first solve the following equation:
\begin{equation}
\label{1st_equation}
\varphi'' \!-\! \frac{2}{\,\tau^2\,}\varphi=0\hp ,\,
\end{equation}
with the initial conditions $\varphi_0^{}\!=\!h_0^{}$ and $\varphi'_0\!=\!\frac{2}{\,3\,}h_0^{}$\hp.\ 
We thus derive the following solution:
\begin{equation}
\label{1st_solution}
\varphi(\tau)=\frac{1}{\,9\,}h_0\tau^2.
\end{equation}

As the conformal time increases, the term $\lambda \varphi^3$ becomes increasingly significant.\ 
When $\lambda \varphi(\tau)^3 \!=\! \frac{2}{\,\tau^2\,}\varphi(\tau)\hp$, 
we identify the conformal time cutoff $\tau_\text{cut}^{}$ as follows:
\begin{equation}
 \label{def_of_taucut}
\tau_\text{cut}=3^{2/3}2^{1/6}\lambda^{-1/6}h_0^{-1/3}.
\end{equation}
Then, we deduce 
$\varphi_{\rm{cut}}^{}\!\equiv \varphi(\tau_\rm{cut}^{})\!=\!\frac{1}{\,9\,}h_0^{}\tau_\rm{cut}^2$ and $\varphi'_\text{cut}\!\equiv\!\varphi'(\tau_\text{cut})\!=\!\frac{2}{\,9\,}h_0^{} \tau_\text{cut}^{}$.\ 
Note that the solution \eqref{1st_solution} is valid when $\tau\!\leqq\!\tau_\text{cut}^{}$.

At the late-time stage $\tau \!\gg\! \tau_\text{cut}^{}$, 
the term $\lambda \varphi^3$ dominates and the term $-\frac{2}{\tau^2}\varphi$ may be neglected.\ 
In consequence, we simplify Eq.\eqref{eof_varphi} as follows:
\begin{equation}
\label{2nd_equation}
\varphi''(\tau)+\lambda \hp\varphi^3(\tau)=0 \hp .
\end{equation}
We can define a conserved quantity as the conformal ``energy'':
\begin{equation}
\frac{1}{\,2\,}\varphi^{\prime\hp 2}+\frac{\,\lambda\,}{4}\varphi^4=E \hp,
\end{equation}
where we could define $E_K^{}\!\!=\hsm\!\frac{1}{2}{\varphi'^2}$ 
as the conformal kinematic energy and $E_V^{}\!\!=\hsm\!\frac{\lambda}{4}\varphi^4$ as the conformal potential energy.\ 
This equation implies that the field $\varphi$ oscillates within the potential $\frac{\,\lambda\,}{4}\varphi^4$ 
without attenuation or damping.\ The solution to Eq.\eqref{2nd_equation} is given by an elliptic sine function, 
which can be approximated as follows:
\begin{equation}
\label{2nd_form}
\varphi(\tau)=\varphi_{\max}^{}
\cos\!\hsm\left[\!\frac{\,\Gamma^2(3/4)\,}{\sqrt{\pi\,}}
\sqrt{\lambda\,}\hp\varphi_{\max}^{}(\tau\!-\!\tau_\rm{cut}^{})\!+\hsm\theta \right] \!,
\end{equation}
where $\varphi^{}_{\max}$ is the oscillation amplitude and $\theta$ is the phase, 
both of which are determined by initial conditions.\ 
Matching the ``energy'' at the time $\tau_\text{cut}^{}$, 
we could determine the phase $\theta$ and the oscillation amplitude $\varphi_{\max}^{}\,$ as follows:
\begin{equation}
\begin{cases}
\dis\frac{\,E_K^{}\,}{\,E_V^{}\,} = 4 = \tan^2\!\theta\hp,
\\[1mm]
\dis 
E=E_K^{}\!+\!E_V^{}=\frac{\,5\lambda\,}{4}\varphi_\rm{cut}^4
\!=\hsm\frac{\,\lambda\,}{4}\varphi_{\max}^4 \hp,
\end{cases}
\hspace*{-4mm}
\Longrightarrow~~  
\begin{cases}
\theta=-\arctan 2\simeq -1.1 \hp,
\\[1mm]
\dis 
\varphi_{\max}^{}=2^{1/3}3^{-2/3}5^{1/4}\!\hsm\(\!\frac{\,h_0^{}\,}{\lambda}\!\)^{\!\!1/3} .
\end{cases}
\end{equation}
Then, we derive an analytic solution as a piecewise function,  
\begin{equation}
\varphi(\tau)=\left\{\begin{array}{ll}
\!\!\frac{1}{\,9\,}h_0^{} \tau^2,\,  & \hspace*{4mm} 
\tau \!\leqq\! \tau_\rm{cut}^{}\hp, 
\\[1.5mm]
\dis 
\!\!\varphi_{\max}^{}\hsm\cos\!\hsm\left[\hsm\frac{\,\Gamma^2(3/4)\,}{\sqrt{\pi}}\sqrt{\lambda\,}
\varphi_{\max}^{}(\tau\!-\!\tau_\rm{cut}^{}) \!+\!\theta \right]\hsm\!, &
\hspace*{4mm} 
\tau\!>\!\tau_\rm{cut}^{} \,.
\end{array}\right.
\end{equation}
With these, we derive the solution of Higgs field $h(t)$ 
in terms of physical time $t$ as follows:
\begin{equation}
\label{eq:h(t)}
h(t)=\frac{\,\varphi(3t^{\frac{1}{3}})\,}{a(t)} =\!\left\{
\begin{array}{ll}
\!h_0^{}\hp, & \hspace*{4mm}  t \!\leqq\! t_\rm{cut}^{}\hp,  
\\[1.5mm]
\dis 
\!\!A \bigg(\!\!\frac{~h_0^{}\,}{\lambda}\!\!\bigg)^{\!\!\frac{1}{3}} \hsm t^{-\frac{2}{3}}\hsm 
\cos\!\hsm\left[\hsm\lambda^{\frac{1}{6}}h_0^{\frac{1}{3}}\omega\big(t^{\frac{1}{3}} \!-\! t_\rm{cut}^{\frac{1}{3}}\big)
\!+\hsm \theta \right]\hsm\!, 
& \hspace*{4mm}             t \!>\! t_\rm{cut}^{}\hp, 
\end{array} \right.
\end{equation}
where $t_\rm{cut}^{} \!=\!\(\frac{1}{3}\tau_\rm{cut}^{}\)^{\!3}$, and the coefficients $A$ and $\omega$ are given by 
\beqs 
\begin{align}
A &= 2^{1/3}3^{-2/3}5^{1/4}\simeq 0.9\hp,
\\
\omega & =\frac{\,\Gamma^2(3/4)\,}{\sqrt{\pi}}2^{1/3}3^{1/3}5^{1/4}\simeq 2.3 \hp. 
\end{align}
\eeqs 
Using the definition of $\tau_\rm{cut}^{}$ in Eq.\eqref{def_of_taucut},
we could convert the term $\lambda^{\frac{1}{6}}h_0^{\frac{1}{3}}\omega\hp t_\rm{cut}^{\frac{1}{3}}$ of Eq.\eqref{eq:h(t)} 
into a constant phase,
\begin{equation}
\lambda^{\frac{1}{6}}h_0^{\frac{1}{3}}\omega\hp t_\rm{cut}^{\frac{1}{3}}
= \frac{1}{3}\lambda^{\frac{1}{6}}h_0^{\frac{1}{3}}\omega\hp\tau_\rm{cut}^{} = 3^{-\frac{1}{3}}2^{\frac{1}{6}}\omega \,.
\end{equation}
Finally, we further derive Eq.\eqref{eq:h(t)}  as follows: 
\begin{equation}
\label{h0 theoretical solution}
h(t)=\left\{
\begin{array}{ll}
\! h_0^{}\,,  & \hspace*{4mm}    t \!\leqq\! t_\rm{cut}^{}\hp,  
\\[1.5mm]
\dis\!\! A\!\(\!\frac{\,h_0^{}\,}{\lambda}\!\)^{\!\!\frac{1}{3}} \!t^{-\frac{2}{3}}\hsm
\cos\!\!\left[\lambda^{\frac{1}{6}}h_0^{\frac{1}{3}}\omega\, t^{\frac{1}{3}}
\!+\hsm \theta^{\prime}\right]\hsm\!,
& \hspace*{4mm}  
t \!>\! t_\rm{cut}^{}\hp,
\end{array}\right.
\end{equation}
where the phase 
$\hp\theta^{\prime}\!=\!-\lambda^{1/6}h_0^{1/3}\omega\hp t_\text{cut}^{1/3}\hsm+\theta\simeq\! -2.9\hp$.\ 
Recall that we employ the factor $\frac{\,3\,}{2}H_\rm{inf}^{}$ to define dimensionless 
physical quantities in Eq.\eqref{eq:dimensionless}.\ 
We can recover the dimensions of relevant physical variables by the following rescaling:
\begin{align}
& t\to \big(\!\Fr{3}{2}H_\text{inf}\big)\hp t\hp,~~~~
t_0^{}\to  \big(\!\Fr{3}{2}H_\rm{inf}^{}\big)\hp t_0^{}\hp,~~~~
t_\rm{cut}^{}\to \big(\!\Fr{3}{2}H_\rm{inf}^{}\big)\hp t_\rm{cut}^{}\hp,
\nn\\[1mm]
& h(t)\to h(t)/\hsm\big(\!\Fr{3}{2}H_\rm{inf}^{}\big),~~~~
h_0^{}\to h_0^{}/\hsm\big(\!\Fr{3}{2}H_\rm{inf}^{}\big),
\end{align}
under which the form of the formula \eqref{h0 theoretical solution} remains unchanged.

\section{\hspace*{-1.5mm}Non-Gaussianity from Three-Point Correlation Functions}
\label{app:B}

The Planck Collaboration gives a list of primordial non-Gaussianity measurements with various shapes\,\cite{Planck9}.\ 
For the present study, we will use the shape templates of local-type, equilateral-type, and orthogonal-type non-Gaussianity measurements.\  
	
For the local-type non-Gaussianity, we compute the three-point correlation function:
\begin{equation}
\begin{aligned}
& \langle \Phi_{\mathbf{k}_1}^{}\!\Phi_{\mathbf{k}_2}^{}\!\Phi_{\mathbf{k}_3}^{}\rangle'_{\rm{local}}
=B_{\Phi}^{\rm{local}}\!\(k_{1}, k_{2}, k_{3}\)\! 
\\ 
& 
=2 A^{2} f_{\mathrm{NL}}^{\mathrm{local}}\!		
\biggl(\!\frac{1}{\,k_{1}^3k_{2}^3\,}\!+\!\frac{1}{\,k_{2}^3 k_{3}^3\,}\!+\!\frac{1}{\,k_{3}^3k_{1}^3\,}\!\biggr)
\!\equiv 2 A^{2} f_{\mathrm{NL}}^{\mathrm{local}} F^\text{local}(k_1^{},k_2^{},k_3^{}) \hp, 
\end{aligned}
\end{equation}
where $\Phi$ denotes the Bardeen gravitational potential, which is related to the comoving curvature perturbation $\zeta$ via 
$\Phi\!\equiv\!\frac{3}{\,5\,}\hp\zeta$ on superhorizon scales.\ 
The power spectrum $P_{\Phi}^{}\!=\!\langle\Phi_{\mathbf{k}_1}^{}\!\Phi_{\mathbf{k}_2}^{}\rangle'$ is given by 
$P_{\Phi}^{}\!=\!{A}/{k^3}$, where $A$ is the normalization constant.\ 
Given $\Phi\!=\! \frac{3}{\,5\,}\zeta$ and 
$P_{\Phi}^{}\!=\!\langle \Phi_{\mathbf{k}_1}^{}\!\Phi_{\mathbf{k}_2}^{}\rangle'
\!=\!{A}/{k^{3}}$ (under the approximation $n_s^{}\!\simeq\! 1$), 
we derive the normalization constant $A\!=\!2\pi^2(\frac{3}{5})^2P_{\zeta}^{}\hp$.\  
In the above formula and hereafter, we have defined
$\langle \Phi_{\mathbf{k}_1}^{}\!\Phi_{\mathbf{k}_2}^{}\!\Phi_{\mathbf{k}_3}^{}\rangle'$
as the 3-point correlation function without including 
the $\delta$ function of momentum conservation, 
$\langle \Phi_{\mathbf{k}_1}^{}\!\Phi_{\mathbf{k}_2}^{}\!\Phi_{\mathbf{k}_3}^{}\rangle
\!=\! (2\pi)^3\delta^3\hsm(\mathbf{k}_{1}\!+\!\mathbf{k}_{2}\!+\!\mathbf{k}_3)
\langle \Phi_{\mathbf{k}_1}^{}\!\Phi_{\mathbf{k}_2}^{}\!\Phi_{\mathbf{k}_3}^{}\rangle'$.  

\vs 

For the equilateral-type non-Gaussianity, the three-point correlation function is computed as follows:
\begin{equation}
\begin{aligned}
&\langle \Phi_{\mathbf{k}_1}^{}\!\Phi_{\mathbf{k}_2}^{}\!\Phi_{\mathbf{k}_3}^{}\rangle'_{\rm{equil}}
=B_{\Phi}^{\rm{equil}}(k_{1}^{}, k_{2}^{}, k_{3}^{}) 
\\ 
& = 6 A^{2}\hsm f_{\mathrm{NL}}^{\mathrm{equil}} 		
\biggl[\hsm -\hsm\frac{1}{\,k_{1}^3 k_{2}^3\,}\!-\!\frac{1}{\,k_{2}^3 k_{3}^3\,}\!-\!\frac{1}{\,k_{3}^3 k_{1}^3\,}
\!-\!\frac{2}{\,(k_{1}k_{2}k_{3})^2\,}\!+\!\(\!\frac{1}{\,k_1^{}k_{2}^2k_{3}^3\,}\!+\!5\,\rm{perm.}\!\)\!\biggr]{}
\\
&\equiv 6 A^{2} f_{\mathrm{NL}}^{\mathrm{equil}} F^\text{equil}(k_1^{},k_2^{},k_3^{})\hp.
\end{aligned}
\end{equation}
For the orthogonal-type non-Gaussianity, we can derive the three-point correlation function as follows:
\begin{equation}
\begin{aligned}
&\langle \Phi_{\mathbf{k}_1}^{}\!\Phi_{\mathbf{k}_2}^{}\!\Phi_{\mathbf{k}_3}^{}\rangle'_{\rm{ortho}}
= B_{\Phi}^{\rm{ortho}}(k_{1}^{}, k_{2}^{}, k_{3}^{}) 
\\ 
& =6 A^{2} f_{\mathrm{NL}}^{\mathrm{ortho}}		
\biggl[\hsm -\frac{3}{\,k_{1}^3 k_{2}^3\,}\!-\!\frac{3}{\,k_{2}^3 k_{3}^3\,}\!-\!\frac{3}{\,k_{3}^3 k_{1}^3\,}
\!-\!\frac{8}{\,(k_{1}k_{2}k_{3})^2\,}\!+\!\(\!\frac{3}{\,k_{1}^{}k_{2}^2k_{3}^3\,}\!+\!5\,\rm{perm.}\!\)\!
\biggr]{}
\\
&\equiv6 A^{2} f_{\mathrm{NL}}^{\mathrm{ortho}}F^\text{ortho}\hsm (k_1^{},k_2^{},k_3^{})\hp.
\end{aligned}
\end{equation}

\vs 
 
Next, we can use the above templates to analyze a three-point correlation function of the curvature perturbation 
$\langle \zeta_{{\kb_1}}^{}\zeta_{{\kb_2}}\zeta_{{\kb_3}}\rangle'$ through
\begin{equation}
\label{fNL_S}
\begin{aligned}
& \langle \Phi_{{\kb_1}}^{}\Phi_{{\kb_2}}^{}\Phi_{{\kb_3}}^{}\rangle'
=\bigg(\!\frac{\,3\,}{5}\!\bigg)^{\!\!3}\hsm\langle \zeta_{\mathbf{k_1}}\zeta_{\mathbf{k_2}}\zeta_{\mathbf{k_3}}\rangle'
\\
&= 2A^2 \!\left[
f_{\rm{NL}}^{\mathrm{local}} F^\rm{local}(k_1^{},k_2^{},k_3^{}) \!+\!
3f_{\rm{NL}}^{\rm{equil}} F^\rm{equil}(k_1^{},k_2^{},k_3^{}) \!+\! 
3f_{\rm{NL}}^{\rm{ortho}}F^\rm{ortho}(k_1^{},k_2^{},k_3^{}) \hsm\right]\!,  
\end{aligned}
\end{equation}
where $F^\rm{local}(k_1^{},k_2^{},k_3^{})$, 
$F^\rm{equil}(k_1^{},k_2^{},k_3^{})$, and $F^\rm{ortho}(k_1^{},k_2^{},k_3^{})$ 
are functions of the external momenta $(k_1^{},k_2^{},k_3^{})$.\ 
In the present analysis, we find that the three-point correlation function 
$\langle \zeta_{{\kb_1}}^{}\!\zeta_{{\kb_2}}^{}\!\zeta_{{\kb_3}}^{}\rangle'$ 
can be expressed as the sum of these three shape templates with relevant coefficients 
$f_{\rm{NL}}^{}$.

We can directly solve the coefficients $f_{\rm{NL}}^{}$ as follows:
\begin{equation}
\label{fNL_complete_result}
f_{\mathrm{NL}}^{\mathrm{equil}}\hsm =\hsm\Fr{5}{\,12\,}P \hp,~~~~
f_{\mathrm{NL}}^{\mathrm{ortho}} \hsm =\hsm -\Fr{1}{\,12\,}P\hp,~~~~ 
f_{\mathrm{NL}}^{\mathrm{local}} \hsm =\! \Fr{1}{\,2\,}\!\left[P(N_e\!-\!\gamma\!+\!\Fr{\,7\,}{3})\!+\!Q\right]\!, 
\end{equation}
where the quantities $P$ and $Q$ are given by 
\beqs 
\begin{align}
P& = \dis -\frac{\,20\hp z_1^3H^3}{\,3(2\pi)^4 P_{\zeta}^{2}\,}\frac{\,\lambda\hp\bar{h}\,}{\,2H\,} \hp,
\\
Q&=\dis \frac{20\hp z_1^3H^3}{\,3(2\pi)^4 P_{\zeta}^{2}\,}\frac{\,z_2 H\,}{\,4z_1^{}\,} \hp. 
\end{align}
\eeqs 

We may also compare our results with the literature\,\cite{Wands10040818,Ichikawa:2008ne,DeSimone12106618}\cite{Litsa201111649}  
which considered the special case of $R\!=\!1$ and ignored the contribution from Higgs self-coupling.\ 
In this case, given $R\!=\!|z_1|\big({P_{\delta h}^{}}/{P_\zeta^{}}\big)^{1/2}$, 
we obtain $\big({P_{\delta h}^{}}/{P_\zeta^{}}\big)^{1/2}\!\!=\!R\hsm /|z_1^{}|\hp$.\ 
Under the assumption that Higgs fluctuation $\delta h$ is scale invariant, i.e., $P_{\delta h}^{1/2}\!=\!H/(2\pi)$, 
we deduce the following result:
\begin{equation}
\begin{aligned}
f_{\mathrm{NL}}^{\mathrm{local}}&=\!\frac{\,Q\,}{2}\!
= \frac{5z_1^3H^3}{\,6(2\pi)^4 P_{\zeta}^{2}\,}\frac{\,z_2^{}H \,}{z_1^{}}
=\frac{5}{6} \frac{P_{\delta h}^{2}}{P_{\zeta}^{2}} z_1^2 z_2^{}
=R^4\frac{\,5z_2^{}\,}{\,6z_1^2\,} \,.
\end{aligned}
\end{equation}
Using $z_1^{}\!=\!-{\Gamma'_0}/{(6\hp\Gamma_0^{})}\hp$ and 
$z_2^{}\!=\!- \hsm{\big(\Gamma_0^{}\Gamma''_0\!-\!\Gamma'_0 \Gamma'_0\big)}/(6\hp\Gamma_0^2)$ 
as defined in Section\,\ref{sec:4.1}, and the condition that $R\!=\!1$, we thus derive the local non-Gaussianity,
\begin{equation}
f_{\mathrm{NL}}^{\mathrm{local}}=\frac{5z_2^{}}{\,6z_1^2\,}
=5\!\(\!1\!-\hsm \frac{\,\Gamma''_0\Gamma_0^{}\,}{\,(\Gamma'_0)^2\,}\!\) \!,
\end{equation}
which agrees with the literature\,\,\cite{Wands10040818,Ichikawa:2008ne,DeSimone12106618}\cite{Litsa201111649}.

\section{\hspace*{-1.5mm}Analysis of NG Dependence on the Higgs Self-Coupling}
\label{app:C}
\label{app:CC}

The local-type non-Gaussianity $f_{\mathrm{NL}}^{\mathrm{local}}$ in Eq.\eqref{fNL_complete_result} 
consists of two distinct contributions: the Higgs self-coupling term $f_{\mathrm{NL}}^{\mathrm{local}}(\mathrm{HSC})$ 
generated by the three-point correlation function of Higgs self-interactions in Eq.\eqref{3pth_SM0} and the nonlinear term $f_{\mathrm{NL}}^{\mathrm{local}}(\mathrm{NT})$ arising from the substitution of $\delta h(\mathbf{k})$ 
with the nonlinear component $\frac{1}{2}z_2^{}\delta h^2$ in Eq.\eqref{3pth_NL}.\ 
These contributions are expressed as follows:
\begin{subequations}
\begin{align}
f_{\mathrm{NL}}^{\mathrm{local}}(\mathrm{HSC}) 
& = -\frac{5 H_{\mathrm{inf}}^2 }{\,3(2\pi)^4 P_{\zeta}^2\,}
\bigg(\!N_e \!-\! \gamma \!+\! \frac{\,7\,}{3}\hsm\bigg)\hsm z_1^3 \lambda \hp\bar{h} \,, 
\label{HSC} 
\\
f_{\mathrm{NL}}^{\mathrm{local}}(\mathrm{NT}) 
& = \frac{5 H_{\mathrm{inf}}^4}{\,6(2\pi)^4 P_{\zeta}^2\,} z_1^2 z_2^{}\,. 
\label{NT}
\end{align}
\end{subequations}
In this Appendix, we give an analysis to estimate the $\lambda$ dependence of the above non-Gaussianity
contributions $f_{\mathrm{NL}}^{\mathrm{local}}(\mathrm{HSC})$ and
$f_{\mathrm{NL}}^{\mathrm{local}}(\mathrm{NT})$.

\vspace*{1mm}
\subsection*{C.1 \,Dependence of \boldsymbol{$f_{\mathrm{NL}}^{\mathrm{local}}(\mathrm{HSC})$} on Higgs Self-Coupling} 
\vspace*{1.5mm}

The $\lambda$ dependence of $f_{\mathrm{NL}}^{\mathrm{local}}(\mathrm{HSC})$ in Eq.\eqref{HSC} arises 
through the three parameters $z_1^{}(\lambda)$, $\lambda$, and $\bar{h}(\lambda)\hp$,
\begin{equation}
f_{\mathrm{NL}}^{\mathrm{local}}(\mathrm{HSC}) \propto \lambda\hp z_1^3(\lambda)\hp\bar{h}(\lambda) \,.
\end{equation}
In the above, the root-mean-square value of the Higgs field $\bar{h}$ is given by Eq.\eqref{hsquaremean},  
from which we deduce the following:
\begin{equation}
\label{hbarlambda}
\bar{h} = \left[\hsm\int_{-\infty}^{+\infty} \!\!\! \d h\hp h^2 \rho_{\mathrm{eq}}^{}\hsm (h) \hsm\right]^{\!\!\frac{1}{2}} 
\!\!\simeq 0.363 \bigg(\!\hsm\frac{\,H_{\mathrm{inf}}\,}{\lambda^{1/4}}\!\hsm\bigg) 
\propto\, \lambda^{-\frac{1}{4}} \,.
\end{equation}
The parameter $z_1^{}$ is defined in Eq.\eqref{z1z2def} 
as follows:
\begin{equation}
z_1 = -\frac{1}{\,6\,} \frac{\,\Gamma'_0\,}{\Gamma_0^{}},
\end{equation}
where $\Gamma_0^{}$ 
is defined in Eq.\eqref{eq:Gamma0-p-pp} with 
$h_\text{inf}^{} \!=\hsm \bar{h}\hp$.\ 
Neglecting kinematic factors, we can approximate the total decay rate as in Eq.\eqref{decay_rate},
\begin{equation}
\Gamma_{\text{reh}} \simeq \frac{\,m_{\phi} M^2\,}{4\pi \Lambda^2}\!\! 
\left[\hsm 1 \!+\hsm \frac{1}{4} \!\(\!\hsm\frac{~y_\nu^{}h_{\mathrm{reh}}^{}\,}{M}\!\hsm\)^{\!\!2}\right]\!. 
\label{gammalambda}
\end{equation}

From the semi-analytical solution for the Higgs field $h(t)$ in Eq.\eqref{h0 theoretical solution formal}, 
and after averaging over a sufficiently large volume such that the factor 
$\cos^2(\omega_{\rm{reh}}^{} h_\text{inf}^{{1}/{3}}\!\hsm +\hsm \theta)$ is treated as $\frac{1}{\,2\,}\hp$, 
we find that $h_{\text{reh}}$ scales as follows:
\begin{equation}
h_{\text{reh}}^{}(\lambda) \,\propto\, h_{\text{inf}}^{\frac{1}{3}}\hp \lambda^{-\frac{1}{3}}_{} \,.
\label{hrehlambda}
\end{equation}
Given the condition $y_\nu^{}h_{\text{reh}}/M \!\!\ll\!\! 1$ in the parameter space under consideration, 
the $\lambda$ dependence of $\Gamma_0$ in Eq.\eqref{gammalambda} can be neglected for this estimate.\
The remaining $\lambda$ dependence in $z_1^{}$ arises from $\Gamma'_0\hp$, 
which is the derivative of $\Gamma^{}_\text{reh}$ with respect to $h_\text{inf}^{}\hp$.\ 
Using Eqs.\eqref{gammalambda} and \eqref{hrehlambda} for computing $\Gamma'_0\hp$, 
we obtain the $\lambda$ dependence of $\Gamma'_0$ and $z_1^{}\hp$,
\begin{equation}
\begin{aligned}
\label{z1lambda}
z_1^{} \propto \Gamma'_0(\lambda) 
\propto \left.\frac{\,\d \Gamma^{}_\text{reh}}{\,\d h_\text{reh}^{}\,} 
\frac{\,\partial h_\text{reh}^{}\,}{\,\partial h_\text{inf}^{}\,}
\right|_{h_{\text{inf}}^{}\hp =\hp \bar{h}} 
\propto h_{\text{reh}}(\lambda) \hp \bar{h}^{-\frac{2}{3}}\hp \lambda^{-\frac{1}{3}} 
\propto \lambda^{-\frac{7}{12}} \,,
\end{aligned}
\end{equation}
where in the last step we have used the mean-field approximation 
$h_\text{inf}^{} \!=\hsm \bar{h}\hp$ with Eq.\eqref{hbarlambda}.\ 

\

Given the $\lambda$ dependence of $z_1^{}(\lambda)$ in Eq.\eqref{z1lambda} and 
$\bar{h}(\lambda)$ in Eq.\eqref{hbarlambda}, we can thus deduce the $\lambda$ dependence of 
the non-Gaussianity $f_{\mathrm{NL}}^{\mathrm{local}}(\mathrm{HSC})\hp$, 
\begin{equation}
\begin{aligned}
f_{\mathrm{NL}}^{\mathrm{local}}(\mathrm{HSC}) \,\propto\, \lambda\hp z_1^3(\lambda)\hp\bar{h}(\lambda) 
\propto \lambda^{-1} \,.
\end{aligned}
\end{equation}
This explains why the predicted values of $f_{\mathrm{NL}}^{\mathrm{local}}(\mathrm{HSC})\hp$
reduce by about a factor $\Fr{1}{2}$ as the Higgs self-coupling varies from $\lambda\!=\!0.01$ to
$\lambda\!=\!0.02\hp$, 
as shown in the 5th row of Table\,\ref{tab:2}.\

\subsection*{C.2 \,Dependence of \boldsymbol{$f_{\mathrm{NL}}^{\mathrm{local}}(\mathrm{NT})$} on Higgs Self-Coupling}
\vspace*{1.5mm}

Next, we study the $\lambda$ dependence of  $f_{\mathrm{NL}}^{\mathrm{local}}(\mathrm{NT})$ 
from the nonlinear term (NT) contribution.\ 
From Eq.\eqref{NT}, we see that $f_{\mathrm{NL}}^{\mathrm{local}}(\mathrm{NT})$  depends on $\lambda$ 
through product $z_1^2z_2^{}$.\ 
Since $\lambda$ dependence of $z_1^{}$ is already given by Equation\,\eqref{z1lambda},
we will focus on the analysis of the $\lambda$ dependence of $z_2^{}\hp$.\ 
Eq.\eqref{z1z2def} gives the following expression of $z_2^{}\hp$:
\begin{equation}
z_2^{} = -\frac{1}{\,6\,} \!\hsm\left[\!\frac{\Gamma''_0}{\,\Gamma_0^{}\,} \hsm -\! 
\(\!\!\frac{\,\Gamma'_0\,}{\Gamma_0^{}}\!\!\)^{\!\!2}\right] \!,
\label{z2def}
\end{equation}
which shows that the $\lambda$ dependence of $z_2^{}$ mainly comes from 
$\Gamma''_0(\lambda)$ and $\Gamma'_0(\lambda)$.\ 
The $\lambda$ dependence of $\Gamma'_0$ is given in Eq.\eqref{z1lambda}.\ 
Thus, we analyze the $\lambda$ dependence of $\Gamma''_0$, which can be derived as follows: 
\begin{align}
\Gamma''_0 &= \left.\frac{\,\d^2\Gamma_\text{reh}\,}{\d h_\text{inf}^2}
\right|_{h_\text{inf}\hp =\hp\bar{h}} 
\!= \left.\frac{\,\d^2\Gamma_\text{reh}^{}\,}{\,\d h_\text{reh}^2\,}
\!\(\!\!\frac{\,\partial h_\text{reh}^{}\,}{\,\partial h_\text{inf}^{}\,}\!\!\)^{\hsm\!\!2}
\right|_{h_\text{inf}\hp =\hp \bar{h}} \!+\! 
\left.\frac{\,\d\Gamma_\text{reh}^{}\,}{\,\d h_\text{reh}\,}
\!\hsm\(\!\!\frac{\,\partial^2 h_\text{reh}^{}\,}{\partial h_\text{inf}^2}\!\!\)^{}\!
\right|_{h_\text{inf}\hp =\hp\bar{h}} 
\nn\\
&= \left.\left[\hsm C_1^{}\hsm\Big(h_\text{inf}^{-\frac{2}{3}}\lambda^{-\frac{1}{3}}\hsm\Big)^{\hsm\!2} 
\!\!+ C_2^{}\hp h_\text{reh}^{}\hp h_\text{inf}^{-\frac{5}{3}}\lambda^{-\frac{1}{3}}\hsm\right]
\!\right|_{h_\text{inf}\hp =\hp\bar{h}}^{} 
\,\propto\, \lambda^{-\frac{1}{3}} \hp,
\end{align}
where the coefficients $C_1^{}$ and $C_2^{}$ are independent of 
the Higgs self-coupling $\lambda\hp$, and
Eqs.\eqref{hrehlambda}, \eqref{gammalambda} and \eqref{hbarlambda} are used.\ 

From Eq.\eqref{gammalambda}, we can deduce the following 
scaling behaviors:
\begin{align}
\frac{\,\Gamma''_0\,}{\Gamma_0^{}} \sim\! \frac{1}{\,M^2\,}\hp, \hspace*{8mm}
\(\!\hsm\frac{\,\Gamma'_0\,}{\Gamma_0^{}}\!\!\)^{\!\!2} \!\sim\! 
\frac{1}{\,M^2\,}\frac{\,h_\text{reh}^2\,}{M^2}\,.
\end{align}
Since $h_\text{reh}^{}/M \!\ll\! 1\hp$, 
we can neglect the $\lambda$ dependence from the second term $(\Gamma'_0/\Gamma_0)^2\hp$ of Eq.\eqref{z2def} 
in comparison to its first term $\Gamma''_0/\Gamma_0^{}\,$.\ 
Hence, we can extract the $\lambda$ dependence of $z_2^{}$ as follows:
\begin{equation}
	\begin{aligned}
		\label{z2lambda}
		z_2(\lambda) \propto \Gamma''_0 \propto \lambda^{-\frac{1}{3}}.
	\end{aligned}
\end{equation}

Using Eqs.\eqref{z1lambda} and \eqref{z2lambda}, we can derive the $\lambda$ dependence of 
$f_{\mathrm{NL}}^{\mathrm{local}}(\mathrm{NT})$ as follows:
\begin{equation}
f_{\mathrm{NL}}^{\mathrm{local}}(\mathrm{NT}) \,\propto\, z_1^2 \hp z_2^{}
\,\propto \lambda^{-\frac{3}{2}} \,.
\end{equation}

In summary, we have estimated the $\lambda$ dependence of both contributions
$f_{\mathrm{NL}}^{\mathrm{local}}(\mathrm{HSC})$ and
$f_{\mathrm{NL}}^{\mathrm{local}}(\mathrm{NT})$ as follows:
\beq
\label{eq:fNL-lambda}
f_{\mathrm{NL}}^{\mathrm{local}}(\mathrm{HSC}) \propto \lambda^{-1}, 
\hspace*{8mm}
f_{\mathrm{NL}}^{\mathrm{local}}(\mathrm{NT}) \propto \lambda^{-\frac{3}{2}}.
\eeq
where we have used the condition $y_\nu^{} h_{\text{reh}}^{}/M \!\ll\! 1$ and have neglected 
kinematic factors in the inflaton decay rate $\Gamma_{\text{reh}}^{}$.\  
Hence, summed contribution to the non-Gaussianity, 
$f_{\mathrm{NL}}^{\mathrm{local}} \!=\! f_{\mathrm{NL}}^{\mathrm{local}}(\mathrm{HSC}) 
\!+\! f_{\mathrm{NL}}^{\mathrm{local}}(\mathrm{NT})$, 
decreases monotonically as the Higgs self-coupling $\lambda$ increases.\  
These analytical behaviors agree well with the numerical results presented in Table\,\ref{tab:2} and Fig.\,\ref{fig:8}.\
We also note that the above scaling behaviors of Eq.\eqref{eq:fNL-lambda} are derived by using the mean-field approximation
which are valid for the parameter space under consideration, but not for arbitrarily large or small coupling $\lambda\hp$.

\section{\hspace*{-1.5mm}Schwinger-Keldysh Propagators of Higgs Field}
\label{SK path integral for massless scalar}
\label{app:DD}
\vs 

The Klein-Gordon equation in the de Sitter spacetime 
for the massless scalar field is given as follows:
\begin{equation}
\Box u_{\mathbf{k}}^{} = 
\ddot{u}_{\mathbf{k}}^{}\hsm +\hsm 3 H \dot{u}_{\mathbf{k}}^{}
\!+\!\frac{\mathbf{k}^{2}}{\,a^{2}(t)\,} u^{}_{\mathbf{k}}=0 \,,
\end{equation}
where $H\hsm\!=\!{\dot{a}}/{a}\hp$ is constant and $a(t)\!=\!e^{Ht}$.\
We may consider the conformal coordinates with $\d\tau \!=\!\d t/{a(\tau)}$ and  
$a\!=\!-(H\tau)^{-1}\hsm$, and re-express the Klein-Gordon equation as follows:
\begin{equation}
	u''_{\mathbf{k}}-\frac{2}{\tau}u'_{\mathbf{k}}+\mathbf{k}^{2}u_{\mathbf{k}}=0 \hp.
\end{equation}
The normalized mode function of the massless scalar field in de Sitter spacetime with the Bunch-Davies vacuum 
can be solved as
\begin{equation}
u_{\mathbf{k}}^{}(\tau)=\frac{H}{\,\sqrt{2k^3\,}\,}(1\!+\hsm \ii\hp k\tau)e^{-\ii\hp k\tau} \,,
\end{equation}
where we denote $u'\!=\!{\d u}/{\d \tau}$ and $u''\!=\!{\d^2 u}/{\d \tau^2}$.\

\vs 
 
In the Schwinger-Keldysh (SK) path integral 
formalism\,\cite{Weinberg2005}\cite{Chen170310166}, 
the bulk-to-bulk propagators are defined as follows:
\beqs 
\begin{align}
G_{++}^{}(\mathbf{k};\tau_1^{},\tau_2^{}) & =  G_{>}^{}(\mathbf{k};\tau_1^{},\tau_2^{})\theta(\tau_1^{}\!-\!\tau_2^{})
+G_{<}^{}(\mathbf{k};\tau_1^{},\tau_2^{})\theta(\tau_2^{}\!-\!\tau_1^{}) \hp, 
\\
G_{+-}^{}(\mathbf{k};\tau_1^{},\tau_2^{}) & = G_{<}^{}(\mathbf{k};\tau_1^{},\tau_2^{}) \hp, 
\\
G_{-+}^{}(\mathbf{k};\tau_1^{},\tau_2^{}) & = G_{>}^{}(\mathbf{k};\tau_1^{},\tau_2^{}) \hp, 
\\
G_{--}^{}(\mathbf{k};\tau_1^{},\tau_2^{}) & =  G_{<}^{}(\mathbf{k};\tau_1,\tau_2^{})\theta(\tau_1^{}\!-\!\tau_2^{})\!+\!
G_{>}^{}(\mathbf{k};\tau_1^{},\tau_2^{})\theta(\tau_2^{}\!-\!\tau_1^{})\hp, 
\end{align}
\eeqs 
where the propagators $G_{>}^{}$ and $G_{<}^{}$ are given by
\beqs 
\begin{align}
G_{>}^{}(\mathbf{k};\tau_1,\tau_2) &= u_{\mathbf{k}}(\tau_1)u_{\mathbf{k}}^*(\tau_2)
=\frac{\,H^2\,}{\,2k^3\,}\left[1\!+\hsm \ii\hp k(\tau_1\!-\!\tau_2)\!+\!k^2\tau_1\tau_2\right]\!
e^{-\ii\hp k(\tau_1^{}\hsm -\tau_2^{})}\hp,
\\
G_{<}^{}(\mathbf{k};\tau_1,\tau_2) &= u_{\mathbf{k}}^*(\tau_1)u_{\mathbf{k}}^{}
(\tau_2^{}) = G_{>}^{*}(\mathbf{k};\tau_1^{},\tau_2^{}) \hp. 
\end{align}
\eeqs 

Additionally, the bulk-to-boundary propagator is a special propagator in which external legs terminated at the final slice, i.e.,   $\tau\!=\!\tau_f^{}\rightarrow0^-$,
\begin{equation}
G_{\pm}^{}(\mathbf{k},\tau) =  G_{\pm+}^{}(\mathbf{k};\tau,\tau_f) \hp,
\end{equation}
which means that the bulk-to-boundary propagator ($\tau\!=\!\tau_1^{}\!<\!\tau_2^{}\!=\!\tau_f^{}\rightarrow 0^{-}$) 
is defined based on the above bulk-to-bulk propagator.\ 
The bulk-to-boundary propagator includes one ``plus-type'' $G_{+}^{}(\mathbf{k},\tau)$ and 
one ``minus-type'' $G_{-}^{}(\mathbf{k},\tau)$,
\begin{subequations}
\begin{eqnarray}
\parbox{40mm}{\includegraphics{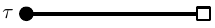}}\hspace*{-5mm}~&=& G_{+}\left(\mathbf{k},\tau\right)~,\\
\parbox{40mm}{\includegraphics{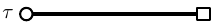}}\hspace*{-5mm}~&=& G_{-}\left(\mathbf{k},\tau\right)~,
\end{eqnarray}
\end{subequations}
where the square at one end of the propagator indicates its boundary point ($\tau_f^{}\hsm\to\hsm 0^-$), 
and a black dot and a white dot denotes ``plus-type'' and ``minus-type'', respectively.\ 
In addition, a shaded dot (often representing a vertex) indicates that the possibilities 
from both the plus- and minus-type propagators 
should be summed up,
\begin{equation}
\parbox{40mm}{\includegraphics{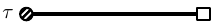}}\hspace*{-5mm}~
=G_{+}^{}(\mathbf{k},\tau) \hsm +\hsm G_{-}^{}(\mathbf{k},\tau) \hp.
\end{equation}
The plus- and minus-type bulk-to-boundary propagators take the following forms:
\beqs 
\begin{align}
G_{+}^{}(\mathbf{k},\tau) &=
\frac{H^2}{\,2k^3\,}\!\hsm\left[1\!-\!\ii\hp k(\tau\!-\!\tau_f^{})\!+\! k^2\tau\tau_f^{}\right]\!
e^{\ii\hp k(\tau-\tau_f^{})}
\!\simeq\! \frac{H^2}{\,2k^3\,}\!\left[1\!-\!\ii\hp k\tau\right]\!e^{\ii\hp k\tau}\hp,
\\
G_{-}^{}(\mathbf{k},\tau) &=
\frac{H^2}{\,2k^3\,}\!\hsm\left[1\!+\!\ii\hp k(\tau\!-\!\tau_f^{})\!+\! k^2\tau\tau_f^{}\right]\!
e^{-\ii\hp k(\tau-\tau_f^{})}
\!\simeq\! \frac{H^2}{\,2k^3\,}\!\left[1\!+\!\ii\hp k\tau\right]\!e^{-\ii\hp k\tau}\hp.
\end{align}
\eeqs

\vspace*{1.5mm}
\section{\hspace*{-1.5mm}Probability Distribution Function of Higgs Field}
\label{app:D}
\label{app:EE}
\label{Probability distribution for a field}

For the sake of convenience, we abbreviate the Hubble parameter during the inflation $H_\text{inf}^{}$ as $H$, namely,  
$H_\text{inf}^{}\!=\! H\hp$.\ 
A scalar field $\phi(\mathbf{x},t)$ with only 1 degree of freedom can be represented 
as the mode functions of long-wavelength part $\phi_L^{}(\mathbf{x},t)$ and 
short-wavelength part $\phi_S^{}(\mathbf{x},t)$\,\cite{STAROBINSKY1982175}\cite{Starobinsky1994}:
\begin{align}
\phi(\mathbf{x},t) & = 
\phi_L^{}(\mathbf{x},t) + \phi_S^{}(\mathbf{x},t)
\nn\\
&= \phi_L^{}(\mathbf{x},t)+\!\int\!\!\!\frac{\rm{d}^3k}{\,(2\pi)^{3}\,}
\theta\big(k\!-\!\epsilon a(t)H\big)
\!\!\left[a_{\mathbf{k}}^{}\phi_{\mathbf{k}}^{}(t)e^{-\ii\hp\mathbf{k}\cdot\mathbf{x}}
\!+\hsm a^{\dag}_{\mathbf{k}}\phi^{\ast}_{\mathbf{k}}(t) e^{\ii\hp\mathbf{k}\cdot\mathbf{x}}\right] \!.
\end{align}
The short-wavelength modes $\phi_{\mathbf{k}}^{}(t)$ are initially sub-horizon.\ 
Over the time, as the Universe expands, these modes are stretched and eventually cross the physical cutoff 
$\epsilon\hp a(t) H$, transitioning into the super-horizon modes $\phi_L^{}$.\ 
The super-horizon modes $\phi_L$ can be effectively treated as a classical stochastic field, 
obeying the Langevin equation: 
\begin{equation}
\dot{\phi}_L^{}\hsm (\mathbf{x},t)=-\frac{1}{3H}\frac{\partial V}{\partial \phi_L^{}} \!+\!f(\mathbf{x},t) \hp.
\end{equation}
In addition to the force from the potential of the field, the long-wavelength modes are also driven 
by an effective stochastic ``force'' $\!f\hp$,  
which is generated by freezing out the short-wavelength modes, 
\begin{equation}
f({\bf x}, t)=\int\!\!\!\frac{\d^3k}{\,(2\pi)^3\,}
\delta\big(k\!-\!\epsilon\hp a(t)H\big)\epsilon\hp a(t)H^2\!\left[a_{\mathbf{k}}^{} \phi_{\mathbf{k}}^{}(t) e^{-\ii\mathbf{k}\cdot\mathbf{x}}\!+\hsm a_{\mathbf{k}}^\dagger \phi_{\mathbf{k}}^{\ast}(t) 
e^{\ii\mathbf{k}\cdot\mathbf{x}}\right]\!.
\end{equation}
In order to obtain the Fokker-Planck equation for the one-point probability distribution function, 
we should derive the two-point correlation function of the stochastic ``force'' $f$.\ 
For this, we could derive the following relation for a canonical massless field $\phi_{\mathbf{k}}^{}$ 
in the late-time limit:
\begin{align}
& \left\langle\! \(\!a_{\mathbf{k}_1}^{}\!\phi_{\mathbf{k}_1}^{}(t_1^{})
e^{-\ii\hp\mathbf{k}_1^{}\cdot\mathbf{x}_1^{}} \!+\! a_{\mathbf{k}_1}^\dagger\!\phi_{\mathbf{k}_1}^{\ast} \!(t_1)e^{i\mathbf{k}_1\cdot\mathbf{x}_1}\!\)\!\!
\(\!a_{\mathbf{k}_2}\hsm\phi_{\mathbf{k}_2}^{}(t_2^{})e^{-\ii\hp\mathbf{k}_2^{}\cdot\mathbf{x}_2^{}}
\!+\! a_{\mathbf{k}_2}^\dagger\! \phi_{\mathbf{k}_2}^{\ast} \!(t_2^{})
e^{\ii\hp\mathbf{k}_2^{}\cdot\mathbf{x}_2^{}}\!\) \!\right\rangle
\nn\\
& = (2\pi)^3\delta^3\hsm (\mathbf{k}_1^{}\!-\!\mathbf{k}_2^{}) 
e^{\ii\mathbf{k}_1^{}\cdot \mathbf{x}_{12}^{}} \frac{H^2}{\,2k_1^3\,} .
\end{align}
Then, we can derive the two-point correlation function of $f\hp$ as follows:
\begin{equation}
\begin{aligned}
& \langle f({\bf x}_1, t_1) f({\bf x}_2, t_2) \rangle
=\!\int\!\!\!\frac{\mathrm{d}^3k_1}{\,(2\pi)^3\,}\!\hsm\(
\prod_{i=1}^{2}\!\delta(k_1^{}\!-\hsm \epsilon\hp a(t_i^{})H)\hp\epsilon\hp a(t_i^{})H^2\hsm\!\)
\!\!\frac{H^2}{\,2k_1^3\,}\hp e^{\ii\hp\mathbf{k}_1^{}\cdot \mathbf{x}_{12}^{}}
\\
&= \frac{1}{\,4\pi^2\,} \!\int_0^{+\infty}\!\!\!{\d}k_1^{}\!\(
\prod_{i=1}^{2}\!\delta(k_1^{}\!-\hsm \epsilon\hp a(t_i^{})H)\hp\epsilon\hp a(t_i^{})H^2\hsm\!\) 
\!\!\frac{\,H^2}{\,k_1^{}}j_0^{}(k_1^{}x_{12}^{})
\\
&= 
\frac{\,H^5\epsilon\hp j_0^{}\big(\epsilon a(t_1^{}) Hx_{12}^{}\big)\delta\big(\epsilon a(t_1^{})H\!-\!\epsilon a(t_2^{})H\big)\,}
{4\pi^2a(t_1^{})} \prod_{i=1}^{2}\! a(t_i) \,,
\end{aligned}
\end{equation}
where $\mathbf{x}_{12}^{}\!=\!{\bf x}_1^{}\!-\!{\bf x}_2^{}$,  $x_{12}^{}\!=\!|\mathbf{x}_{12}^{}|$, and $\,j_0^{}(z)\!=\!(\sin z)/{z}$ is the spherical Bessel function of zeroth order.\ 
Using the property of the $\delta$ function, 
we can simplify the above formula for 
the two-point correlation function of the stochastic ``force'':
\begin{equation}
\langle f({\bf x}_1^{}, t_1^{}) f({\bf x}_2^{}, t_2^{})\rangle 
= \frac{\,H^3}{\,4\pi^2\,}  j_0^{}\big(\epsilon a(t_1^{}) Hx_{12}^{})
\delta\hsm (t_1^{}\!-\!t_2^{}\big) \hp.
\end{equation}

If we take the one-point limit $\mathbf{x}_2^{}\!\to\!\mathbf{x}_1^{}$ in the position space with 
$j_0^{}(\epsilon aHx_{12}^{})\!\to\!1\hp$, the one-point probability distribution function 
for the field $\phi$ obeys the Fokker-Planck equation:
\begin{equation}
\label{1dimfkqua}
\begin{aligned}
\hspace*{-2.5mm}
\frac{\,\partial\rho(\phi,t)\,}{\partial t} =
\frac{1}{\,3H\,}\!\left\{\!\rho(\phi,t)\frac{\,\partial^2V(\phi)\,}{\partial \phi^2}
\!+\!\frac{\,\partial V(\phi)\,}{\partial\phi} 
 \frac{\,\partial \rho(\phi,t)\,}{\partial \phi}\!\hsm\right\}
\hsm +\hsm\frac{\,H^3}{\,8\pi^2\,}\frac{\,\partial^2\rho(\phi,t)\,}{\partial \phi^2}
\hp.~
\end{aligned}
\end{equation}
To solve Eq.\eqref{1dimfkqua}, we decompose the solution $\rho(\phi,t)$ in terms of the eigenfunctions 
$\{\Psi_n\}$ with eigenvalues $\{\Lambda_n\}$,
\begin{equation}
\label{1dimeigenfunction}
\rho(\phi,t) = e^{-v}\sum^{\infty}_{n=0}\!a_n^{}\hsm \Psi_n(\phi)e^{-\Lambda_n(t-t_0)} \,,
\end{equation}
where $v$ is defined as
\begin{equation}
v(\phi) \equiv \frac{\,4\pi^2V\hsm (\phi)\,}{3H^4} \,,
\end{equation}
and the coefficients $a_n^{}$ could be given by the initial condition of $\rho(t\!=\!t_0^{})\hp$, 
\begin{equation}
a_n^{} = \int\!\!\d\phi\, \rho(\phi,t_0)\hp e^{v(\phi)}\Psi_n(\phi) \hp .
\end{equation}
Then, we derive the corresponding equation for the eigenfunctions 
$\Psi_n$ and eigenvalues $\Lambda_n$ by substituting  Eq.\eqref{1dimeigenfunction} into Eq.\eqref{1dimfkqua}:
\begin{equation}
\label{1dim_eigen_equation}
\left[\!\frac{\partial^2}{\,\partial\phi^2\,} \!+\! \frac{\,\partial^2v\,}{\,\partial\phi^2\,}
\!-\!\(\!\frac{\,\partial\hp v\,}{\partial\phi}\!\)^{\!\!2}\right]\!\!\Psi_n(\phi) 
= - \frac{\,8\pi^2\,}{\,H^3\,}\Lambda_n\Psi_n(\phi) \hp,
\end{equation}
which is a typical Sturm-Liouville equation.\ 
Thus, all of the eigenvalues $\Lambda_n$ are non-negative and 
the eigenfunctions $\Psi_n(\phi)$ are orthonormal functions,
\begin{equation}
\label{normalization}
\int_{-\infty}^{+\infty}\!\!\d\phi\, \Psi_{n}^{}(\phi)\Psi_{n^{\prime}}^{}(\phi) =\delta_{n,n^{\prime}}^{} \hp.
\end{equation}
They satisfy the completeness condition,
\begin{equation}
\label{completeness_condition}
\sum_n\!\Psi_n(\phi)\Psi_n(\phi_0)=\delta(\phi\!-\!\phi_0^{}) \hp .
\end{equation}
There is a special eigenfunction with the eigenvalue $\Lambda_0^{}\!=\!0\hp$, 
\begin{equation}
\Psi_0^{}(\phi)=N_0^{}\hp e^{-v(\phi)} , 
\end{equation}
where $N_0^{}$ is the normalization constant.\ 
Thus, Eq.\eqref{1dimeigenfunction} can be reexpressed as follows: 
\begin{equation}
\rho(\phi,t)=\Psi_0(\phi)\sum^{\infty}_{n=0}\!b_n\Psi_n(\phi)e^{-\Lambda_n(t-t_0)} ,
\end{equation}
where coefficients $b_n^{}$ are also the normalization factors analogous to $a_n^{}$, 
defined as $b_n\equiv a_n/N_0\hp$.\ If the inflation lasts long enough, 
the system quickly approaches its equilibrium state.\ 
The only term that remains in the above summation is $\Psi_0(\phi)$ with $\Lambda_0\!=\!0\hp$.\ 
Thus, the equilibrium probability distribution $\rho_{\text{eq}}$ is derived as
\begin{equation}
\label{equilibrium_distribution}
 \rho_{\text{eq}}^{}(\phi)=\Psi_0(\phi)^2 = N_0^{2}\exp\hsm\!\(\!-\frac{\,8\pi^2V(\phi)\,}{\,3H^4\,}\hsm\!\)\!,
\end{equation} 
where the equilibrium probability distribution satisfies the normalization condition,
\begin{equation}
    \int\!\! \td\phi\, \rho_\text{eq}^{}\hsm(\phi) = 1 \hp.
\end{equation}

\vspace*{5mm}

\addcontentsline{toc}{section}{References}

\end{document}